\documentclass[12pt,leqno,twoside]{article}

\usepackage{amssymb} 
\usepackage{epsf} 

\font\teneufm=eufm10 at 12pt 
\font\seveneufm=eufm7 scaled\magstep1
\font\fiveeufm=eufm5 scaled\magstep1
\newfam\eufmfam
\textfont\eufmfam=\teneufm
\scriptfont\eufmfam=\seveneufm
\scriptscriptfont\eufmfam=\fiveeufm

\newcommand{\emps}[1]{\underbar{\em#1\/}} 
\newcommand{\la}{\langle }
\newcommand{\ra}{\rangle}
\newcommand{\f}{\varphi}
\newcommand{\Reals}{{\mathbb{R}}} 

\reversemarginpar

\begin{document}

\title{Visualizing some ideas about G\"odel-type rotating universes.}

\author{N\'emeti, I.,  Madar{\'a}sz, J. X., Andr\'eka, H.\ and Andai, A.}
\date{\today .}
\maketitle
\bigskip

Some kinds of physical theories describe what our universe looks
like. Other kinds of physical theories describe instead what the
universe could be like independently of the properties of the actual
universe. This second kind aims for the ``basic laws of physics'' in
some sense which we will not make precise here (but cf.\ e.g.\
Malament~\cite[pp.98-99]{Mal84}). The present paper belongs to the
second kind. Moreover, it is even more abstract than this, namely it
aims for visualizing or grasping some mathematical or logical
aspects of what the universe could be like.
\bigskip

The first six pages of this material are of a
``science-popularizing'' character in the sense that first we recall
a space-time diagram from Hawking-Ellis~\cite{Hawel} as ``God-given
truth'', i.e.\ we do not explain why the reader should believe that
diagram. Then we derive carefully in an easily understandable visual
manner an exciting, exotic consequence of that diagram: time-travel.
This applies to the first six pages. The rest of this work is of a
more ambitious
character. {\it The reader does not have to believe anything}%
\footnote{Not even the diagram recalled from
Hawking-Ellis~\cite{Hawel} in Figure~\ref{godel-fig} or any of the
statements made in the first six pages.}. We do our best to make the
paper self-contained and explain and visualize most of what we say.
\bigskip

In more detail, this work consists of
Sections~\ref{time-section}-\ref{literature-section}.
Section~\ref{time-section} (p.\pageref{time-section}) is the just
mentioned ``popular'' part. Section~\ref{spiral-section}
(p.\pageref{spiral-section}) lays the foundation for discussing
rotating universes. E.g.\ it shows how to visualize such
space-times. The space-time built up in this section is called the
``Naive Spiral world''. Section~\ref{folia-section}
(p.\pageref{folia-section}) is about non-existence of a natural
``now'' in G\"odel's universe GU. Section~\ref{dervish-section}
(p.\pageref{dervish-section}) introduces co-rotating coordinates
``transforming the rotation away''. Section~\ref{tilting-section}
(p.\pageref{tilting-section}) refines the G\"odel-type universe
(obtained in Section~\ref{spiral-section}).
Section~\ref{refined-section} (p.\pageref{refined-section})
illustrates a fuller view of the refined version of GU.
Section~\ref{gyroscope-section} (p.\pageref{gyroscope-section})
re-coordinatizes the refined GU in order that the so-called
gyroscopes do not rotate in this coordinatization.
 Section~\ref{literature-section}
(p.\pageref{literature-section}) gives connections with the
literature. E.g.\ it presents detailed computational comparison with
the space-time metric in G\"odel's papers.
Section~\ref{technical-section} (p.\pageref{technical-section})
contains technical data about how we constructed the figures
illustrating G\"odel's universe.
\newpage


\section{Prelude: Some facts from the literature and how they imply
time-travel.} \label{time-section}

\begin{figure}[hbtp]
\setlength{\unitlength}{0.67 truemm} \small
\begin{center}
\begin{picture}(250,200)(0,0)
\put(130,160){\makebox(0,0)[l]{$p'$}}
\put(60,176){\makebox(0,0)[l]{\shortstack[l]{Matter world-line\\
$(r,\varphi)$ constant}}}
\put(132,183){\makebox(0,0)[b]{\shortstack[l]{$r=0$\\ (coordinate
axis)}}}
\put(148,177){\makebox(0,0)[lt]{\shortstack[l]{$p'$'s future null cone\\
(refocuses at $p''$)}}}

\put(152,147){\makebox(0,0)[lb]{$p$'s null cone refocuses at $p'$}}
\put(152,132){\makebox(0,0)[lb]{Null geodesics}}
\put(195,129){\makebox(0,0)[lb]{\shortstack[l]{Caustic on $p$'s\\
future null cone}}}
\put(183,97){\makebox(0,0)[l]{\shortstack[l]{Null cone\\ tangent
to\\ circle}}} \put(201,68){\makebox(0,0)[l]{$L$}}
\put(174,75){\makebox(0,0)[t]{\shortstack[l]{Null cone\\includes
circle}}} \put(180,47){\makebox(0,0)[l]{$t=0$}}
\put(140,52){\makebox(0,0)[lt]{\shortstack[lt]{$p$'s future\\ null
cone}}} \put(129,18){\makebox(0,0)[l]{$t$}}
\put(132,12){\makebox(0,0)[l]{$\varphi$}}
\put(142,4){\makebox(0,0)[l]{$r$}}
\put(105,45){\makebox(0,0)[rt]{\shortstack[l]{$r<\log(1+\sqrt{2})$\\
(closed spacelike\\ curve)}}}
\put(34,85){\makebox(0,0)[t]{\shortstack[l]{$\quad r>\log(1+\sqrt{2})$\\
(closed timelike\\ curve)}}}
\put(58,78){\makebox(0,0)[lt]{\shortstack[l]{$r=\log(1+\sqrt{2})$\\
(closed null curve)}}} \put(129,40){\makebox(0,0)[l]{$p$}}
\put(125,98){\makebox(0,0)[rt]{O}}
\put(67,135){\makebox(0,0)[b]{\shortstack[l]{Null cone\\
includes\\circle}}}
\put(101,138){\makebox(0,0)[b]{\shortstack[l]{Null cone\\ tangent to\\
circle}}} \epsfysize =  200 \unitlength \epsfbox{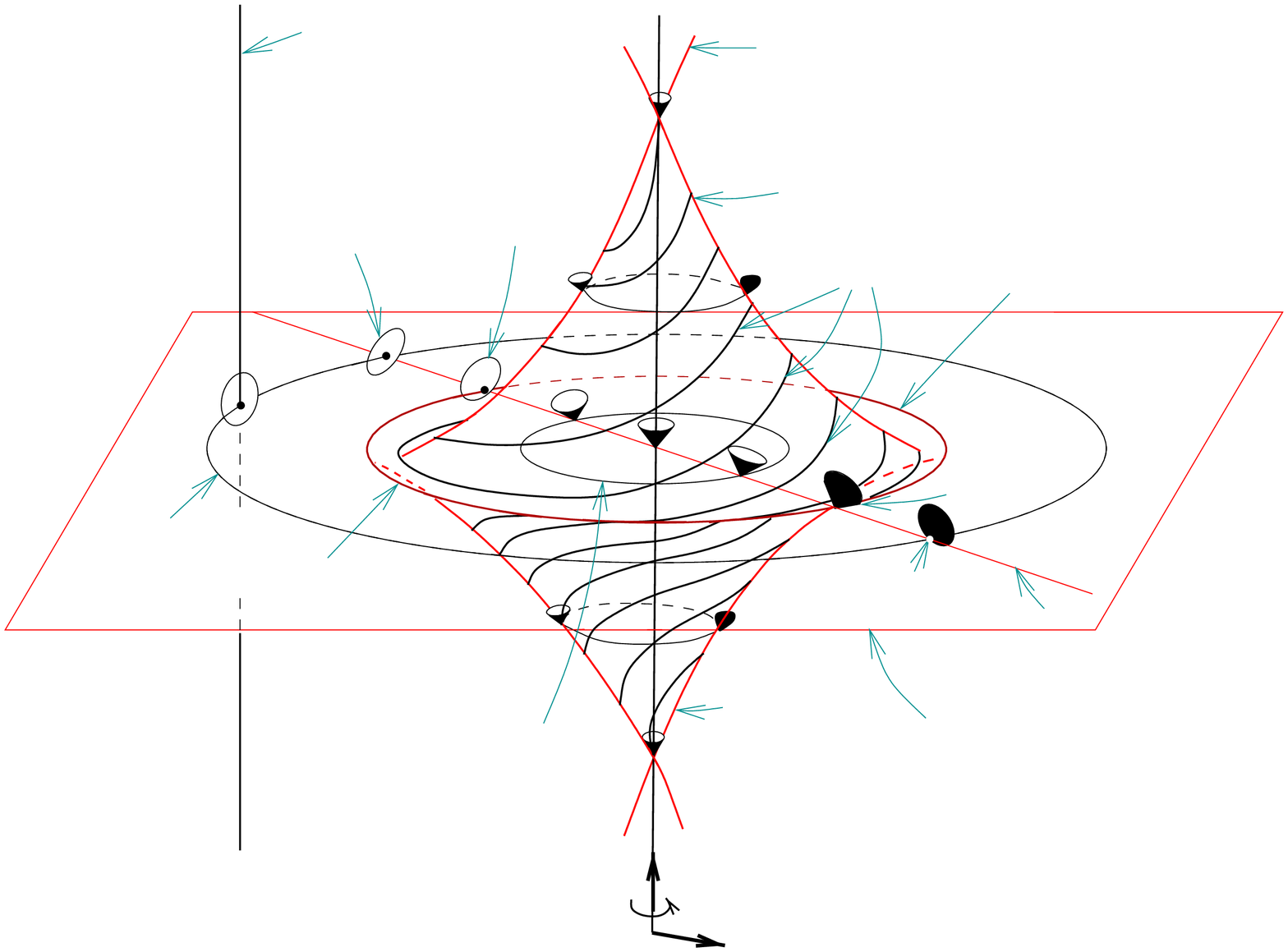}
\end{picture}
\end{center}
\caption[haho]{\label{ujgodel-fig} G\"odel's universe
\label{godel-fig} in co-rotating cylindric-polar coordinates
$\langle t,r,\varphi\rangle$. Irrelevant coordinate $z$ suppressed.
Light-cones (null-cones) and photon-geodesics indicated. Light-cone
{\em opens up} and tips over as $r$ increases (see line $L$)
resulting in closed time-like curves (CTC's). Drag effect (of
rotation) illustrated. Photons emitted at  $p$ spiral out, reach CTC
and reconverge at $p'$. This is a slightly corrected version of
Figure~31 in Hawking-Ellis~\cite[p.169]{Hawel} (cf.\
p.\pageref{corr-p}). (null cone = light-cone, null curve = photon
curve)}
 \end{figure}

\begin{figure}[!hp]
\setlength{\unitlength}{1.2 truemm} \small
\begin{center}
\begin{picture}(130,150)(0,0)
\put(21,139){\makebox(0,0)[l]{\shortstack[l]{matter
world-line\\$(r,\varphi)$ constant}}}
\put(36,112){\makebox(0,0)[b]{\shortstack[l]{light-cone\\ tangent to
\\circle}}}
\put(47,128){\makebox(0,0)[r]{\shortstack[r]{$r=0$\\ (coordinate
axis)}}} \put(87,134){\makebox(0,0)[l]{$p$'s light-cone refocuses at
$p'$}} \put(100,118){\makebox(0,0)[b]{photon geodesics}}
\put(119,103){\makebox(0,0)[b]{\shortstack[l]{$r$ = critical\\
(closed photon
\\curve)}}}
\put(115,51){\makebox(0,0)[t]{\shortstack[l]{light-cone\\ includes\\
circle}}}
\put(92,50){\makebox(0,0)[t]{\shortstack[l]{light-cone\\tangent to\\
circle}}} \put(64,16){\makebox(0,0)[l]{$t$}}
\put(69,7){\makebox(0,0)[l]{$\varphi$}}
\put(78,2){\makebox(0,0)[lb]{$r$}}
\put(13,53){\makebox(0,0)[t]{\shortstack[l]{closed time-like\\ curve
(a CTC)}}} \put(45,35){\makebox(0,0)[r]{$p$'s future light-cone}}
\put(67,23){\makebox(0,0)[l]{$p$}}
\put(67,143){\makebox(0,0)[l]{$p'$}} \epsfysize =  150 \unitlength
\epsfbox{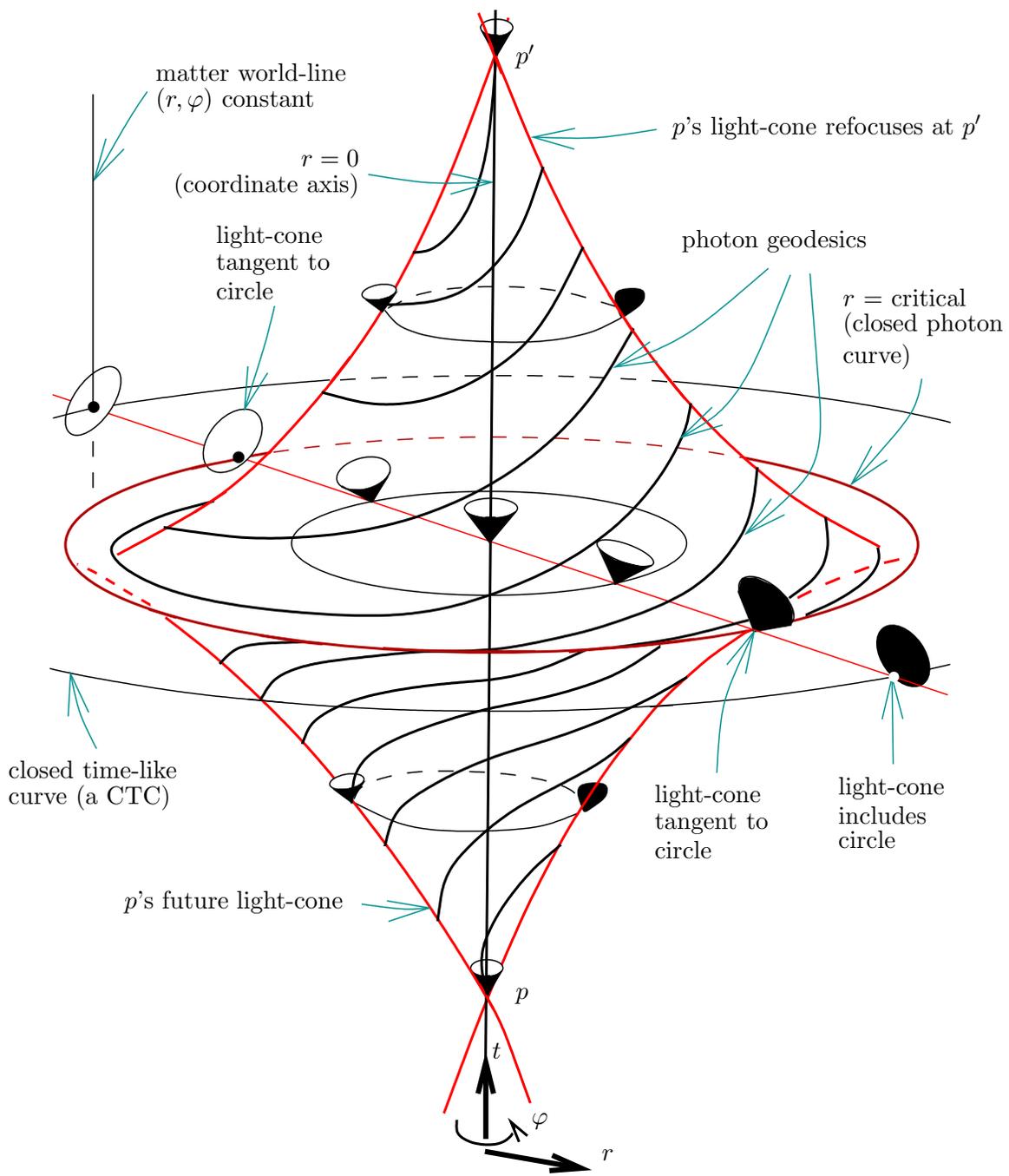}
\end{picture}
\end{center}
\caption[haho]{\label{godelnagy-fig} A closer look at G\"odel's
universe.}
 \end{figure}

The following series of figures represent G\"odel's famous rotating
universe. One of the many interesting features of G\"odel's universe
is that it contains closed time-like curves (CTC's for short), i.e.\
it permits ``time-travel''. In the following figures we use
geodesics and light-cones in the spirit of e.g.\ \cite[sections
3.1-3.3]{AMN07} for visualizing G\"odel's universe together with
some of its main features. For these notions cf.\ p.\pageref{gr-p}
herein. In Figures~\ref{godel-fig},\ref{godelnagy-fig}
\underbar{null-geodesic} \index{null-geodesic} is the same as
photon-like geodesic and ``null-cone'' \index{null-cone} is the same
as light-cone in the present paper.

\vfill\eject\newpage

\vfill\eject\newpage

\begin{figure}[!hbtp]
\setlength{\unitlength}{0.67 truemm} \small
\begin{center}
\begin{picture}(250,310)(0,0)
\epsfysize = 310  \unitlength \epsfbox{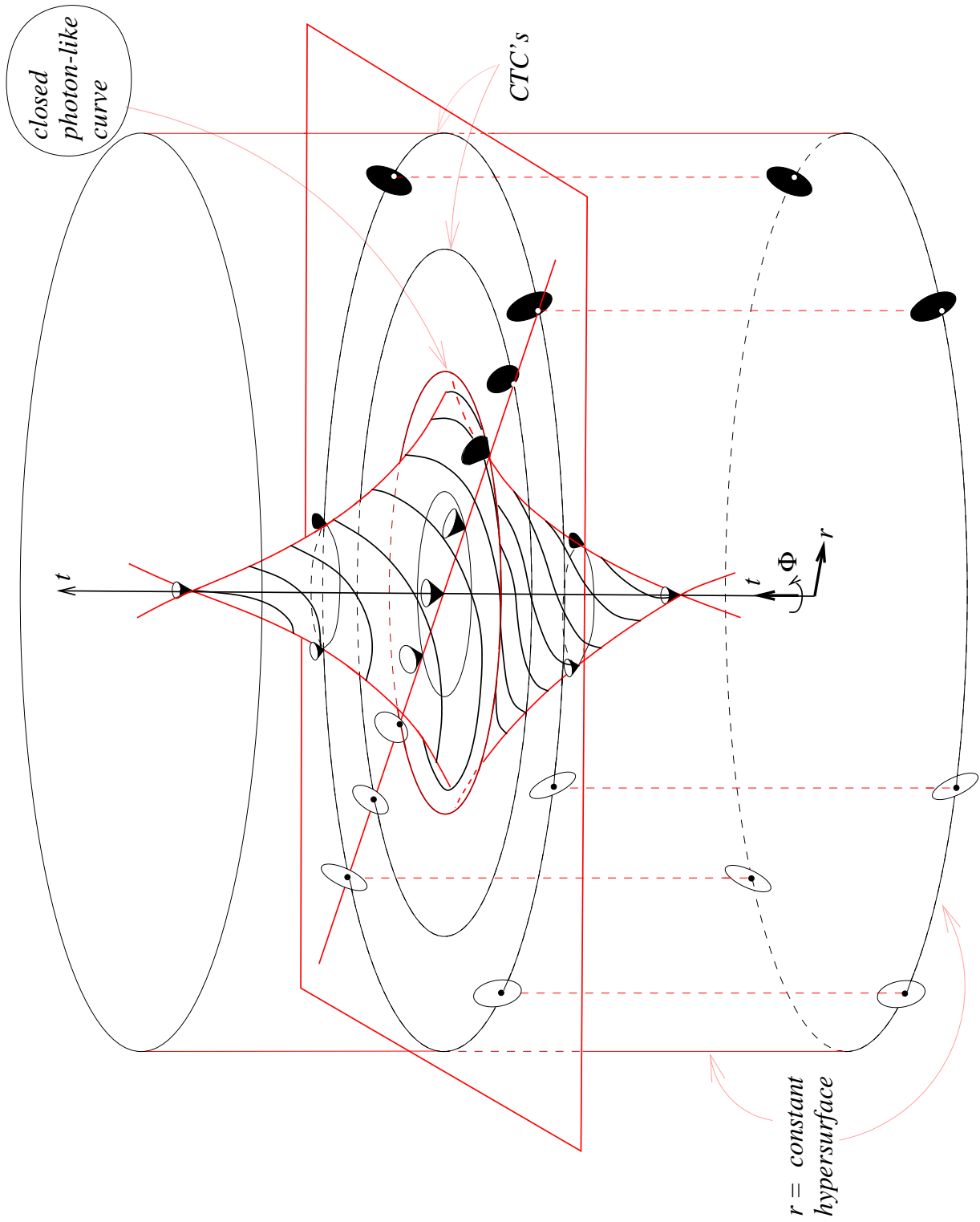}
\end{picture}
\end{center}
\caption[godel5]{\label{god5-fig}G\"odel's universe as on previous
figure but with an ``$r$=constant'' (and $z$=constant) hypersurface
indicated. This hypersurface is parallel with the $t$-axis.
Throughout this work, z=constant. I.e.\ throughout we suppress the
irrelevant spatial coordinate $z$. In
Figures~\ref{god5-fig}-\ref{gode1-fig}, $\Phi$ is the same as
$\varphi$ in the rest of the paper.}
 \end{figure}

\vfill\eject\newpage

\begin{figure}[!hbtp]
\setlength{\unitlength}{0.67 truemm} \small
\begin{center}
\begin{picture}(250,310)(0,0)

\epsfysize = 310  \unitlength \epsfbox{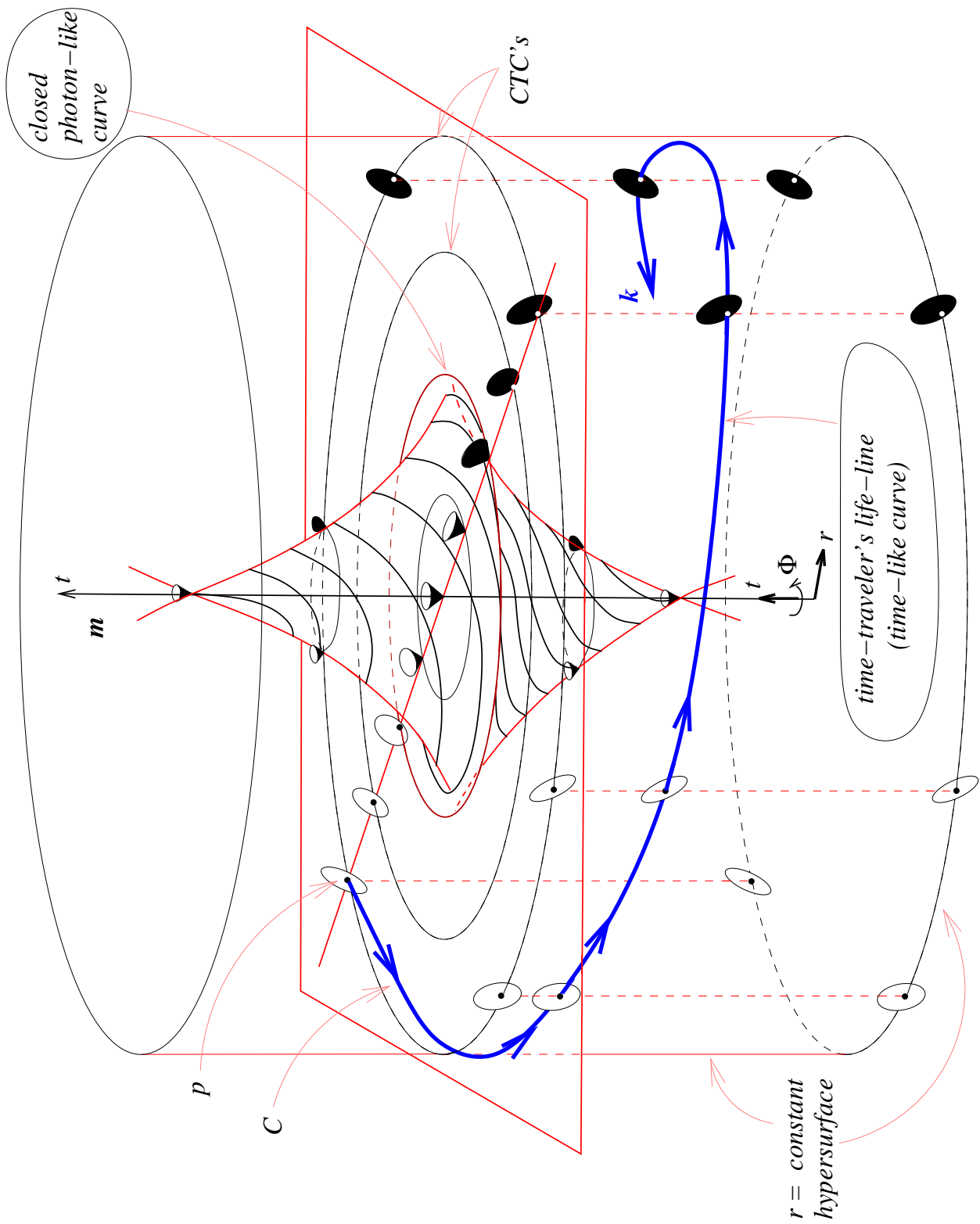}
\end{picture}
\end{center}
\caption[godel4]{\label{god4-fig}G\"odel's universe with a
time-traveler's (time-like) life-line indicated. The time-traveler's
acceleration is bounded (but cannot be zero). The time-like curve
$C$ stays always inside the light-cones and spirals back to the past
as $m$ observes it. This is possible because the light-cones far
away from the $t$-axis are so much tilted that they reach below the
horizontal plane. See the explanation on p.\pageref{time-expl}.}
\end{figure}

\begin{figure}[!hbtp]
\setlength{\unitlength}{0.67 truemm} \small
\begin{center}
\begin{picture}(250,310)(0,0)

\epsfysize =   310\unitlength \epsfbox{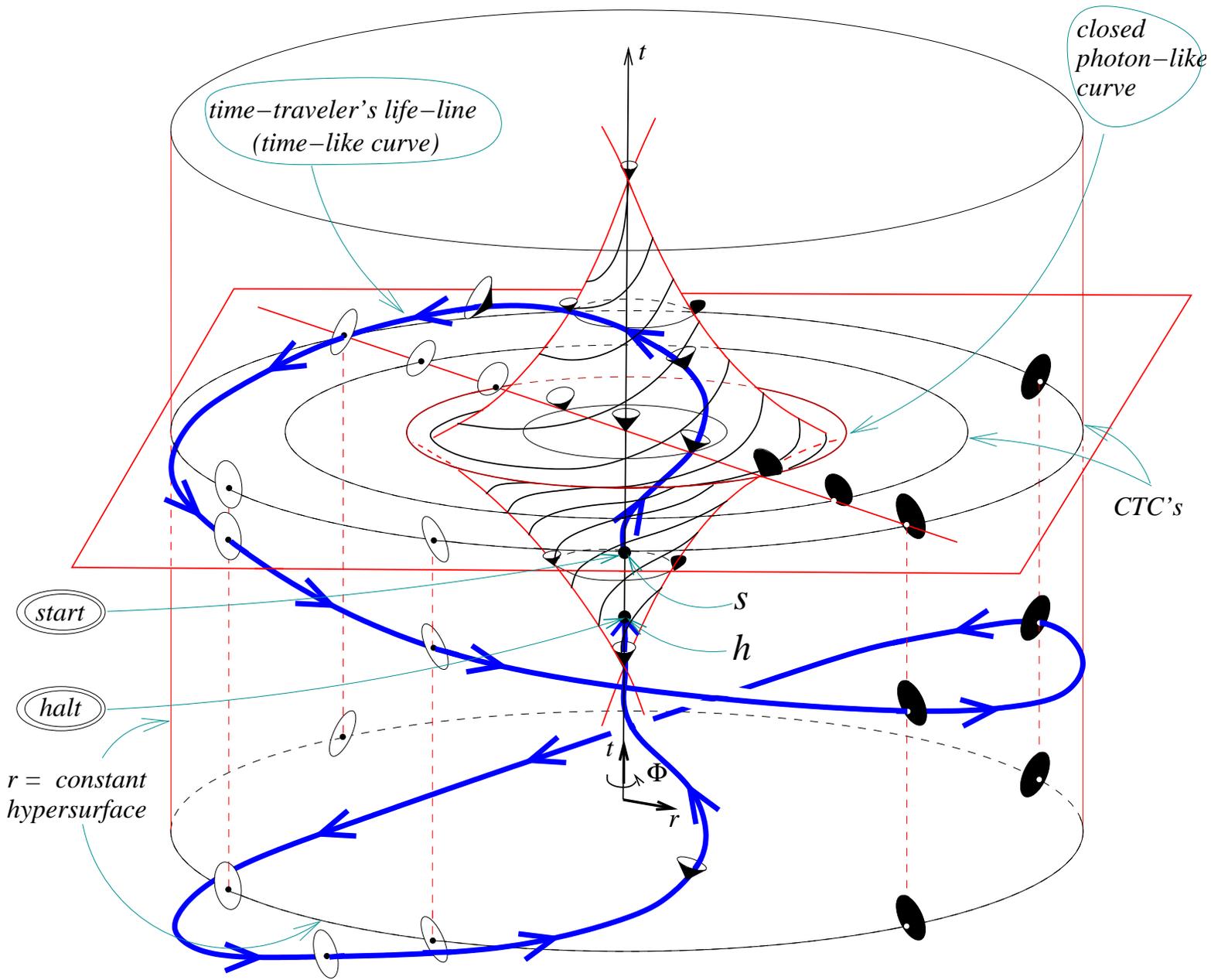}
\end{picture}
\end{center}
\caption[haho]{\label{timetrav} \label{gode1-fig} Time-traveler
starting at time $s$ and arriving at time $h$, where $h$ is earlier
than $s$.}
 \end{figure}

\vfill\eject\newpage

\bigskip
\noindent \emps{Explanation for
Figures~\ref{god4-fig},\ref{timetrav}:} \label{time-expl}
Figures~\ref{god4-fig},\ref{timetrav} illustrate the time-travel
aspect in G\"odel's universe. Assume observer $m$ lives on the time
axis $\bar t$. Assume $p$ is a point far enough from $\bar t$. I.e.\
the radius $r$ of $p$ is large enough. Then at $p$ the light-cones
are so much tilted that a time-like curve $C$ can spiral back into
the past as observed by $m$. $C$ involves only bounded acceleration.
An observer, say $k$, can live on $C$. Then in $m$'s view, $k$ moves
towards the past. Moreover, $k$ can go back to the past as far as he
wishes.

It is an entertaining exercise to prolong curve $C$ such that it
starts at $s\in\bar t$ and ends at $h\in\bar t$ such that $h\prec
s$, i.e.\ $h$ is in the past of $s$, see Figure~\ref{timetrav}. Then
our observer $k$ can start its journey at $s$, spiral outwards to
radius $r$, then spiral back along $C$ and then spiral inwards to
$h$. Then $k$ can wait on the time axis $\bar t$ to meet itself at
point $s$. We leave the details to the reader, but see
Figure~\ref{timetrav}.

Cf.\ also Figure 28 on p.113 in Horwich~\cite{Horw}, which we
include below.

\begin{figure}[!hbtp]
\setlength{\unitlength}{1.2 truemm} %
\begin{center}
\begin{picture}(134,115)(0,0) 
\epsfysize = 115\unitlength   
\epsfbox{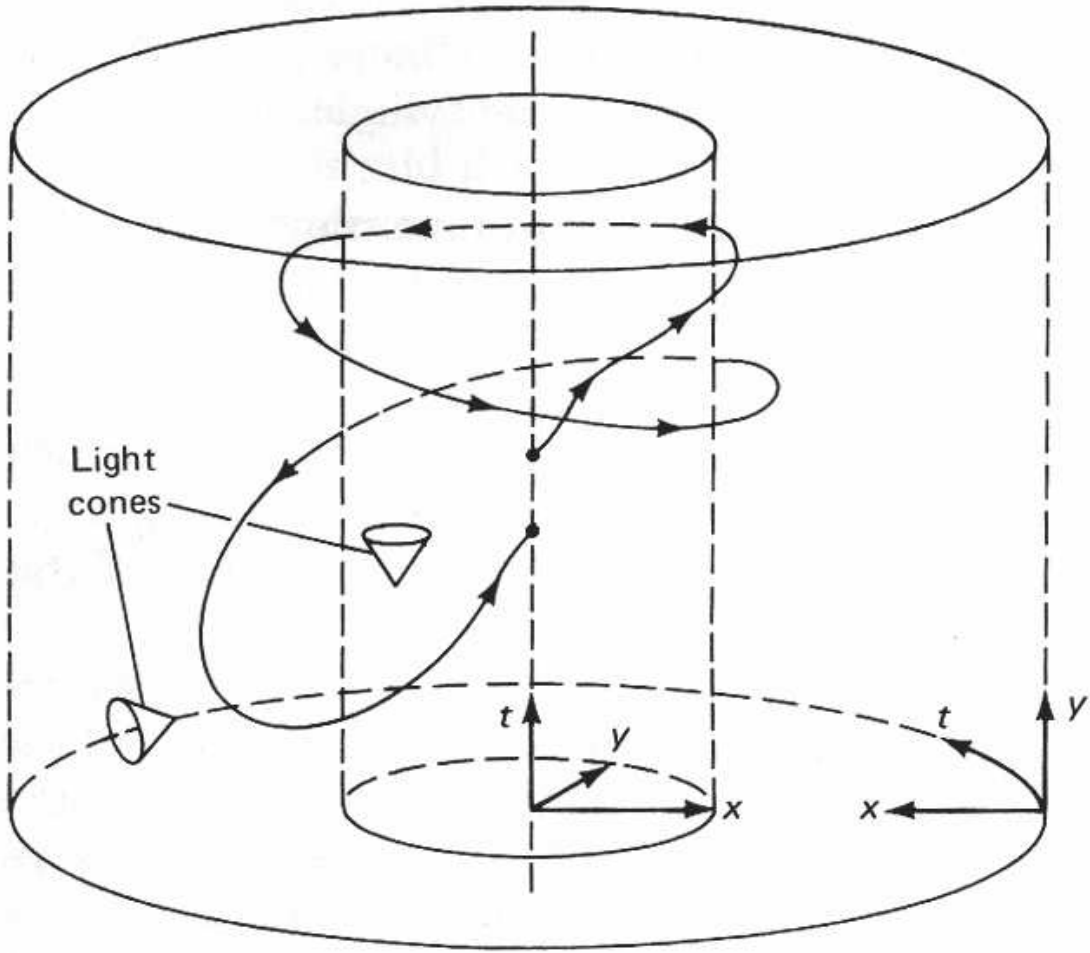}
\end{picture}
\end{center}
\caption{\label{Horwich-fig} Figure from Horwich~\cite[Figure 28
(p.113)]{Horw}.}
 \end{figure}

\vfill\eject\newpage

\section{Preparation for constructing G\"odel style rotating universes. The Naive Spiral World.}
\label{spiral-section}

\vspace*{36pt}

In this part we populate Newtonian space with massive observers
$m_i$ for $i\in I$ which carry equal mass and are evenly distributed
(where we understand ``even" in the common sense). We will call
these $m_i$'s {\em distinguished observers} or
{\em mass-carriers} or {\em galaxies}%
\footnote{We use the world ``galaxy'' only in a metaphorical sense
and it means nothing more than our distinguished observers carrying
mass. Cf.\ Rindler~\cite[p.203]{Rin} for more on our usage for
galaxies.} . Then we rotate this inhabited space around the $z$
axis. The galaxy in the origin is called $m_0$. We will make sure
that nothing happens in the direction $z$, therefore we can suppress
direction $z$ in our pictures and discussion. So space-time becomes
three-dimensional with axes $t,x,y$. We concentrate on the
$xy$-plane inhabited by the galaxies (or distinguished observers)
$m_i$. We rotate this plane of galaxies around the origin, i.e.\
around $m_0$. The rotation is rigid, i.e.\ the distances between the
galaxies do not change. The {\em angular velocity} of this rotation
is denoted by $\omega$. We call the plane inhabited by the $m_i$'s
the {\em universe}. Hence $\omega$ is called the angular velocity of
the universe. The rotation takes place in a Newtonian inertial frame
of reference.%
\footnote{Here we use the expression ``inertial frame of reference"
in the most classical (Newtonian) way, namely as it was given by L.\
Lange in 1885: ``A reference frame in which a mass point thrown from
the same point in three different (non co-planar) directions follows
rectilinear paths each time it is thrown, is called an inertial
frame."} The angular velocity $\omega$ is chosen such that the
resulting centrifugal force exactly balances the gravitational
attraction between the $m_i$'s. This is possible, cf.\ G\"odel's
paper \cite[second half of p.270]{Go96} for a proof. (Cf.\
\cite[pp.261-289]{Go96} for more detail.)

So our first pictures will show space-time diagrams in which the
life-lines%
\footnote{What we call life-line is called world-lines in most of
the literature of general relativity.} of the galaxies $m_i$ appear
as spirals around the $t$-axis (which happens to be the life-line of
$m_0$). An extra feature is that, similarly to G\"odel's papers, we
assume the existence of certain kinds of {\em cosmic compasses}. Our
cosmic compasses need not agree with what are called gyroscopes in
physics. For the time being cosmic compasses constitute only certain
conventions. Equivalently, they can be regarded as distinguished
{\em local coordinate frames} or ``local coordinate systems'' for
our distinguished observers or mass-carriers (the $m_i$'s). These
local frames need not be inertial. For the time being we do not
associate any tangible or
observational physical meaning to our compasses and local frames.%
\footnote{What they represent is mainly a logical ``stage'' in our
construction of rotating universes. Though, in principle we could
associate (a fairly complicated) observational meaning to them. We
do not go into this here.} In Section~\ref{gyroscope-section} we
will turn our attention to gyroscopes and local inertial frames,
too.

We assume that all the $m_i$'s agree with each other in that they
have two cosmic compasses for carrying the original spatial
directions $x$ and $y$ of our original Newtonian inertial reference
frame with which we began our construction. This makes them
equivalent (with each other) in the sense that any of them, say $m$,
may think that he is at the center, he is not rotating and it is the
rest of the observers who are rotating around $m$.
\bigskip

This \label{gr-p} paper is based on general relativity but we do not
assume that the reader is familiar with the details of general
relativity. What we do assume is familiarity with (i) the basics of
special relativity and (ii) awareness of some of the basic
principles of general relativity explained in items (1)-(2) below.
All this can be found in \cite{AMN07}. All what we need to know
about special relativity in this paper can be found in
\cite[sections 2.1-2.4]{AMN07}. What we need to know about general
relativity theory, summarized in items (1)-(2) below, can be found
in \cite[sections 3.1-3.3]{AMN07}.

(1) General relativity assumes that {\em special relativity holds
locally}. This means, roughly, that in a general relativistic
space-time, every point (event) is ``surrounded'' by a small, local
coordinate frame (LF for short) and in each LF special relativity
holds in some sense (cf.\ e.g.\ Rindler~\cite{Rin} for a simple
explanation of this). The LF's are local in the topological sense
that space-time $M$ comes together with a topology and then LF's are
local in the sense that the ``closer" we go to the point $p\in M$
the more accurately the local special relativity frame LF describes
the behavior of light-signals and moving bodies. (For a precise
formulation see \cite[sec.3.3, e.g., Def.3.3]{AMN07}.)

In the case of G\"odel's universe, $M$ together with this topology
is just the original (Newtonian) space-time $\Reals^4$. Thus, in the
case of G\"odel's universe $\langle M,\dots\rangle$ a single
``global" coordinate system can cover the whole of $M$. This means
that there exist coordinatizations $Co:\Reals^4\longrightarrow M$
with $Co$ a bijection which satisfy some natural requirements which
we do not list here. E.g.\ $Co$ involves one ``time coordinate'' and
three ``space coordinates'', hence at first glance it looks similar
to the familiar coordinatization of Newtonian space-time or special
relativity. Further, one of the space coordinates turns out to be
irrelevant, hence $Co:\Reals^4\longrightarrow M$ will admit a
3-dimensional representation (via suppressing the irrelevant
coordinate). So in our pictures there will be {\em one big
coordinate system} $Co$ covering the whole picture and there will be
{\em many small coordinate systems} representing the LF's or other
local coordinate systems. The big coordinate system represents the
whole of our manifold $M$ to be described.

When we describe a space-time $M$, the key ingredient is specifying
how the little LF's are glued together to form the whole of $M$. We
will do this by specifying a (fairly arbitrary) coordinatization $C$
of $M$ and then to each point $p\in M$ we describe how the LF at $p$
is fitted into $M$ at point $p$.%
\footnote{The effect is somewhat similar to an Escher painting,
e.g.\ he glues little birds together and there emerges an over-all
pattern which has nothing to do with birds.} When specifying which
LF is glued to what point, we use the coordinate system $C$ as a
tool for communication. Most of the time we will use geometric
constructions for presenting the above data. In such a picture, the
LF at $p$ is represented by drawing the {\em light-cone} at $p$
together with the {\em unit vectors}
 $\langle t_p, x_p,y_p\rangle$ of the LF at $p$. Sometimes we
 indicate only the future light-cones, sometimes we indicate both
 the future and the past light-cones. Most of the time we indicate
 the local simultaneity of the LF, too.%
 \footnote{To specify the LF, it is enough to specify the unit vectors
 $\langle t_p, x_p, y_p\rangle$. These determine the light-cones and the local simultaneity.
 However, the latter are very helpful in visualizing the space-time, that's why we
 indicate them in the pictures.}
These pictures, beginning with Figure~\ref{spi-fig}, represent {\em
precise geometrical constructions}, hence they intend to specify the
space-time in question completely (as opposed to being a mere
``sketch'' conveying only intuitive ideas). In
Sections~\ref{technical-section},\ref{literature-section} which
contain the technical details we present the constructions behind
the pictures together with the metric tensor field of the space-time
in question. (To explain the latter, we note that a model of general
relativity is usually given in the form $\langle M,{\sf g}\rangle$
where $M$ is a manifold and ${\sf g}$ is a tensor field defined on
$M$. We will not need these tensor-fields until
Section~\ref{literature-section}.) We note that ${\sf g}$ can be
reconstructed from the way the LF's are glued together in our
pictures, hence if the reader understands the geometry of these
pictures, he will automatically understand the space-time (or
general relativity model) they represent.
\smallskip

(2) Occasionally we will mention so-called {\em geodesics}.
Geodesics are the general relativistic counterparts of straight
lines of special relativity, in particular, the life-lines of
inertial bodies or freely falling bodies are called geodesics. The
same applies to life-lines of photons. {\em Curves} are understood
in the usual sense, e.g.\ geodesics are special curves. Properties
of curves are generalized from special relativity to general
relativity by saying that curve $\ell$ has property $P$ if it has
$P$ locally (in the sense of special relativity). E.g.\ $\ell$ is
{\em time-like} if for each $p\in\ell$ the LF surrounding $p$
``thinks'' that $\ell$ is time-like in the sense of special
relativity. Similarly for {\em space-like}, {\em photon-like} (and
for other properties of geodesics).

We note that time-like curves are the possible life-lines of
arbitrary bodies, i.e.\ of not necessarily inertial bodies. These
may undergo acceleration. Both geodesics and time-like curves are
curves in the usual sense. A  curve is time-like if it always stays
inside the light-cones. A curve $\ell$ is photon-like if for any
point $p\in\ell$, $\ell$ is tangent to the light-cone at $p$.
\bigskip

\begin{figure}[!hp]
\setlength{\unitlength}{0.2 truemm} \small
\begin{center}
\begin{picture}(769,455)(0,0)

\epsfysize =  455 \unitlength \epsfbox{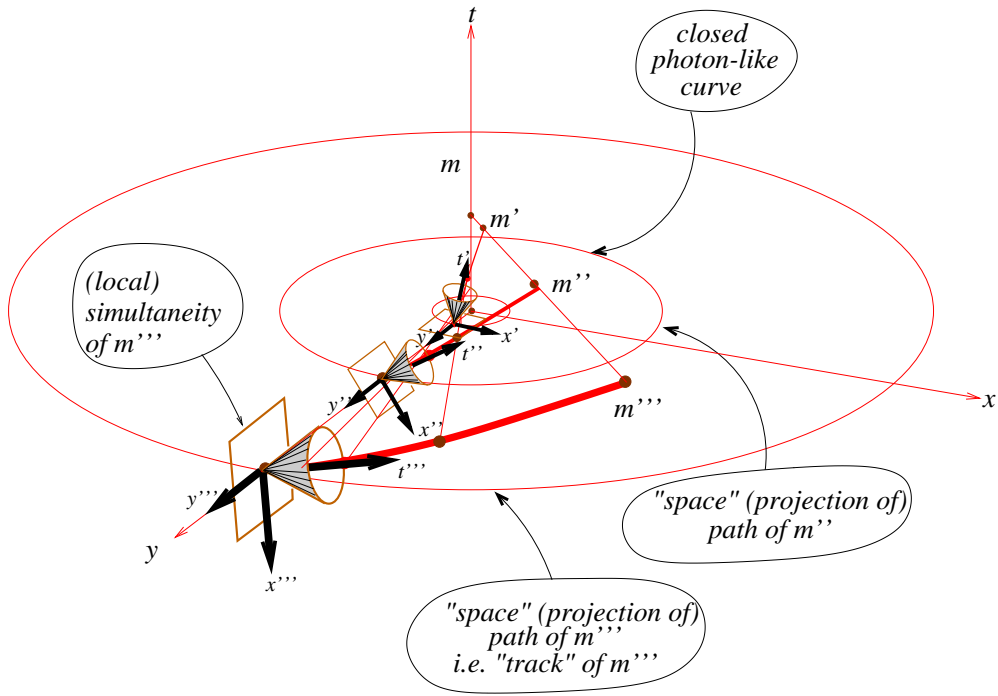}
\end{picture}
\end{center}
\caption{\label{inercelo} Observers $m', m'', m'''$ perform a rigid
rotation around observer $m$. Such observers are the only
mass-carriers in this universe. Because of this rotation, $m'''$
moves so fast that his light-cone tilts over so much that it is
almost horizontal.}
 \end{figure}
\vfill\eject

\begin{figure}[!hp]
\setlength{\unitlength}{0.18 truemm} \small
\begin{center}
\begin{picture}(875,1035)(0,0)

\epsfysize =  1035\unitlength \epsfbox{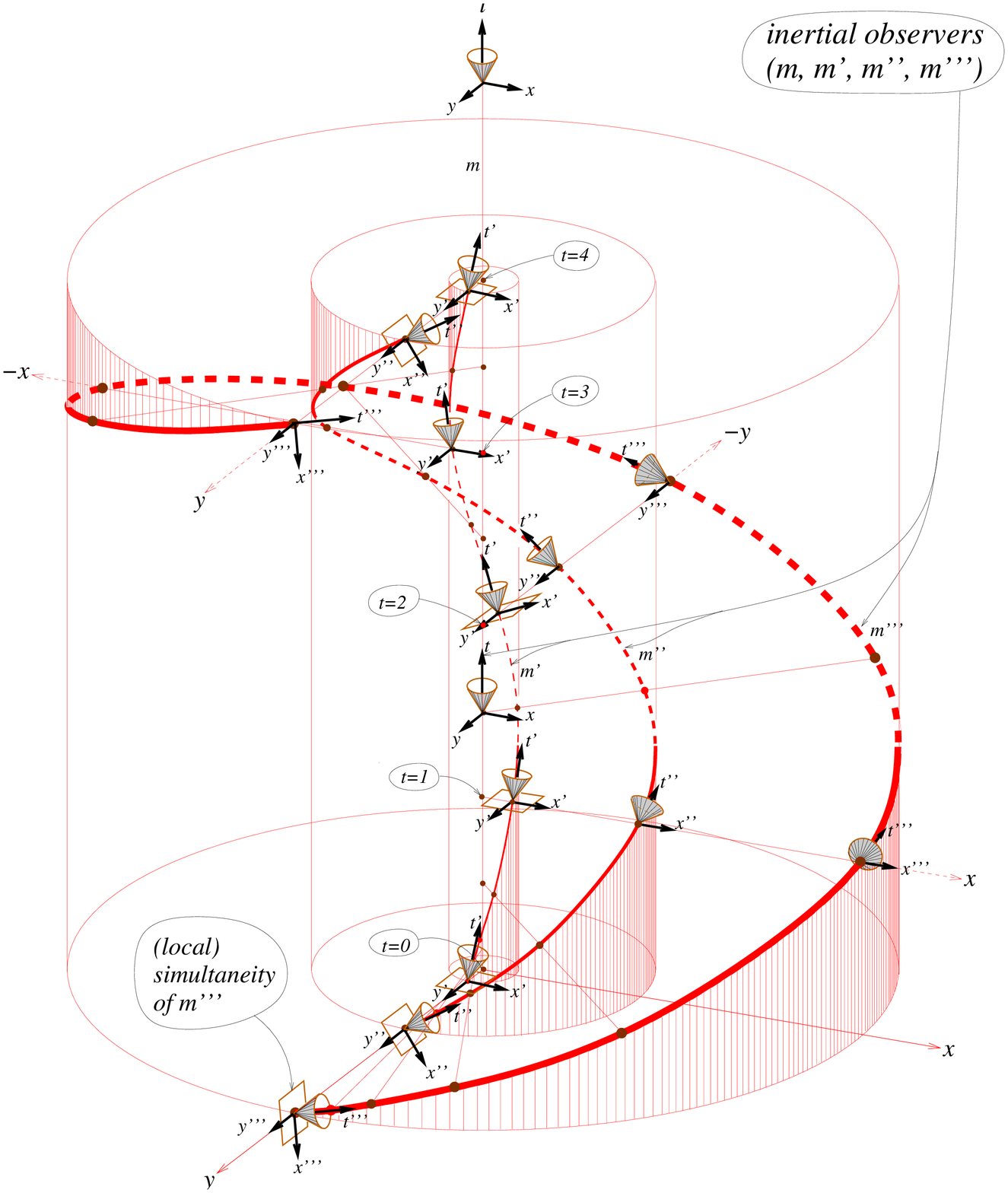}
\end{picture}
\end{center}
\caption[G\"odel's universe GU with emphasis on inertial
observers]{\label{godinerc} G\"odel's Universe with emphasis on
\underbar{inertial} observers instead of  photons (the rotation is
``rigid''). The coordinate system $\la t',x',y'\ra$ of say $m'$ does
\underbar{not} follow the rotation of the matter in this universe.
The life-lines of $m,\ldots,m'''$ are (special) geodesics. $\la
t,x,y\ra$, $\la t',x',y'\ra$ etc.\ are distinguished local
coordinate systems. E.g.\ $\la t'',x'',y''\ra$ is the local
coordinate system of observer $m''$.}
 \end{figure}

\begin{figure}[!hp]
\setlength{\unitlength}{0.13 truemm} \small
\begin{center}
\begin{picture}(827,1590)(0,0)

\epsfysize =  1590\unitlength \epsfbox{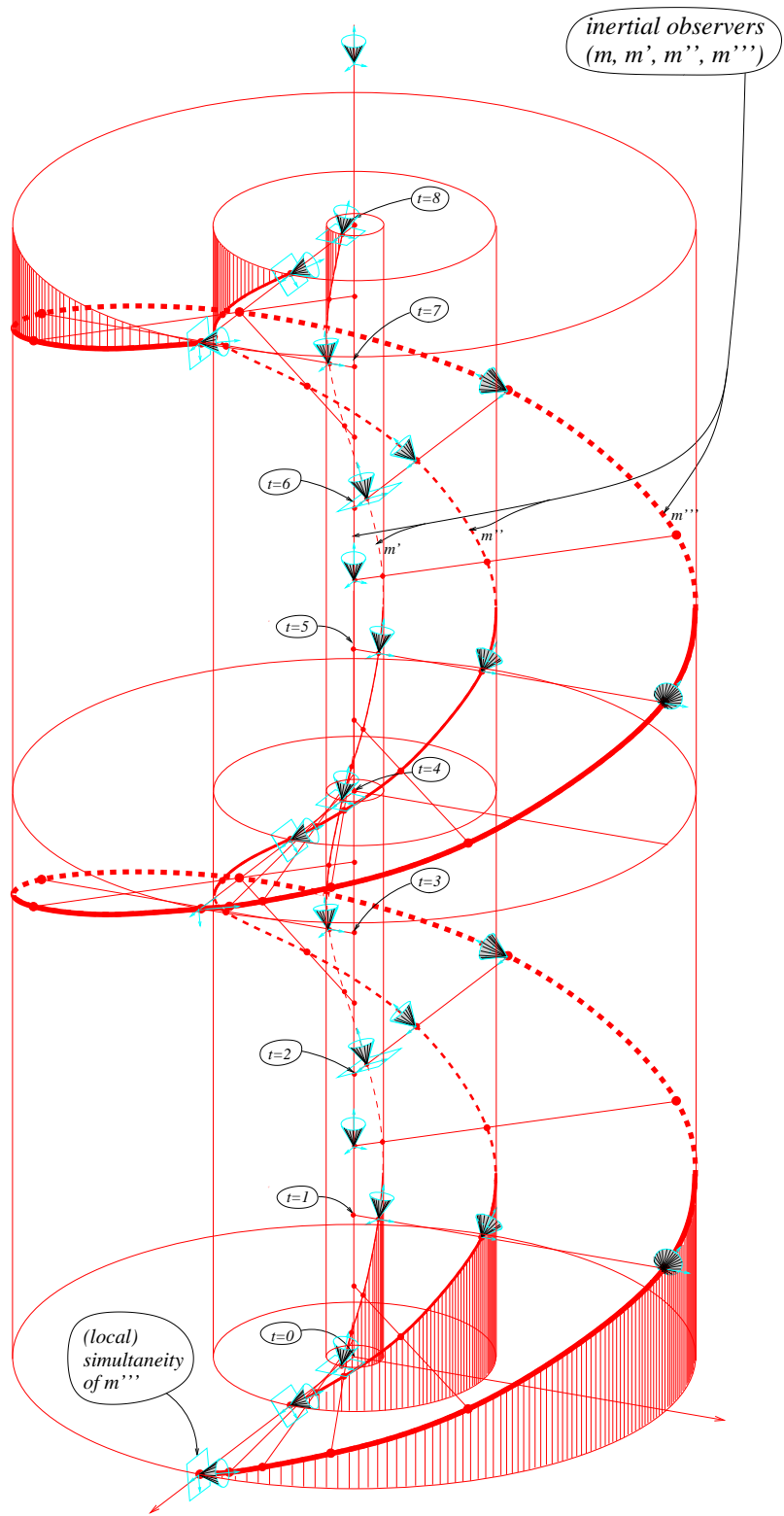}
\end{picture}
\end{center}
\caption{\label{hajnalka} Previous figure copied on top of itself.
It goes on like this in both directions forever. $m',m'',m'''$ are
(time-like life-lines of) observers ``equivalent with'' the observer
$m$ living on $\bar t$.}
 \end{figure}
\vfill\eject

\begin{figure}[!hp]
\setlength{\unitlength}{0.13 truemm} \small
\begin{center}
\begin{picture}(874,1630)(0,0)

\epsfysize =  1630\unitlength \epsfbox{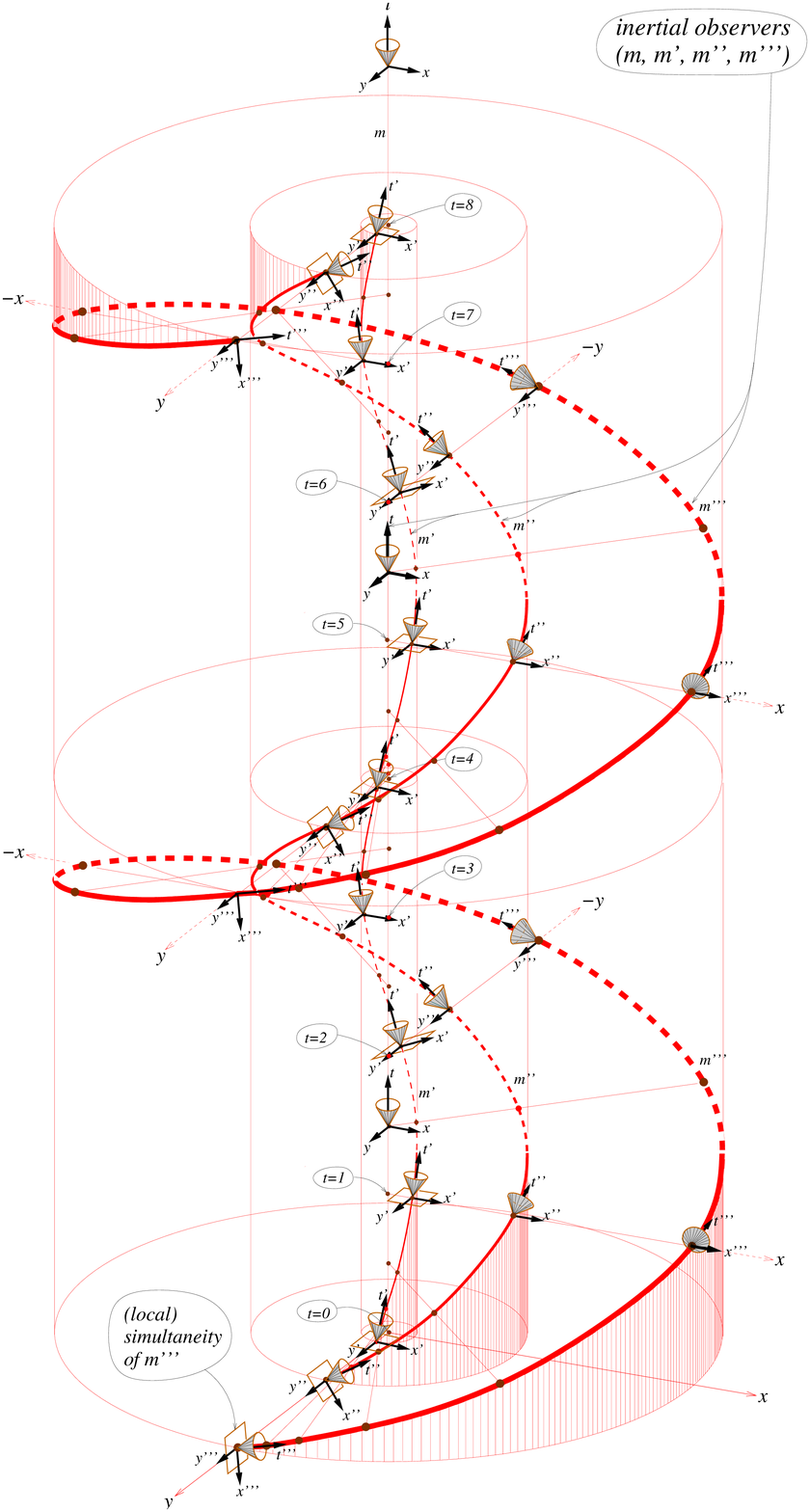}
\end{picture}
\end{center}
\caption{\label{double} Previous figure with non-rotating local
coordinate systems $\la t',x',y'\ra$, $\la t'',x'',y''\ra$ etc.\
emphasized.}
 \end{figure}
\vfill\eject

\begin{figure}[!hp]
\setlength{\unitlength}{0.25 truemm} \small
\begin{center}
\begin{picture}(220,730)(0,10)

\epsfysize =  730 \unitlength 
\epsfbox{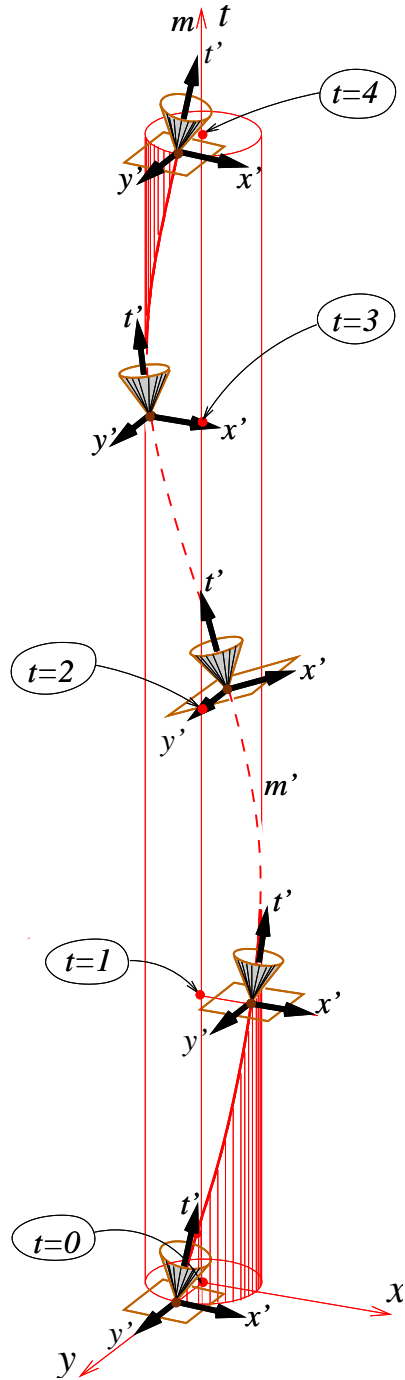}
\end{picture}
\end{center}
\caption{\label{notrot} The coordinate system $\la t',x',y'\ra$ of
say $m'$ does \underbar{not} follow the rotation of the matter in
this universe. The reader is asked to check that in a certain sense
the direction $x'$ remains parallel with the original direction $x$.
This is why $m'$ thinks that $m$ is rotating around $m'$.}
 \end{figure}
\vfill\eject

\begin{figure}[!hp]
\setlength{\unitlength}{0.08 truemm} \small
\begin{center}
\begin{picture}(2111,2476)(0,0)

\epsfysize = 2476  \unitlength \epsfbox{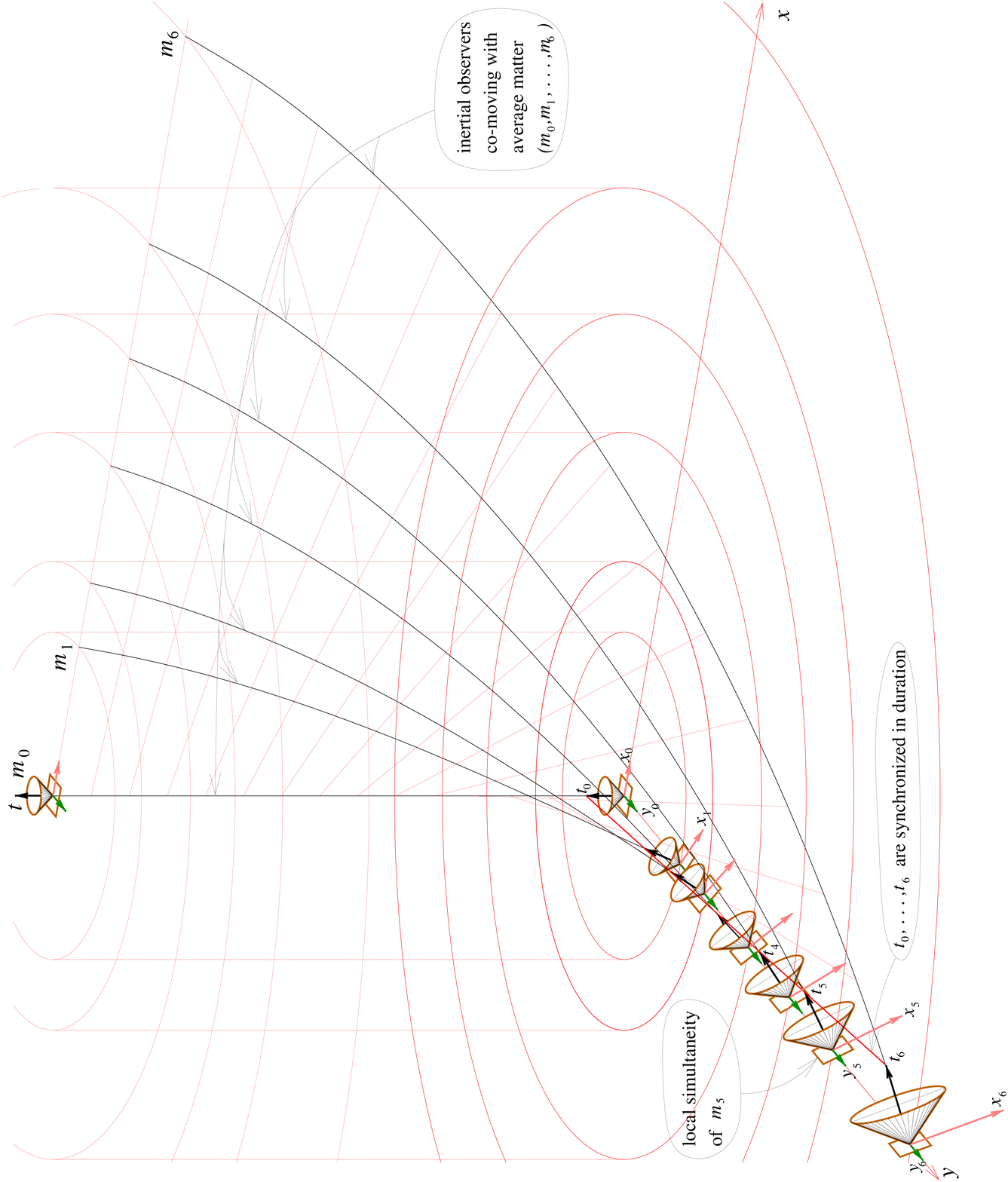}
\end{picture}
\end{center}
\caption{\label{spi-fig}  Each $m_i$ can measure the time needed for
a single turn of the universe. (I.e.\ each $m_i$ can measure the
angular velocity $\omega$ of the universe.) To ensure this we have
to calibrate the $t_i$ vectors of the $m_i$'s such that in $m_0$'s
view the vertical components of all the $t_i$'s are {\em equal} with
that of $t_0$. $\omega=\pi/30$, Map 2 applies. Cf.\
p.\pageref{map2-fig}.}
 \end{figure}


\begin{figure}[!hp]
\setlength{\unitlength}{0.055 truemm} \small
\begin{center}
\begin{picture}(3066,3759)(0,0)

\epsfysize = 3759  \unitlength \epsfbox{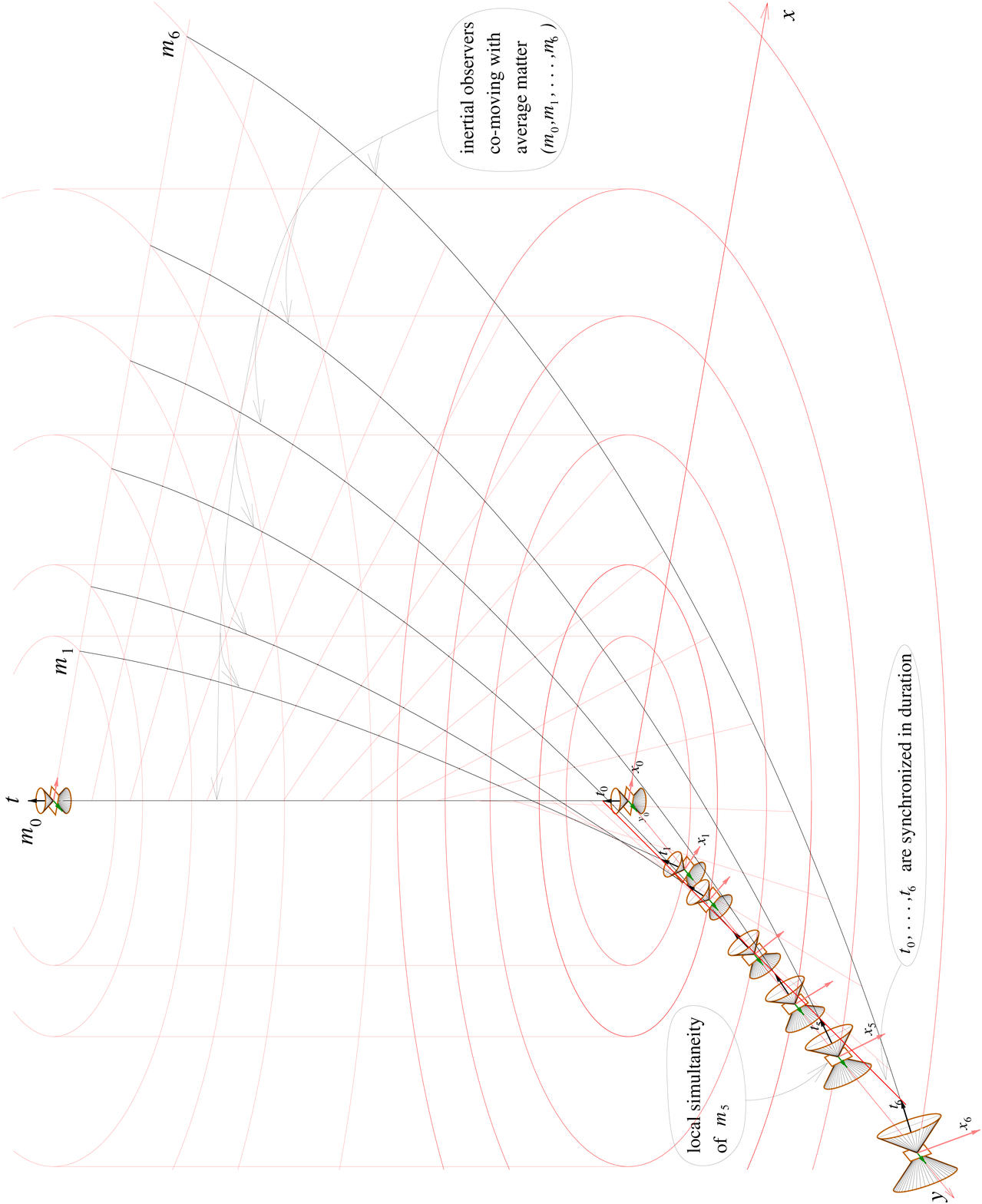}
\end{picture}
\end{center}
\caption{\label{spi1-fig} Previous figure with past-light-cones
indicated. $\omega=\pi/45$, Map 1 applies. Cf.\
p.\pageref{map1-fig}.}
 \end{figure}


\begin{figure}[!hp]
\setlength{\unitlength}{0.085 truemm} \small
\begin{center}
\begin{picture}(1679,2603)(0,0)

\epsfysize = 2603  \unitlength \epsfbox{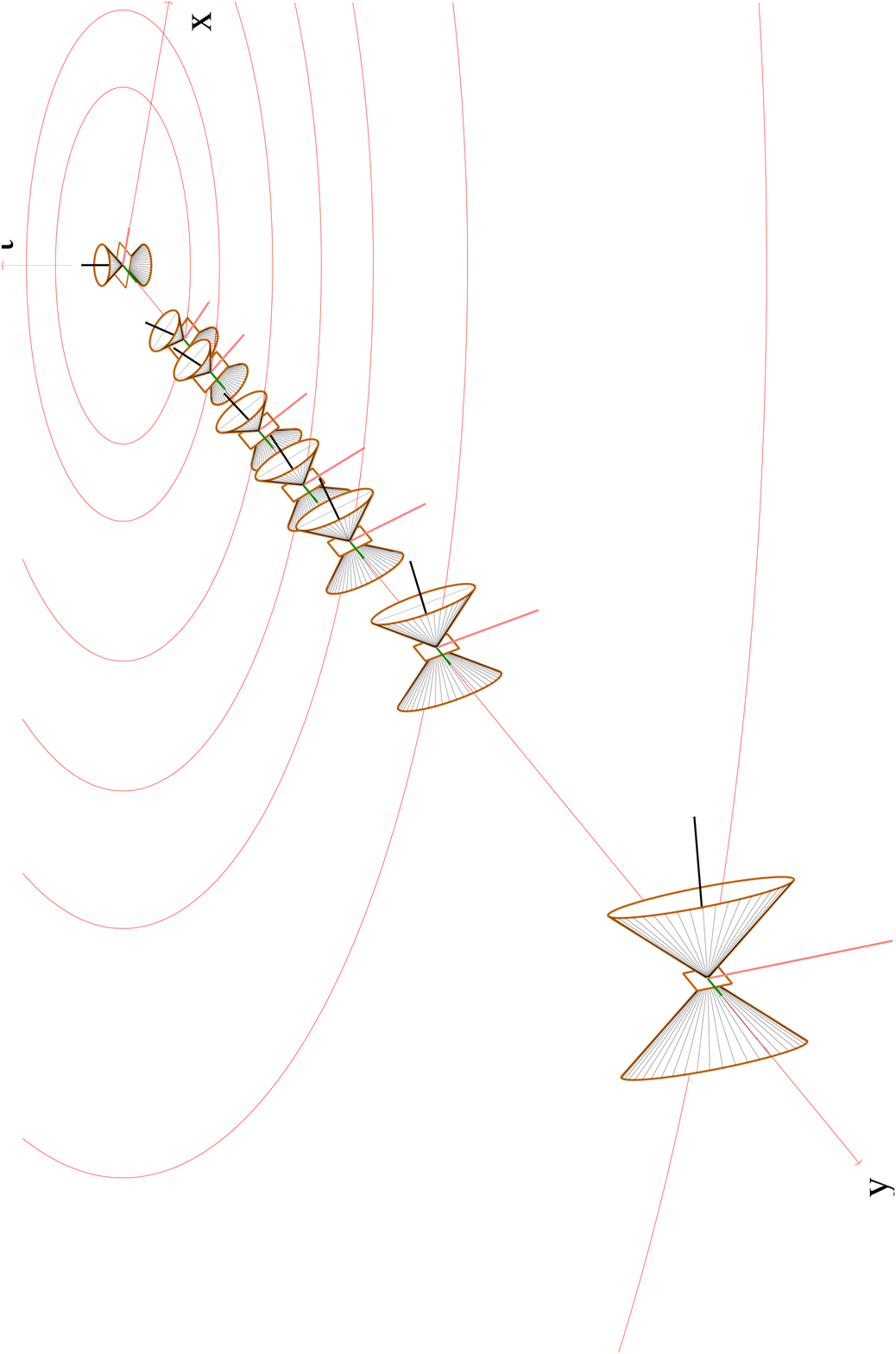}
\end{picture}
\end{center}
\caption{\label{torta2-fig} $\omega=\pi/30$, Map 2 applies. Cf.\
p.\pageref{map2-fig}.}
 \end{figure}

\newpage

G\"odel wanted the distinguished massive observers $m_0,\dots,
m_i,\dots$ of his universe to be equivalent with each other. So far
they are equivalent from the point of view that each of them thinks
that the rest of the universe rotates around himself. This is so
because the local coordinate systems (hence the cosmic compasses) of
the distinguished observers $m_i$ do not rotate, do not follow the
rotation of the universe. At this point we can ensure one more
symmetry property of the $m_i$'s.
Each $m_i$ can measure the time needed for a single turn of the
universe, for example as follows: $m_i$ picks a distinguished
observer, say $m_0$, such that $m_i$'s $y$-compass points in the
direction of $m_0$ at an instant, and then measures the time passed
until his
$y$-compass again points in $m_0$'s direction.%
\footnote{What does it mean that $m_i$'s $y$-compass points in
$m_0$'s direction at some time $t$? We may use the following
definition: there is a curve $\ell$ connecting $m_i$'s life-line
(starting with the event at $t$) with $m_0$'s life-line such that at
each point $p$ of the curve $\ell$ the following holds: $\ell$ lies
in the local simultaneity of the distinguished observer $m$ passing
through $p$ and $m$'s $y$-compass points in $\ell$'s direction in
$p$.}
 This is how $m_i$ can measure the angular velocity
$\omega$ of the universe. To ensure that all the distinguished
observers get the same value for the angular velocity, we have to
calibrate the $t_i$ vectors of the $m_i$'s such that in $m_0$'s view
the vertical components of all the $t_i$'s are {\em equal} with that
of $t_0$. This is ensured in
Figure~\ref{spi-fig}, and from now on we will always ensure this.%
\footnote{This will also ensure that each $m_i$ will measure the
same angular velocity for the universe, no matter which ``partner"
he chooses (in place of $m_0$) for the measurement.} This choice of
the local time-unit vectors ensures also that the local LF's measure
a kind of ``universal time", namely that of the big global reference
frame. However, this ``universal time" does not satisfy natural
requirements about ``time" presented in the next section.

\label{NGU-page} Above we specified the time-unit-vectors of the
local frames. Let us now specify three other unit-vectors at each
point $p$, these will specify the light-cone and the local special
relativity at $p$. All what we say below in specifying the three
unit vectors are meant in the big global reference frame. The
r-unit-vector at $p$ points in the radial direction parallel to the
$xy$-plane and has length 1. The (suppressed) $z$-unit-vector points
in the direction of the (suppressed) $z$-axis and has length 1.
Finally, the last unit-vector is orthogonal to the three
unit-vectors given so far and has the same length as the
t-unit-vector. In the local frame at $p$, these 4 vectors constitute
an orthonormal system. By this, we
specified fully our general relativistic space-time.%
\footnote{The corresponding metric tensor is given in
section~\ref{literature-section}.}

The {\em preliminary} version of G\"odel's universe GU constructed
above and depicted in Figures~\ref{inercelo}-\ref{torta2-fig} will
be referred to as ``Naive GU'' (NGU) or more specifically, ``{\em
Naive Spiral World}''. The reason for this is that so far we have
chosen the simplest possible arrangement of light-cones without
checking whether they will satisfy certain properties we have in
mind. Indeed, Section~\ref{tilting-section} will lead to some
refinement/fine-tuning of the light-cone structure. However, the
Naive GU has many of the desired properties already. Namely, the
life-lines of the galaxies are geodesics, i.e., the distinguished
observers $m_i$ are really inertial observers. The radial straight
lines parallel to the $xy$-plane are all geodesics, too.

\vfill\eject \noindent
\section{\bf Non-existence of a global time in G\"odel's universe.}
\label{folia-section}

\begin{figure}[!ht]
\setlength{\unitlength}{0.25 truemm} 
\bigskip\bigskip
\small
\begin{center}
\begin{picture}(597,781)(0,0)

\epsfysize = 781  \unitlength \epsfbox{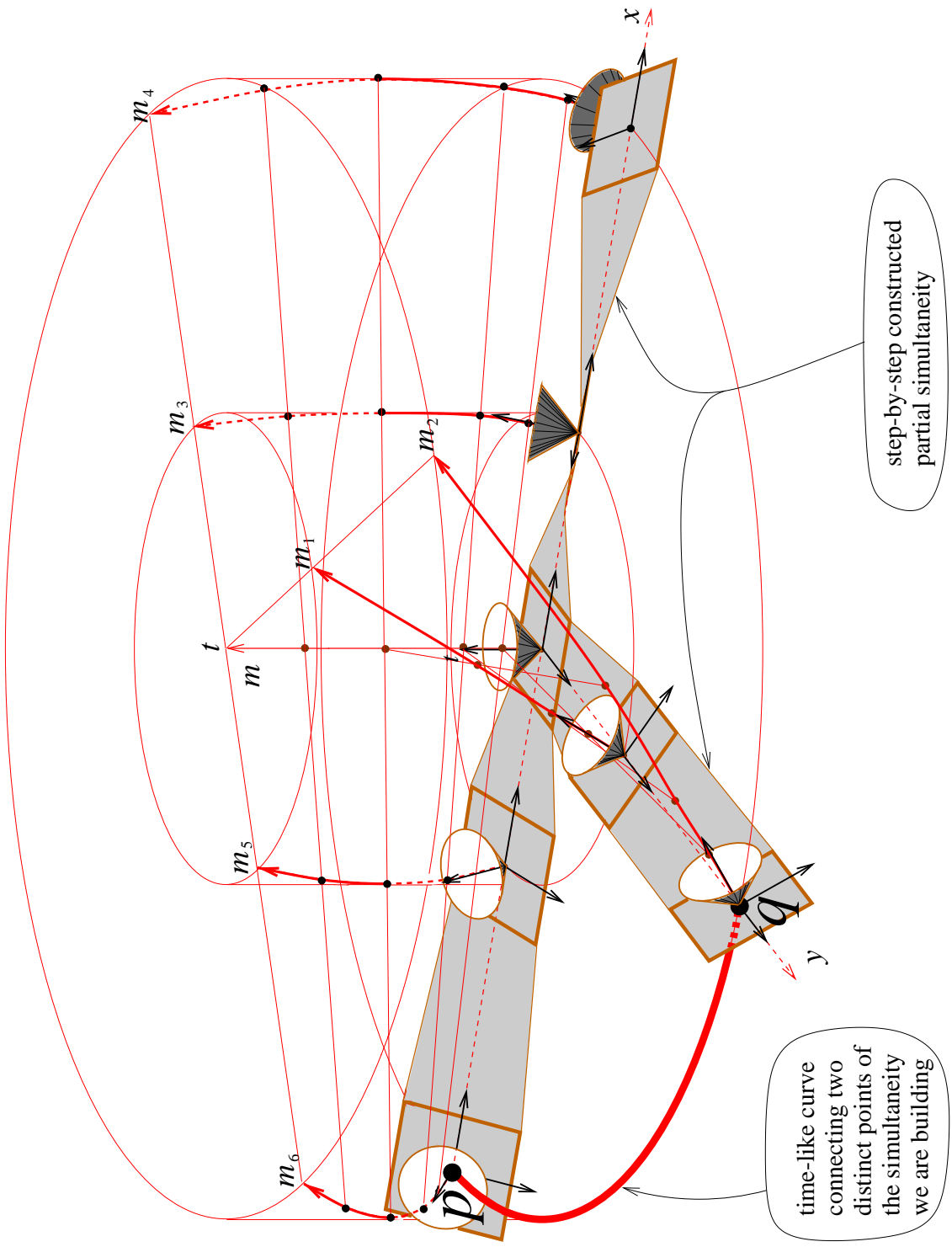}
\end{picture}
\bigskip\bigskip
\end{center}
\caption{\label{nonfolia} Idea of ``non-foliasibility'' of G\"odel's
space-time. I.e.\ nonexistence of a global, natural simultaneity (or
global time) in G\"odel's universe. See explanation on
p.\pageref{folia-expl}.}
 \end{figure}

\begin{figure}[!hbtp]
\setlength{\unitlength}{0.26 truemm} \small
\begin{center}
\begin{picture}(598,740)(0,0)

\epsfysize =  740 \unitlength \epsfbox{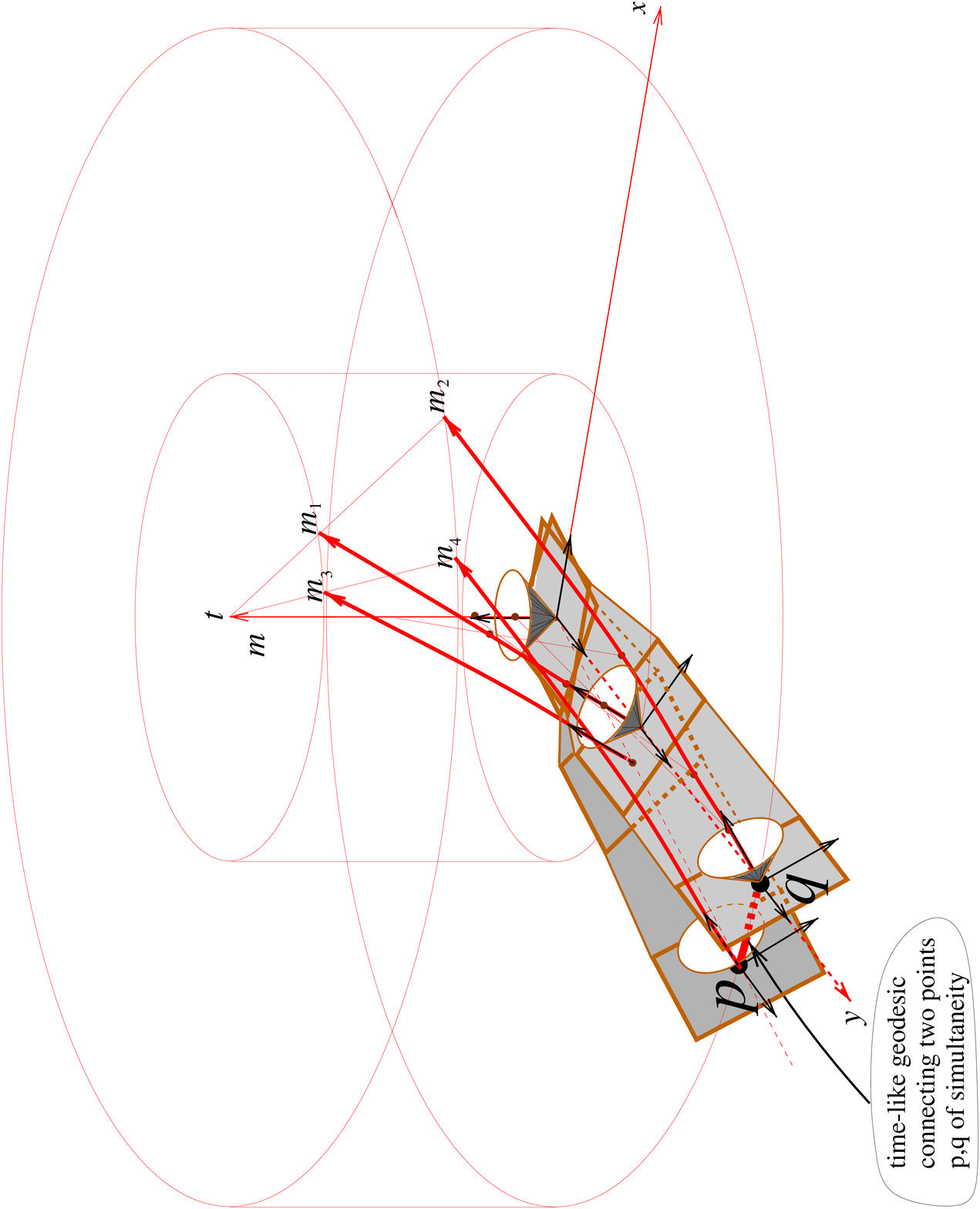}
\end{picture}
\end{center}
\caption{\label{folia} Previous figure but with the two strips of
constructed simultaneity closer to each other, $p,\bar 0$ and
$q,\bar 0$ are still simultaneous. The ``informal logic'' of these
two figures generates a simultaneity connecting all points of
space-time with each other. This is in contradiction with the
intuitive notion of simultaneity.}
 \end{figure}

\noindent \label{folia-expl} Figures~\ref{nonfolia}--\ref{folia}
below form an informal illustration for the idea of
``non-foliasibility'' of G\"odel's universe GU. I.e.\
Figures~\ref{nonfolia}--\ref{folia} intend to illustrate the claim
that there is no global natural simultaneity in GU.

By a \emps{potential simultaneity} of GU we can understand a
hyper-surface $S$ in the usual sense and we can require it to
satisfy conditions like (i)--(vi) below.
\smallskip

\begin{description}
\item[(i)]
$(\forall p,q\in S)[p\ne q\ \Rightarrow\ (\exists$ {\it maximal
space-like geodesic } $\ell)(p,q\in\ell\subseteq S)]$.
\item[(ii)]
$(\forall$ {\it space-like geodesic }$\ell)[$ {\it a nonempty open
segment of $\ell$ lies in }$S\ \ \Rightarrow\ \ \ell\subseteq S]$.
\item[(iii)]
Every maximal time-like geodesic $\ell$ intersects $S$ (i.e.\
$\ell\cap S\ne\emptyset$).
\item[(iv)]
$S$ ``avoids'' the light-cones, i.e.\ no nonempty segment of a
photon-geodesic lies inside $S$. (Note that any open segment of a
geodesic is a geodesic again.)
\item[(v)] there is no time-like curve connecting two points of $S$.
\item[(vi)] there is no time-like geodesic connecting two points of
$S$.
\end{description}
\smallskip

Note that (i)-(iii) are ``closure conditions'', i.e.\ they try to
make $S$ big, while condition (iv) points in the direction that $S$
is only $n-1$--dimensional (in some sense), hence it tries to make
$S$ ``thin'' like a usual surface.

In the pictures we start out from the origin $\bar 0$ and try to
build a simultaneity containing $\bar 0$ first by moving along the
$\bar y$--axis and then by moving along the negative $-\bar
x$--axis. Then we try to combine the two. While the figure does not
prove the nonexistence theorem, it illustrates ideas about its
plausibility.
For more careful formulation and proof of non-existence of global
time in GU cf.\ \cite[p.263 (written by Malament),
pp.269--287]{Go96}, Hawking-Ellis~\cite[p.170]{Hawel}.
Earman~\cite[Lemma 4.1]{Earm} is also (remotely) relevant here, but
it proves less than what G\"odel claims, namely, we do not require
$S$ to satisfy all properties of a Cauchy hypersurface (cf.\
\cite[p.44]{Earm} for definition of Cauchy
hypersurfaces).%
\footnote{ The general relativistic computer constructed in
Etesi-N\'emeti~\cite{EN} (cf.\ also Hogarth~\cite{HogDis},
Earman~\cite{Earm}, N\'emeti-D\'avid ~\cite{ND06}) can be realized
in the G\"odel-type universes, too, because of their special causal
structure. This is interesting because we do not know whether the
GU's enjoy the so called Malament-Hogarth property (in the
literature general relativistic computers are usually constructed in
Malament-Hogarth space-times).}

\bigskip

\vfill\eject\newpage

\vfill\eject\newpage

\vfill\eject\newpage

\section{G\"odel's universe in co-rotating coordinates, ``whirling
dervishes''. Transforming the rotation away.}
\label{dervish-section} \vspace*{36pt}

Gott~\cite[p.91]{Gott} writes ``You could equally well view
G\"odel's universe as static and non-rotating, as long as
self-confessed ``nondizzy observers'' would be spinning like
whirling dervishes with
respect to the universe as a whole.''%
\footnote{G\"odel~\cite[p.271]{Go49} writes: ``Of course, it is also
possible and even more suggestive to think of this world as a rigid
body at rest and of the compass of inertia as rotating everywhere
relative to this body.''} Below we will introduce new coordinates
$\langle T^r, X^r, Y^r, Z^r\rangle$ co-rotating with the matter
content $m_0,\dots,m_i,\dots$ of the universe. In $\langle
T^r,\dots\rangle$ the massive bodies $m_i$ appear as static with
their life-lines vertical lines. We will call $\langle
T^r,\dots\rangle$ ``{\em Dervish World}'' motivated by the above
quotation from Gott. The transformation between the old spiral
coordinates and the new rotating coordinates is elaborated later, on
pp.\pageref{coord1}--\pageref{coord2}.
\bigskip

In the Spiral World, the ``galaxies'' $m_1, m_2,\dots, m_i$ appear
as rotating around $m_0$ in direction $\varphi$ with angular
velocity $\omega$ while their cosmic compasses $x_i, y_i$ appear
fixed (non rotating). As a contrast, the Dervish World shows
$m_1,\dots, m_i$ as motionless, while it shows their cosmic
compasses as rotating in direction $-\varphi$ with angular velocity
$\omega$.
\bigskip

We will indicate on page~\pageref{Mach-page} how this dervish world
can be used to show that GU can be used to demonstrate that General
Relativity (in its present form) does not imply the full version of
Mach's principle.
\bigskip

\vfill\eject\newpage

\begin{figure}[!hp]
\setlength{\unitlength}{2 truemm} 
\begin{center}
\begin{picture}(60,100)(9,-1) 
\epsfysize = 100\unitlength   
\epsfbox{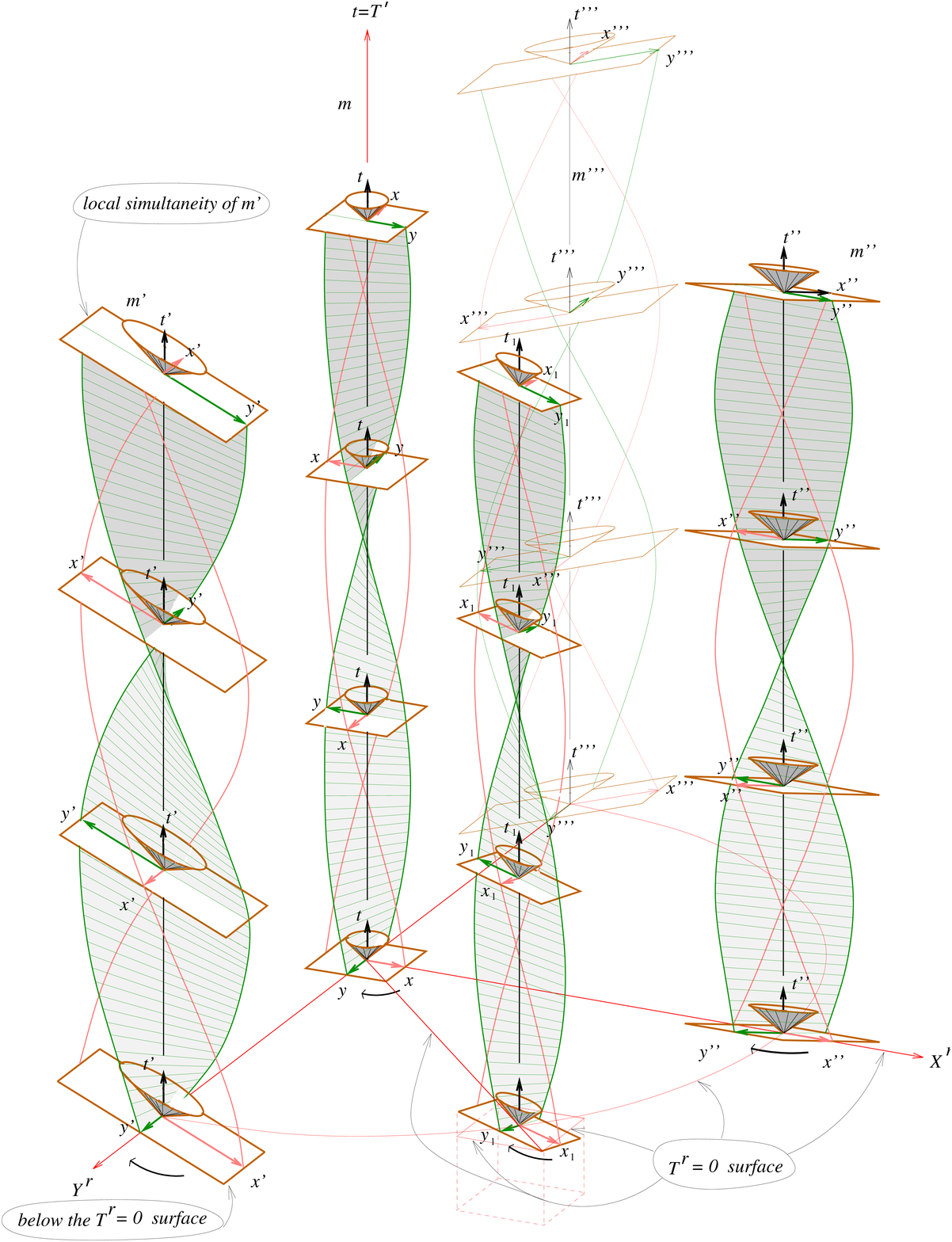}
\end{picture}
\end{center}
\caption[haho]{\label{dervis1}
G\"odel's universe GU in {\em rotating} coordinates $T^r=t,\; X^r,\;
Y^r$. These coordinates co-rotate with GU, hence GU appears as {\em
being at rest}. As a price, the local coordinate systems like $\la
t',x',y'\ra$ appear as rotating backwards (in direction $-\varphi$)
in the new coordinate system. The transformation between the old
spiral coordinates and new rotating ones is elaborated on
p.\pageref{coord1}.}
 \end{figure}

\vfill\eject\newpage

\begin{figure}[!hp]
\setlength{\unitlength}{2 truemm} 
\begin{center}
\begin{picture}(60,100)(9,-5) 
\epsfysize = 100\unitlength   
\epsfbox{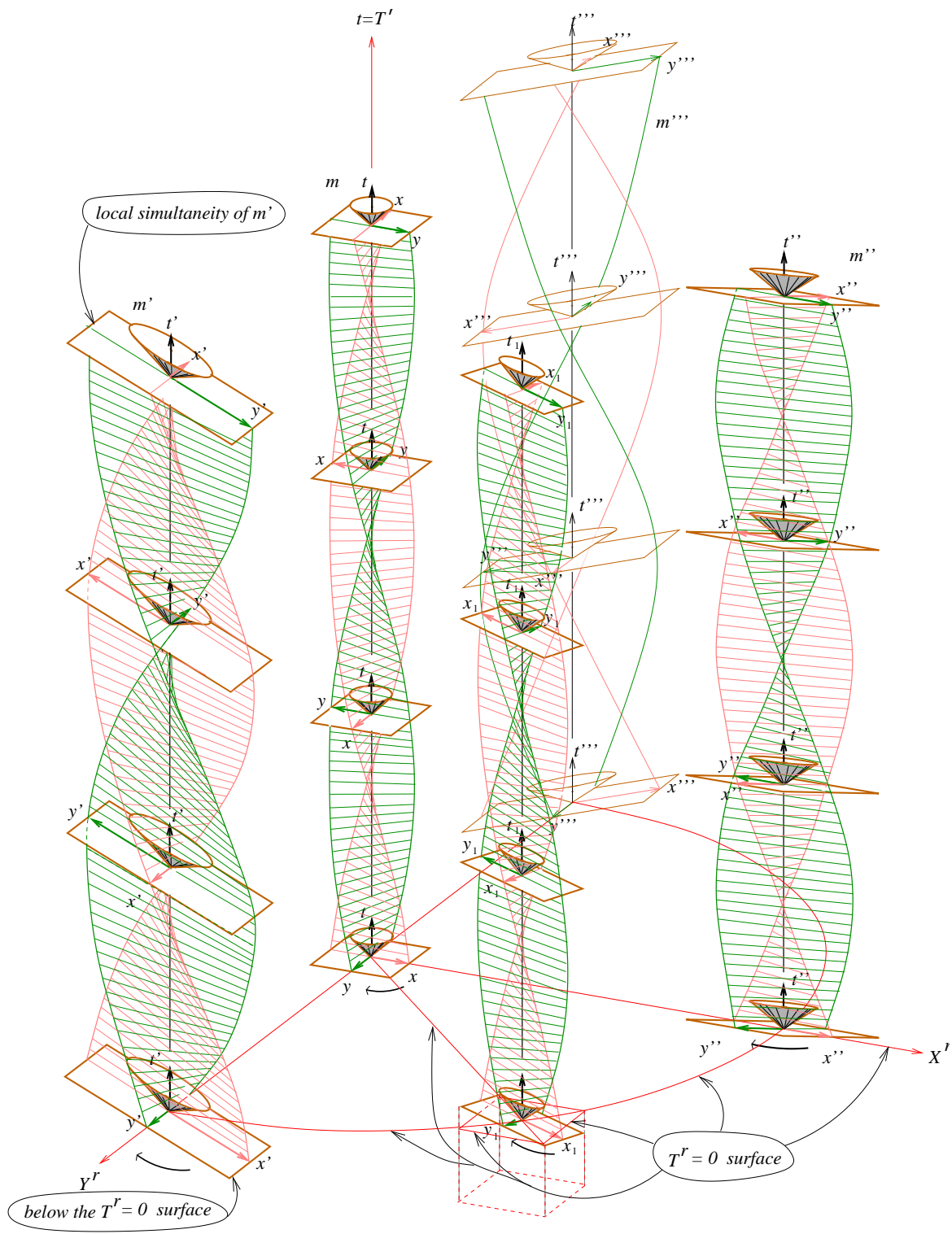}
\end{picture}
\end{center}
\caption[haho]{\label{dervis2}
We have a system of static, non-moving massive observers $m,m',m''$
etc.\ (the same as in Figures~\ref{inercelo}--\ref{hajnalka}) whose
cosmic compasses i.e.\ whose local coordinate systems are spinning
around creating a whirling effect. Gott~\cite[p.91]{Gott} called
these ``whirling dervishes''. This arrangement can be used to show
that Mach's principle is violated.} See p.\pageref{Mach-page} for
explanation.
 \end{figure}

\vfill\eject\newpage

\begin{figure}[!hp]
\setlength{\unitlength}{0.1 truemm} \small
\begin{center}
\begin{picture}(1000,2060)(0,0)
\epsfysize = 2060  \unitlength \epsfbox{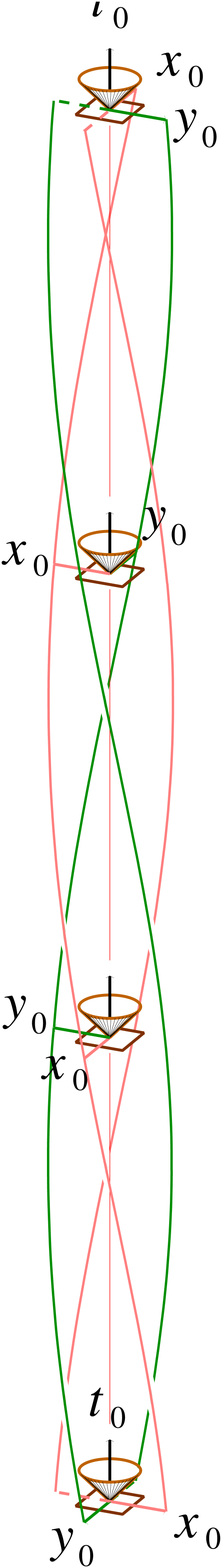}
\end{picture}
\end{center}
\caption{\label{nade0-fig} A typical dervish consisting of massive
observer (or galaxy) $m_0$ and its cosmic compasses $\langle
x_0,y_0,z_0\rangle$. In other words, $m_0$'s dervish is $m_0$'s {\em
local} coordinate system. $\omega=\pi/15$. }
 \end{figure}

\vfill\eject\newpage


\begin{figure}[!hp]
\setlength{\unitlength}{0.062 truemm} \small
\begin{center}
\begin{picture}(2460,3360)(0,0)
\epsfysize =  3360 \unitlength \epsfbox{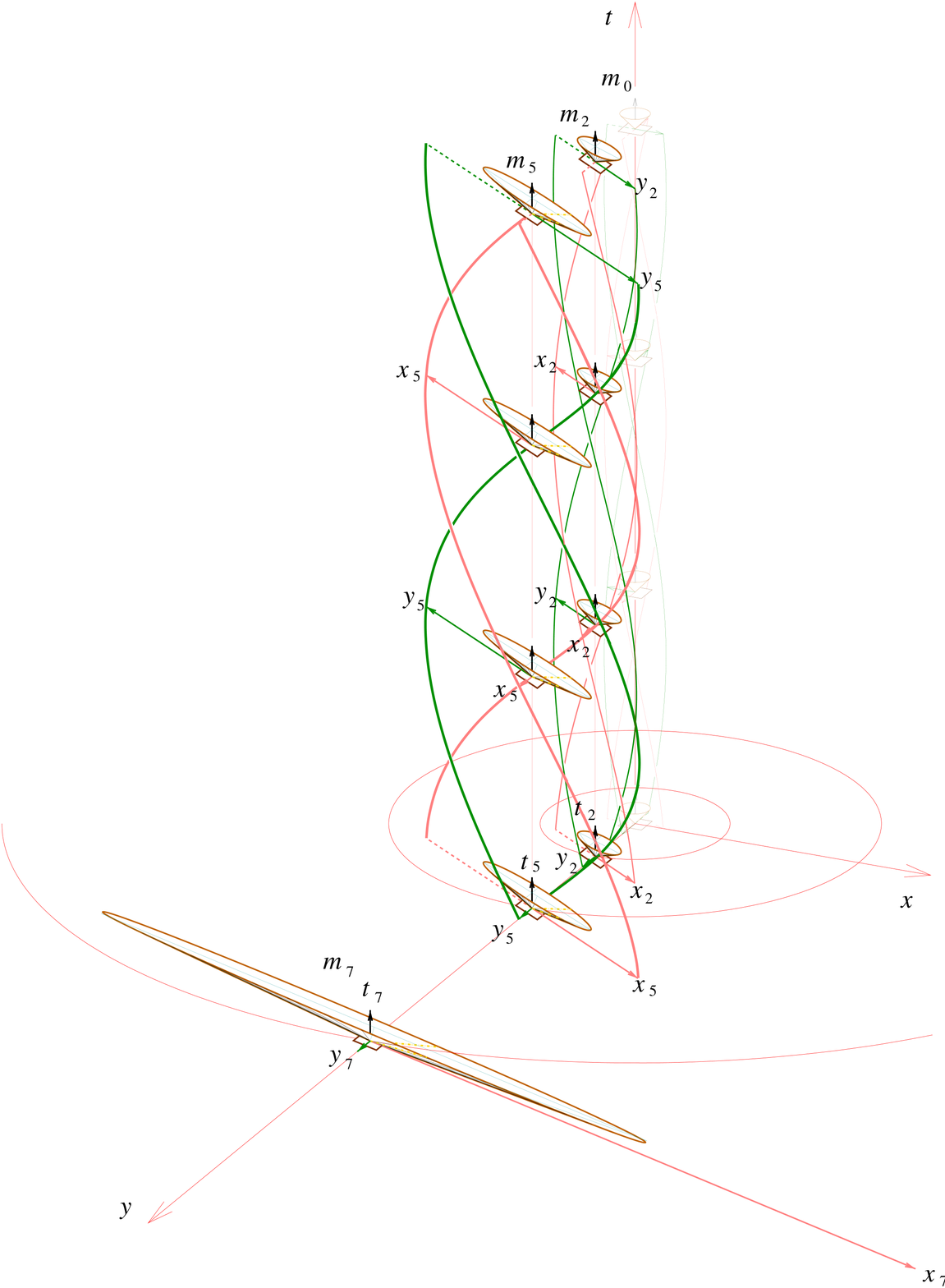}
\end{picture}
\end{center}
\caption{\label{nade1-fig} Dervishes $m_0,\dots,m_7$ involving
greater radiuses, hence more ``violent'' whirling effects.
$\omega=\pi/15$. Re-calibrated version of Map 2 applies, cf.\
p.\pageref{map2-fig}
.}
 \end{figure}

\vfill\eject\newpage


\begin{figure}[!hp]
\setlength{\unitlength}{2 truemm} 
\begin{center}
\begin{picture}(60,100)(9,-5) 
\epsfysize = 100\unitlength   
\epsfbox{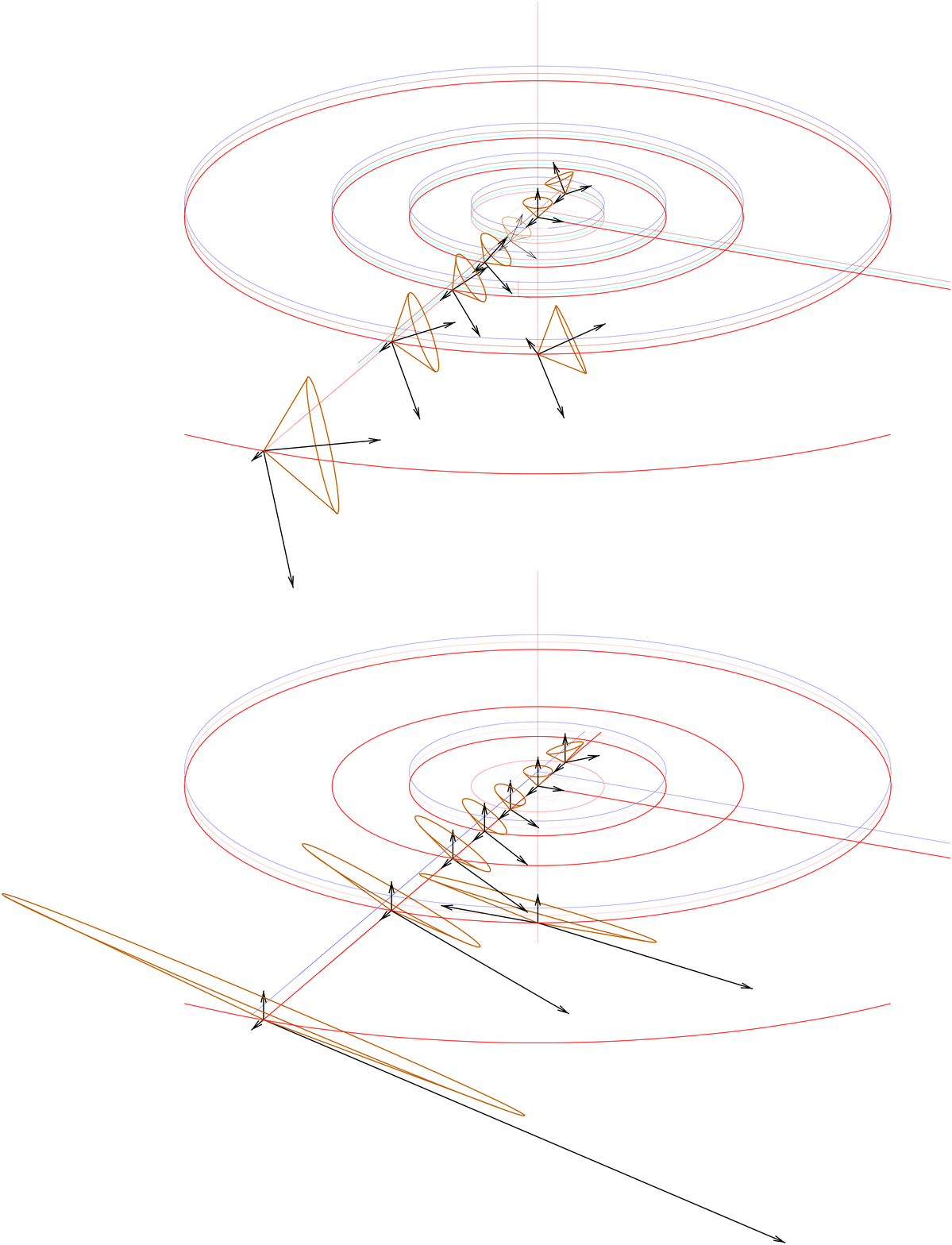}
\end{picture}
\end{center}
\caption{\label{uder1-fig} Light-cones and local unit vectors of
spiral world above, and their counterparts in dervish world $\langle
T^r,\dots, Z^r\rangle$ below. Detailed representation of upper part
is in Figures~\ref{spi-fig}, \ref{spi1-fig}, \ref{torta2-fig} and
that of lower part is in next Figure~\ref{torta-fig}. See also
Figures~\ref{dervis1}-\ref{nade1-fig}. The transformation between
the two worlds is described on
pp.\pageref{coord1}-\pageref{coord2}.}
\end{figure}

\vfill\eject\newpage


\begin{figure}[!hp]
\setlength{\unitlength}{0.085 truemm} \small
\begin{center}
\begin{picture}(1914,2603)(0,0)
\epsfysize = 2603  \unitlength \epsfbox{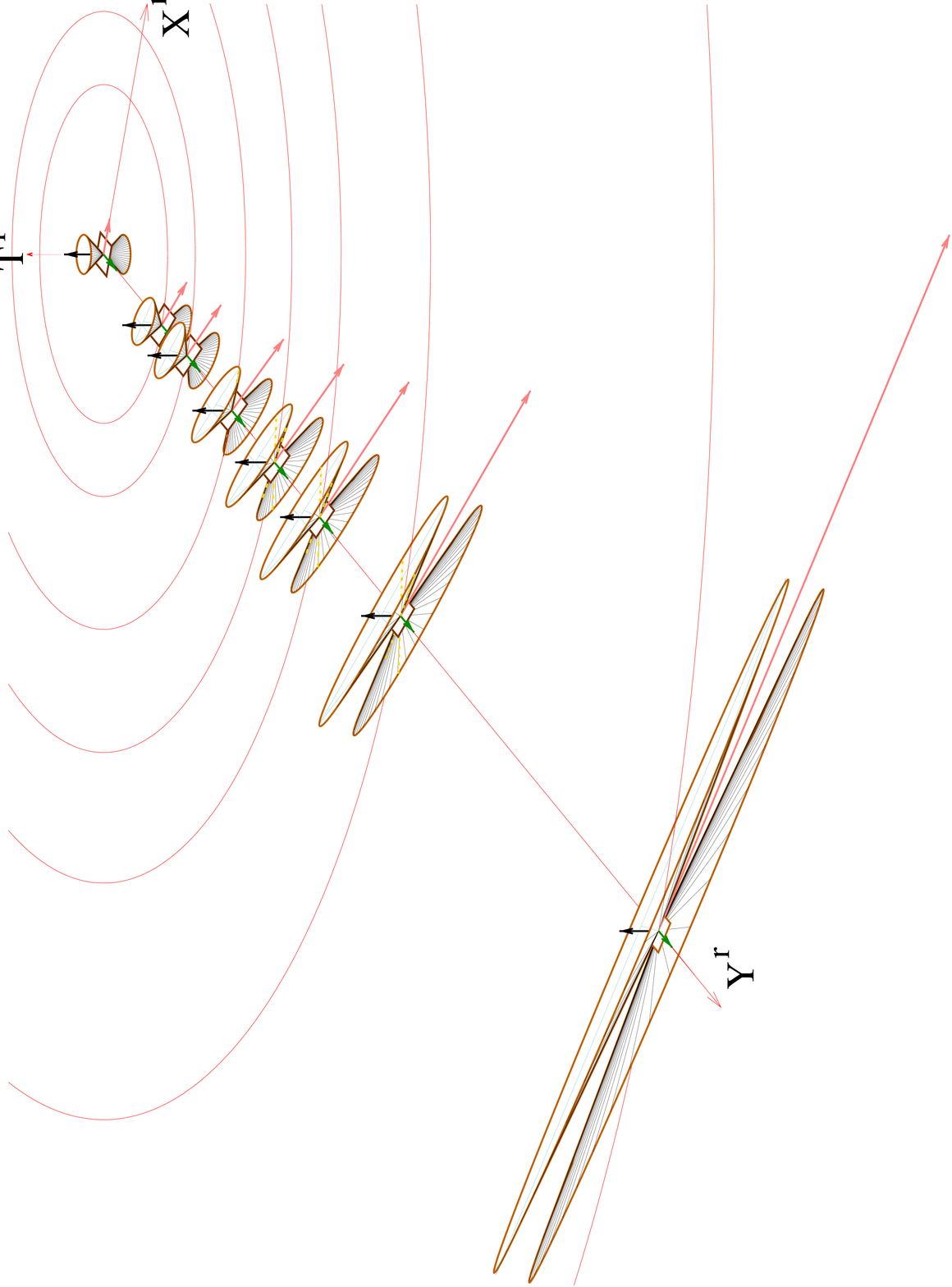}
\end{picture}
\end{center}
\caption{\label{torta-fig} Light-cones with local unit vectors in
dervish world $\langle T^r,\dots\rangle$. $\omega=\pi/30$, Map 2
applies.}
 \end{figure}

\vfill\eject\newpage

\section{Fine-tuning the space-time structure of the Naive GU obtained
so far. Tilting the light-cones.} \label{godel-section}
\label{tilting-section}

First we show two pictures hinting at the fact that the lengths of
unit-vectors etc.\ in our Naive Dervish World might be of
inconvenient proportions.

\begin{figure}[!hp]
\setlength{\unitlength}{0.085 truemm}
\small
\begin{center}
\begin{picture}(1980,2260)(0,0)
\epsfysize =2260   \unitlength \epsfbox{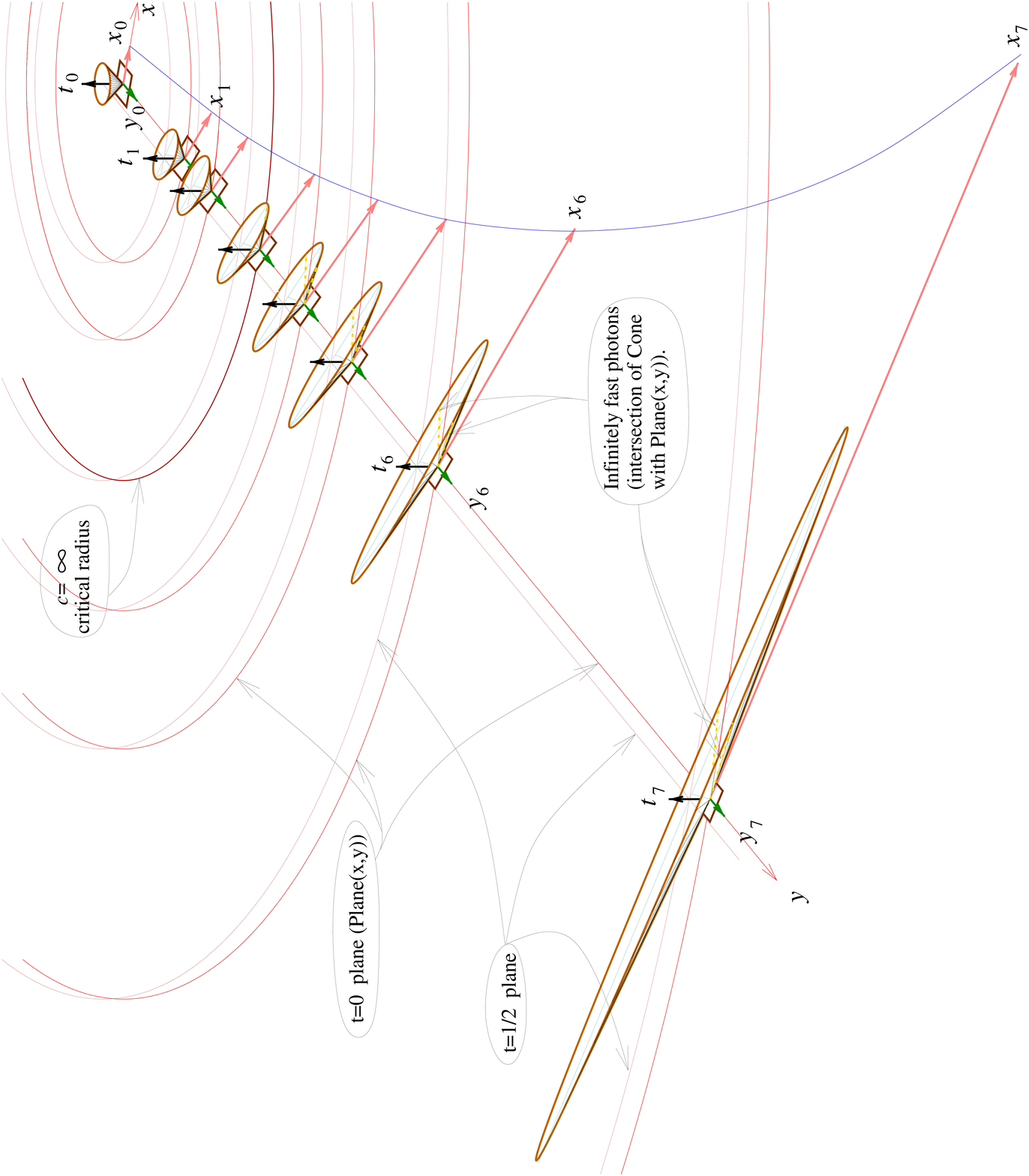}
\end{picture}
\end{center}
\caption{\label{naivtorta1-fig} $\omega=\pi/30$, Map 2 applies.}
 \end{figure}

\begin{figure}[!hp]
\setlength{\unitlength}{0.057 truemm} \small
\begin{center}
\begin{picture}(2940,3340)(0,0)
\epsfysize =  3340 \unitlength \epsfbox{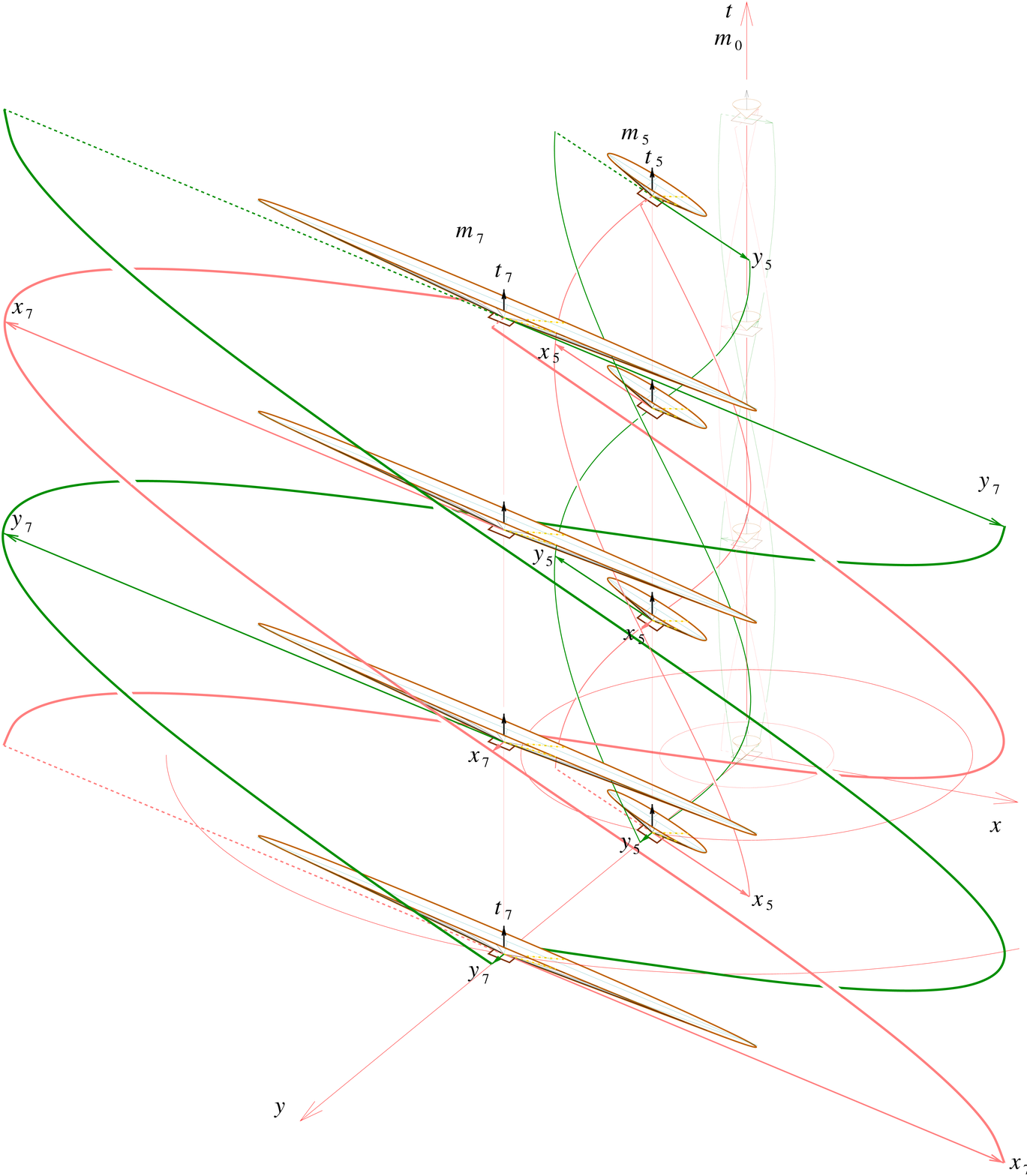}
\end{picture}
\end{center}
\caption{\label{nade2-fig} Whirling dervishes on larger radiuses.
Re-calibrated version of Map 2 applies as follows. $r'(m_i)=2\cdot
r(m_i), v'(m_i) = v(m_i), \omega' = \omega/2$; where $r',v',\omega'$
belong to the present figure while $r,v,\omega$ belong to Map 2.}
 \end{figure}

\newpage

\begin{figure}[!hbtp]
\setlength{\unitlength}{0.044 truemm} \small
\begin{center}
\begin{picture}(3380,4820)(0,0)

\epsfysize = 4820  \unitlength \epsfbox{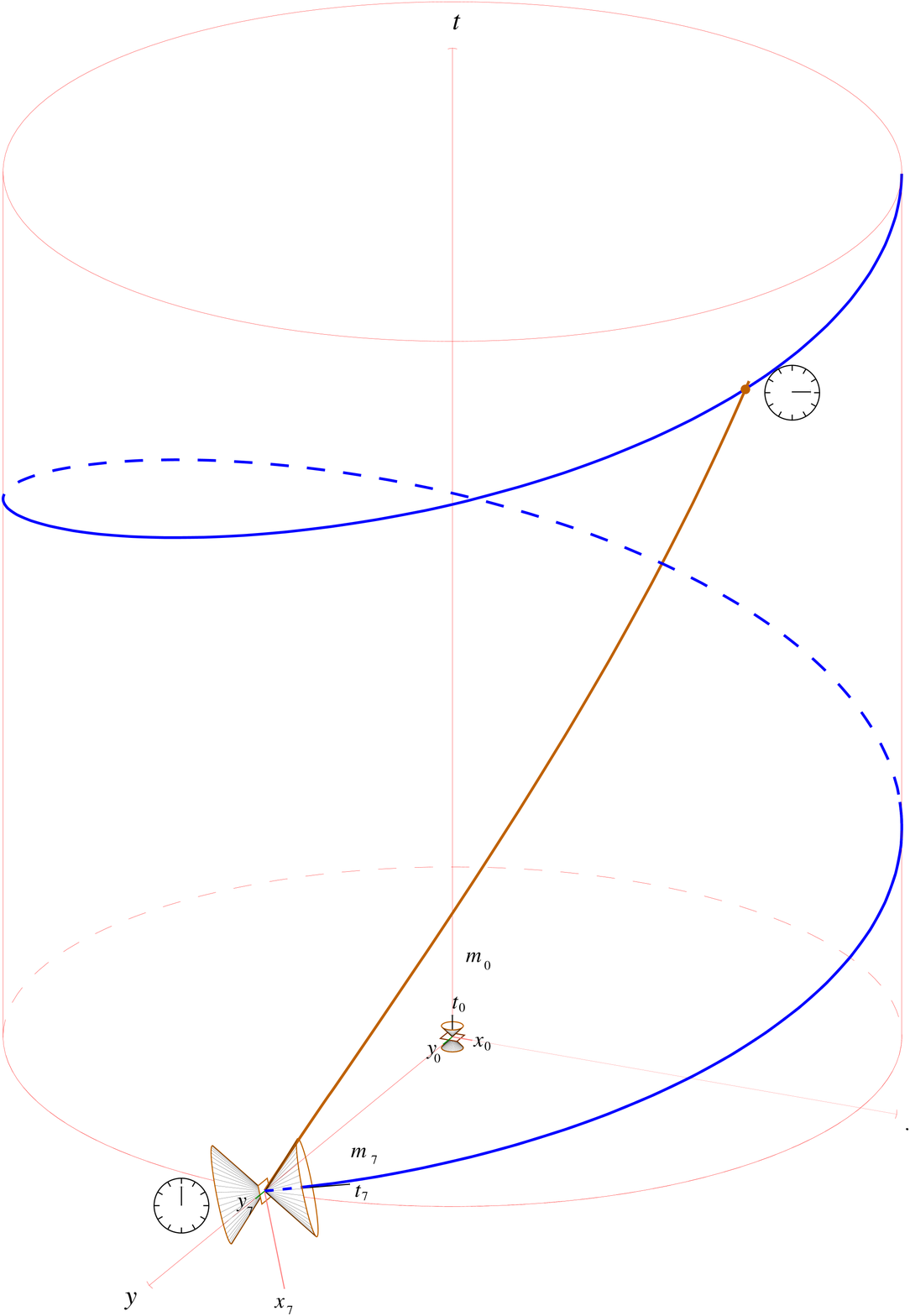}
\end{picture}
\end{center}
\caption{\label{bizony-fig} The time needed for a photon sent out by
$m_7$ and kept with mirrors on a circle around $m_0$ to come back is
a little more than the time needed for the universe to make a turn.}
 \end{figure}

The fact that the $x_i$ vector of $m_i$ has a much longer component
parallel with coordinate $X^r$ than $x_0$ (illustrated in the
previous two figures) is the visual manifestation of the following
fact, seen better in the spiral world. In the spiral world, $m_i$
can send a photon $ph$ upward almost parallel with the $t$ axis such
that $ph$ reaches $m_i$ again in a ``rigidly bounded'' time (an
upper bound is $4\pi/\omega$) where the bound is independent of the
choice of $i$. We choose the path of $ph$ such that its distance
from $m_0$ remains constant(ly the $m_0$--$m_i$ distance). This path
need not be geodesic but as G\"odel wrote, we can use mirrors to
force $ph$ to follow this path. See Figure~\ref{bizony-fig}.

In G\"odel's Universe the return-time of the photons sent around
$m_0$ in a circle of radius $r$ tend to infinity as $r$ tends to
infinity.
\smallskip

Let us see how we can remove this difference with G\"odel's universe
without destroying the logic of our construction. How can we
fine-tune our construction? We are aiming at the ``smallest'' and
simplest change so that the logic of our construction would remain
intact. Changing the length's of the $x_i$ vectors and keeping the
other unit-vectors as they were results in making the light-cones
narrower. Since this will not lead to CTC's, we will ``tilt" the
light-cones, instead. So, in fine-tuning the Naive GU we will speak
about tilting the light-cones, and we will call the new space-time
Tilted GU.

Let us work in the dervish world.
\bigskip

\noindent \underbar{Choice 1} \label{choice1-p}
 We can tilt the light-cones forwards (in the
positive $\varphi$ direction) such that with increasing $r$ (radius)
we also increase the tilting. This can be done in such a manner that
the difference we talked about disappears. The result of such
tilting results a version of NGU represented in
Sections~\ref{tilting-section}-\ref{refined-section}
(Figures~\ref{godtorta2-fig}--\ref{2vis-fig}). The so obtained
tilted universe resembles very closely the universes presented in
G\"odel's papers. (E.g.\ they agree in many structural properties
[in G\"odel's sense].)
\bigskip

\begin{figure}[!ht]
\setlength{\unitlength}{0.7 truemm} 
\begin{center}
\begin{picture}(87,60)(0,0) 
\put(20,60){\makebox(0,0){t}} \put(23,45){\makebox(0,0){$m_i$}}
\put(90,5){\makebox(0,0){$\varphi$}}
\epsfysize = 60\unitlength   
\epsfbox{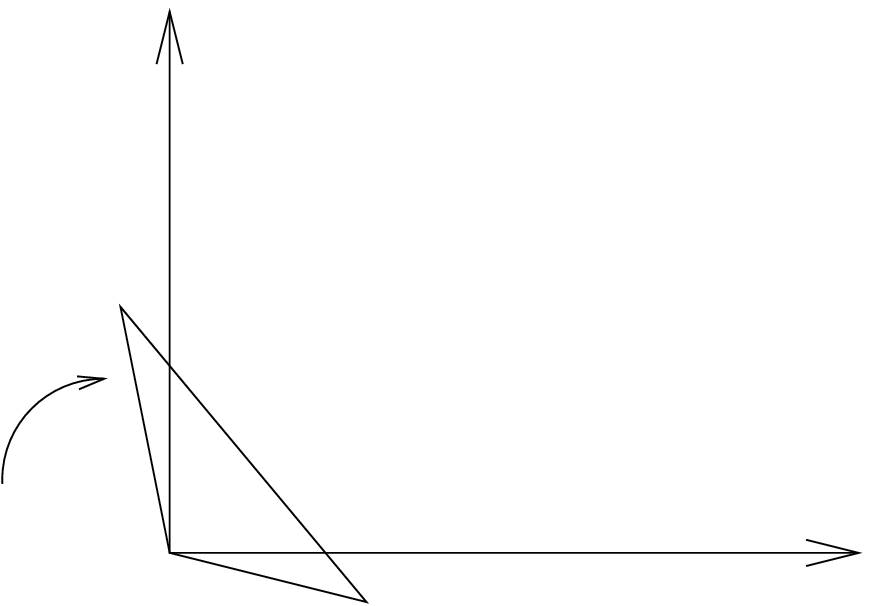}
\end{picture}
\end{center}
\caption{\label{forward-fig} Choice 1 is that we tilt the
light-cones forwards.}
 \end{figure}

\begin{figure}[!ht]
\setlength{\unitlength}{0.7 truemm} 
\begin{center}
\begin{picture}(88,60)(0,0) 
\put(35,60){\makebox(0,0){t}} \put(38,45){\makebox(0,0){$m_i$}}
\put(90,5){\makebox(0,0){$\varphi$}}
\epsfysize = 60\unitlength   
\epsfbox{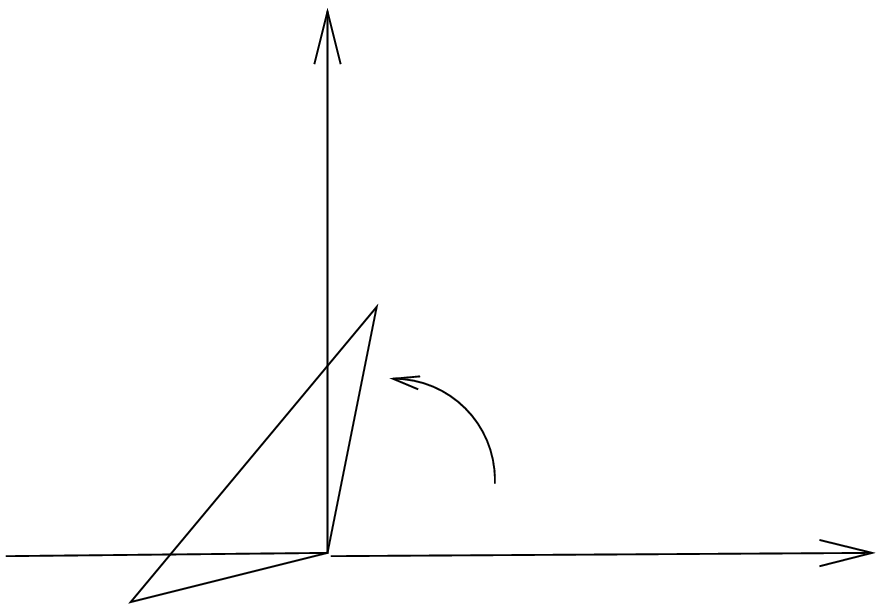}
\end{picture}
\end{center}
\caption{\label{backward-fig} Choice 2 is that we tilt the
light-cones backwards.}
 \end{figure}

\noindent \underbar{Choice 2} We can also tilt the light-cones (in
dervish world) backwards, opposite to the $\varphi$ direction,
carefully enough such that the difference goes away and we do not
induce other undesirable effects. See Figure~\ref{backward-fig}. 
This Choice~2 tilting is just Choice~1 tilting seen from another
coordinate system (namely by using the coordinate transformation
$\varphi\to-\varphi$). Below we will explore Choice~1, and in
Section~\ref{gyroscope-section} (p.\pageref{gyroscope-section}) we
explore Choice~2. We will see that both Choice~1 and Choice~2 have
their advantages.\label{choicevege-p}
\bigskip

From now on, we concentrate on Choice 1.
\bigskip
\vfill\eject

 We will call the tilting in Choice~1 ``{\em
forward-tilting}'', the so obtained dervishes {\em tilted
dervishes}, and the so obtained (tilted) dervish world {\em Tilted
Dervish World} or {\em Choice~1 Dervish World}. Recall that we
describe a simple transformation between the spiral world $\langle
t^s,\dots\rangle$ and the dervish world $\langle t^d,\dots\rangle$
in Section~\ref{technical-section} (p.\pageref{technical-section}).
We use this transformation for transforming the new, tilted universe
from the dervish world to the spiral world. We call the result {\em
Tilted Spiral World} or use simply the adjective ``new spiral'' or
``refined-spiral'' for referring to the so obtained light-cones as
new spiral cones or back rotated ones. The expression ``rotating
back'' or ``back-rotating'' intends to refer to application of the
inverse transformation $\langle
t^d,\dots\rangle\longrightarrow\langle t^s,\dots\rangle$ described
in Section~\ref{technical-section}. In such contexts the inverse
transformation is applied to the result of forward-tilting.

The result of the above outlined forward-tilting is the G\"odel-type
universe which we will describe in more detail in the coming parts.
We will call this space-time Tilted GU (or sometimes new GU).
Instead of defining the tilting of the cones at each point, we will
give details of the tilting for the cones occurring in the figures
only. These tilted light-cones (with local unit-vectors) and their
new spiral versions are depicted and constructed in detail in
Section~\ref{technical-section}. These objects (light-cones, $m_i$'s
etc) are systematically arranged in space-time (i.e.\ are
coordinatized) in Maps 1,2 (pp.\pageref{map1-fig},
\pageref{map2-fig}). These maps also include angular velocities,
tangential velocities.\bigskip

In this section we describe ``Tilted Dervish World'', and in the
next section, Section~\ref{refined-section}, we describe ``Tilted
Spiral World''.
\bigskip
\label{goduniv1-vege}
\newpage

\section*{Tilted dervishes (fuller description of new GU in dervish world).}
\label{fuller-section}

\begin{figure}[!hp]
\setlength{\unitlength}{0.1 truemm} \small
\begin{center}
\begin{picture}(1600,1660)(0,0)
\epsfysize = 1660  \unitlength \epsfbox{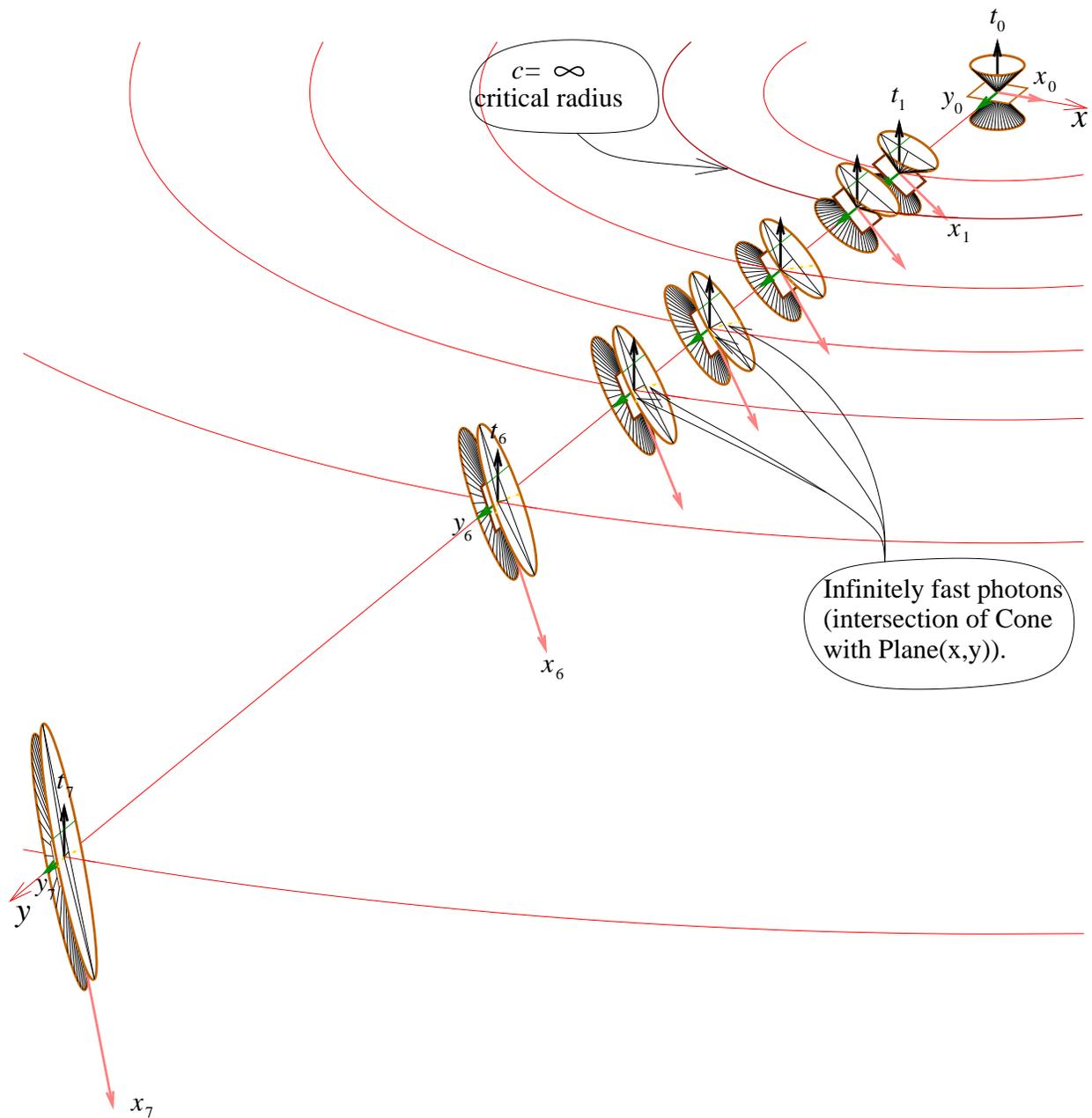}
\end{picture}
\end{center}
\caption{\label{godtorta2-fig} Tilted-dervish universe or Choice~1
Dervish World. Light-cones, local unit-vectors along the $y$-axis.
 $\omega=\pi/30$, Map 2 applies.}
 \end{figure}

\begin{figure}[hp]
\setlength{\unitlength}{0.1 truemm} \small
\begin{center}
\begin{picture}(1600,1760)(0,0)
\epsfysize = 1760  \unitlength \epsfbox{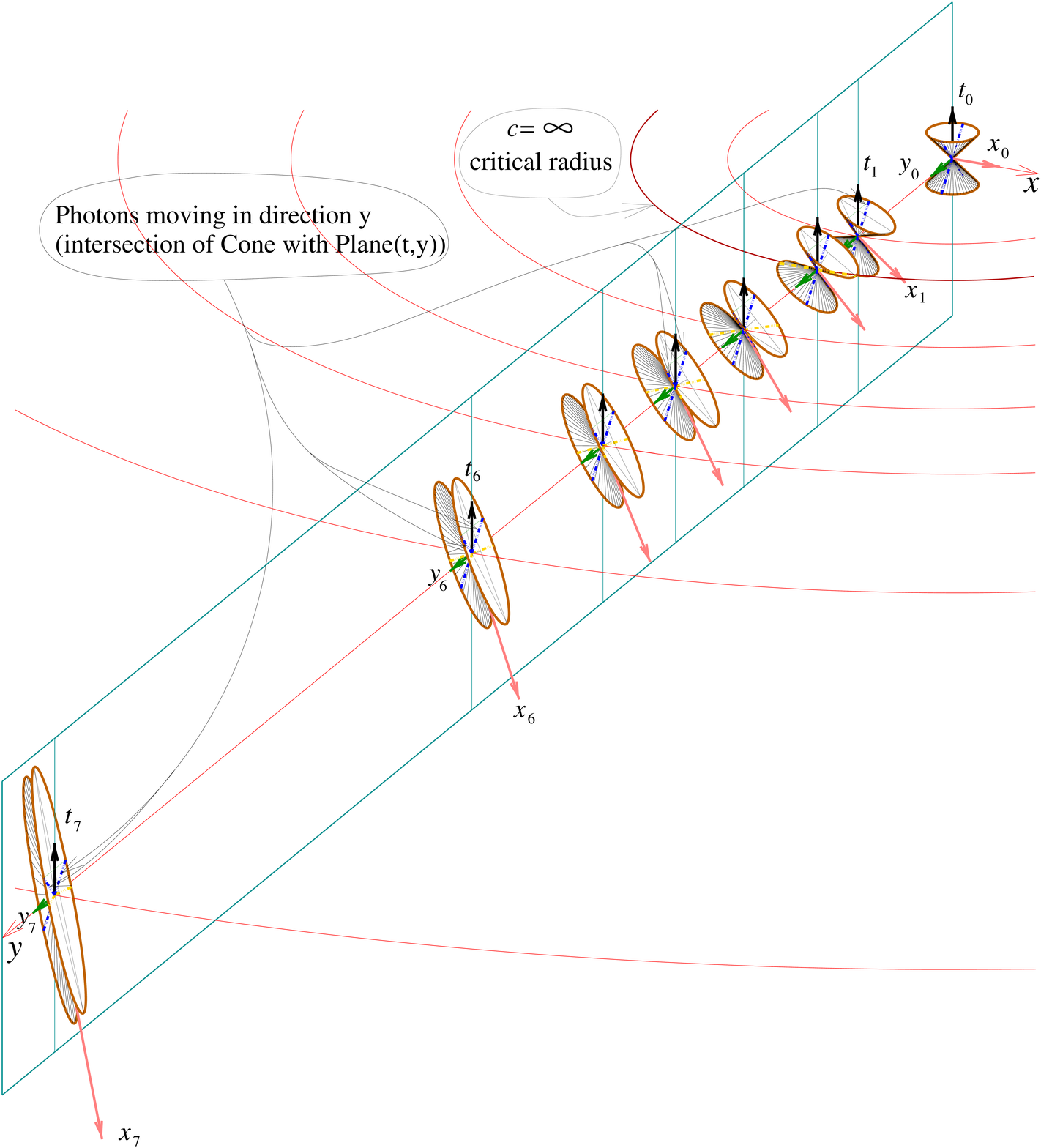}
\end{picture}
\end{center}
\caption{\label{godtorta2a-fig} Tilted Dervish World (Choice 1
Dervish World). $\omega=\pi/30$, Map 2 applies.}
 \end{figure}


\begin{figure}[hp]
\setlength{\unitlength}{0.165 truemm} \small 
\begin{center}
\begin{picture}(1020,1220)(0,0)
\epsfysize = 1220  \unitlength \epsfbox{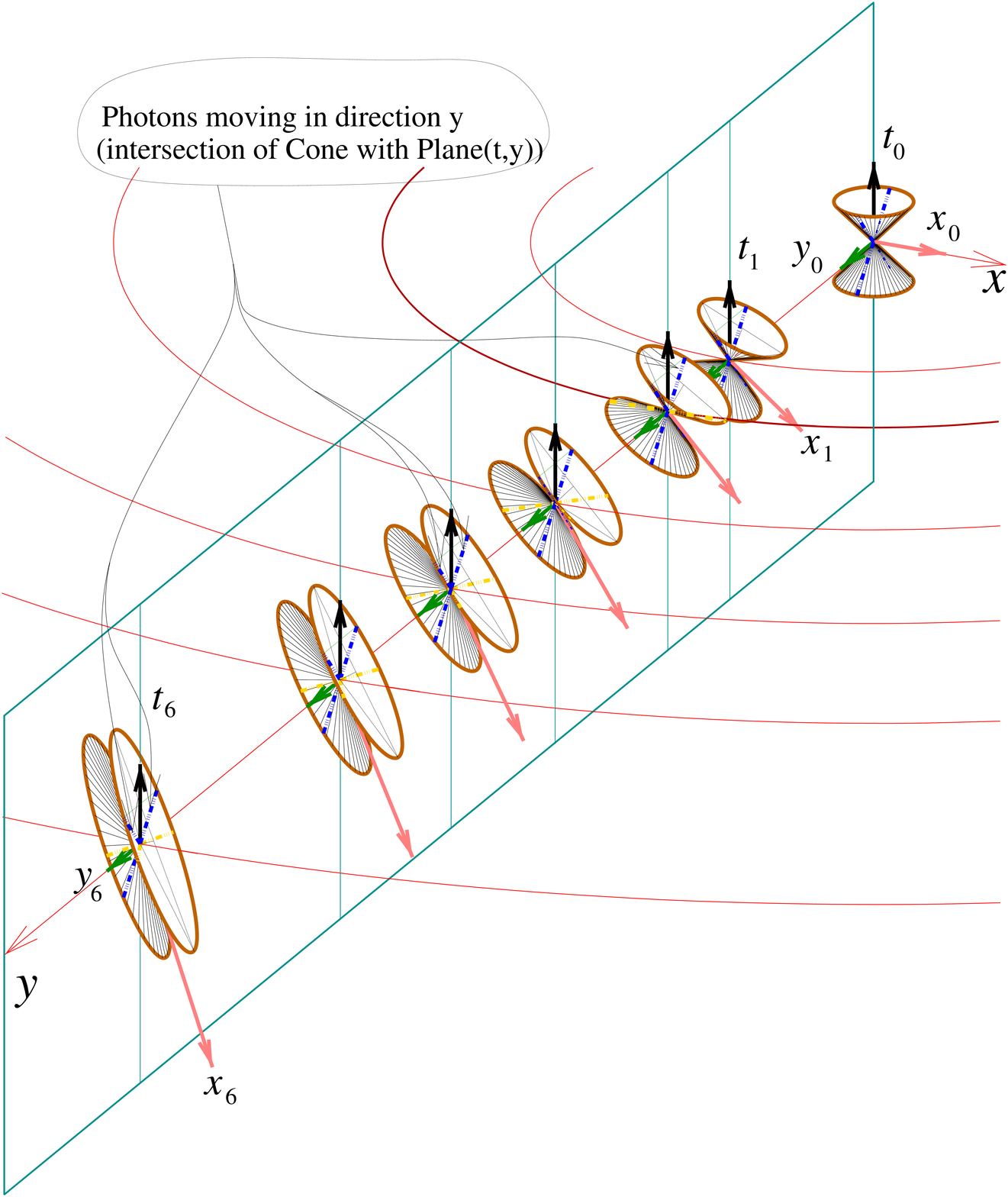}
\end{picture}
\end{center}
\caption{\label{godtorta3a-fig} Tilted Dervish World.
$\omega=\pi/30$, Map 2 applies.}
 \end{figure}


\begin{figure}[hp]
\setlength{\unitlength}{0.1 truemm} \small
\begin{center}
\begin{picture}(1600,1760)(0,0)
\epsfysize = 1760  \unitlength \epsfbox{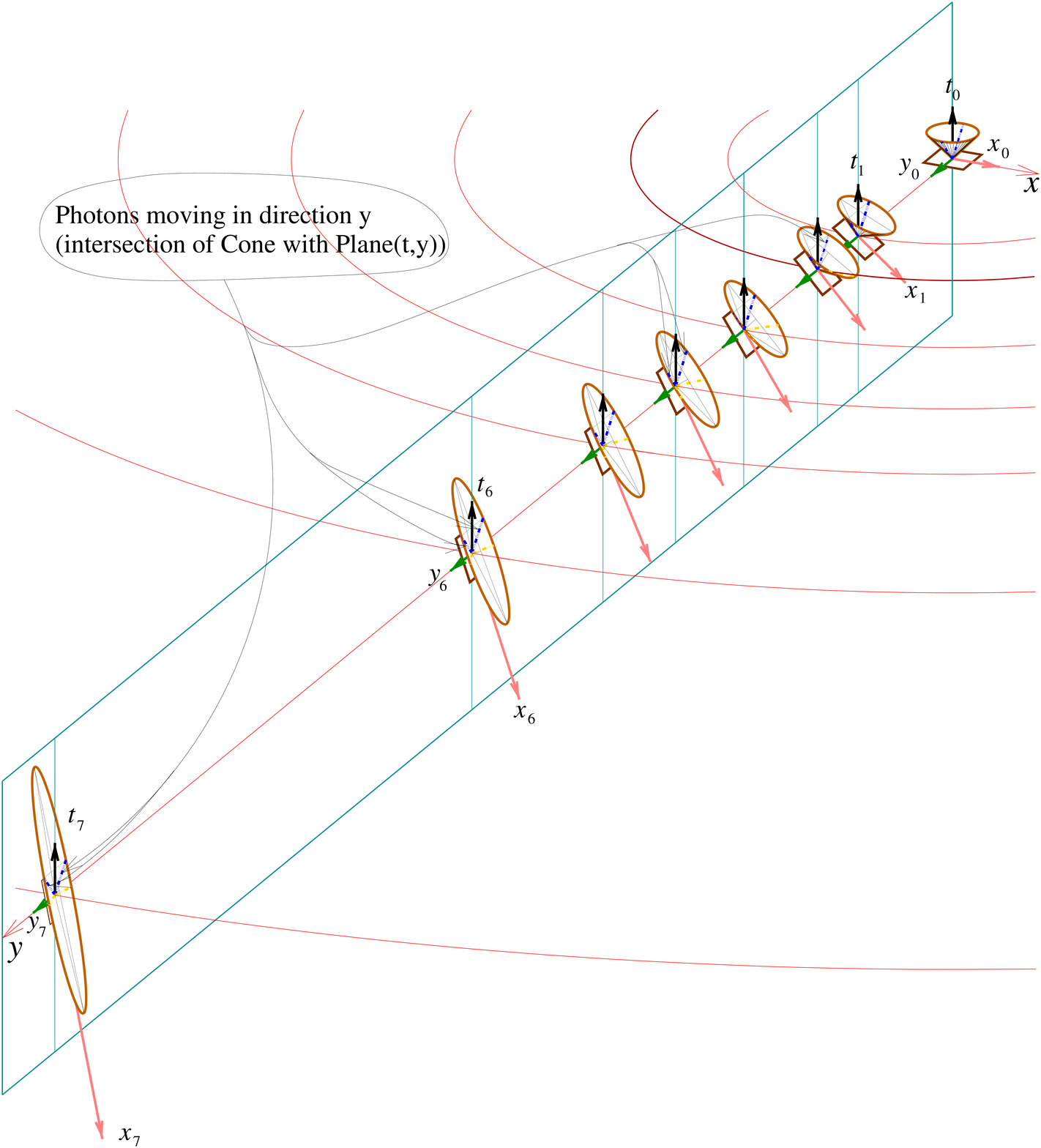}
\end{picture}
\end{center}
\caption{\label{godtorta1-fig} Tilted Dervish World.
$\omega=\pi/30$, Map 2 applies.}
 \end{figure}



\begin{figure}[p]
\setlength{\unitlength}{0.046 truemm} \small
\begin{center}
\begin{picture}(2240,4500)(0,0)
\epsfysize =   4500\unitlength \epsfbox{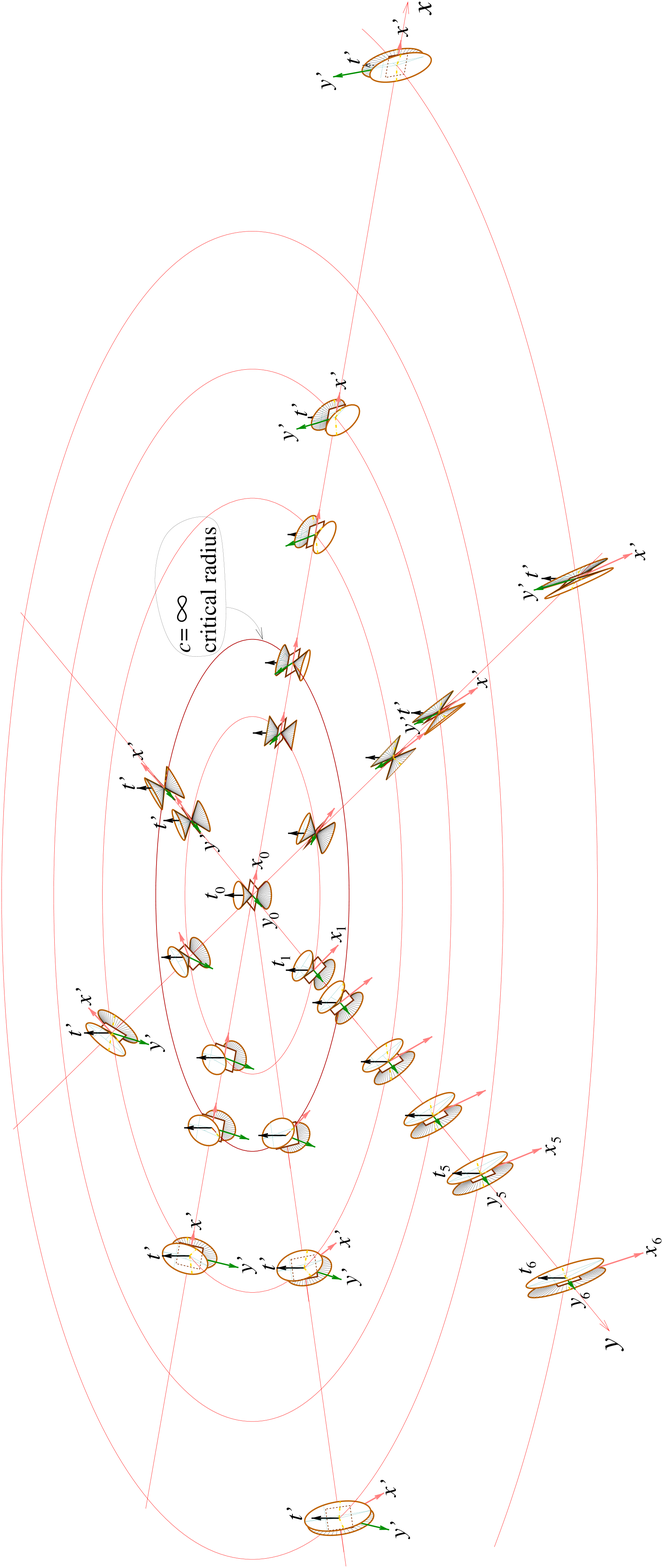}
\end{picture}
\end{center}
\caption{\label{goder0-fig} Tilted-dervish universe. (Choice 1
Dervish World.) Light-cones, local unit-vectors on the $xy$-plane.
 $\omega=\pi/45$, Map 1 applies.}
 \end{figure}


\begin{figure}[p]
\setlength{\unitlength}{0.065 truemm} \small
\begin{center}
\begin{picture}(2020,3260)(0,0)
\epsfysize =3260   \unitlength \epsfbox{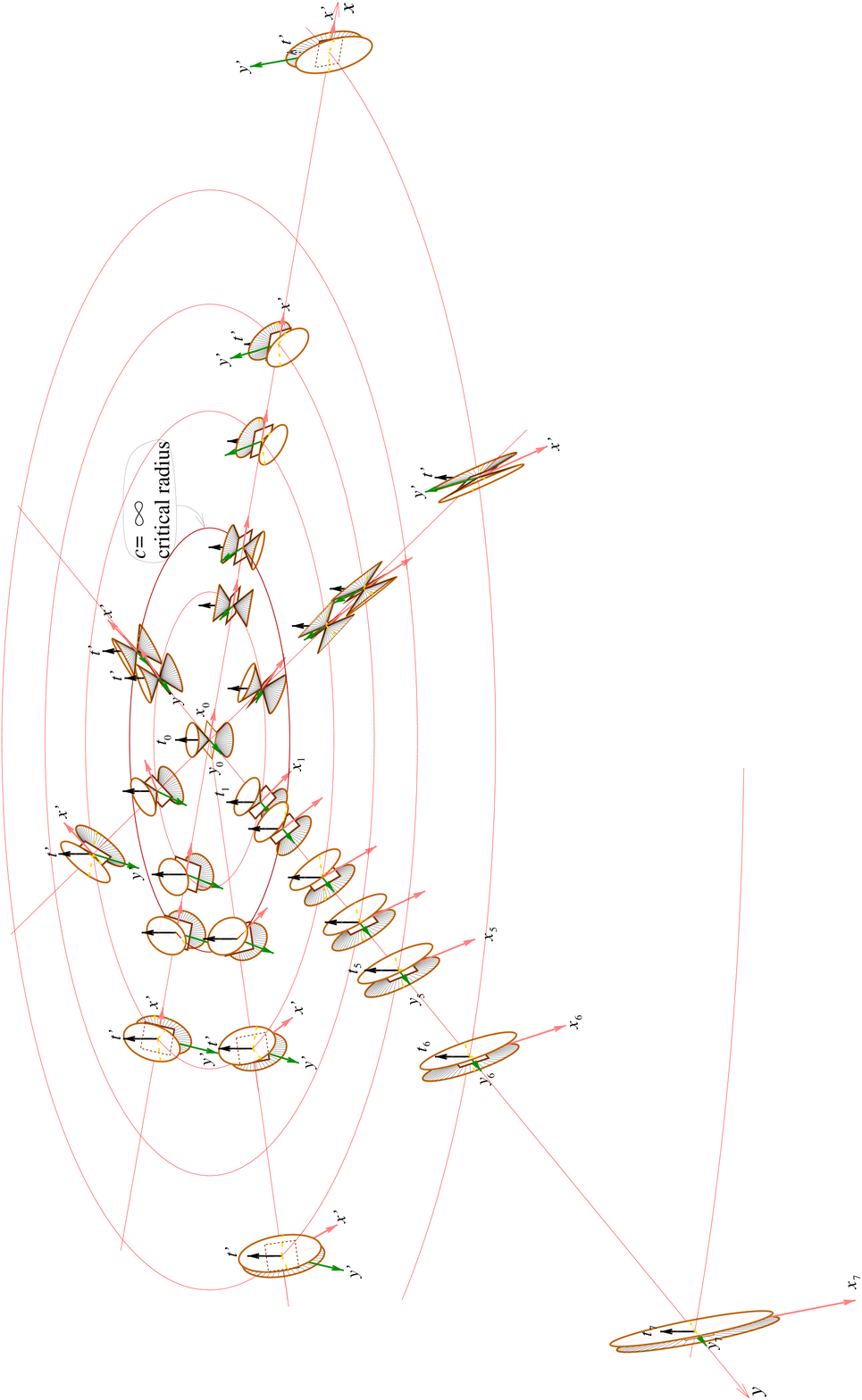}
\end{picture}
\end{center}
\caption{\label{Goder1-fig} Tilted Dervish World. $\omega=\pi/30$,
Map 2 applies.}
 \end{figure}


\begin{figure}[p]
\setlength{\unitlength}{0.085 truemm} \small
\begin{center}
\begin{picture}(1280,2480)(0,0)
\epsfysize = 2480  \unitlength \epsfbox{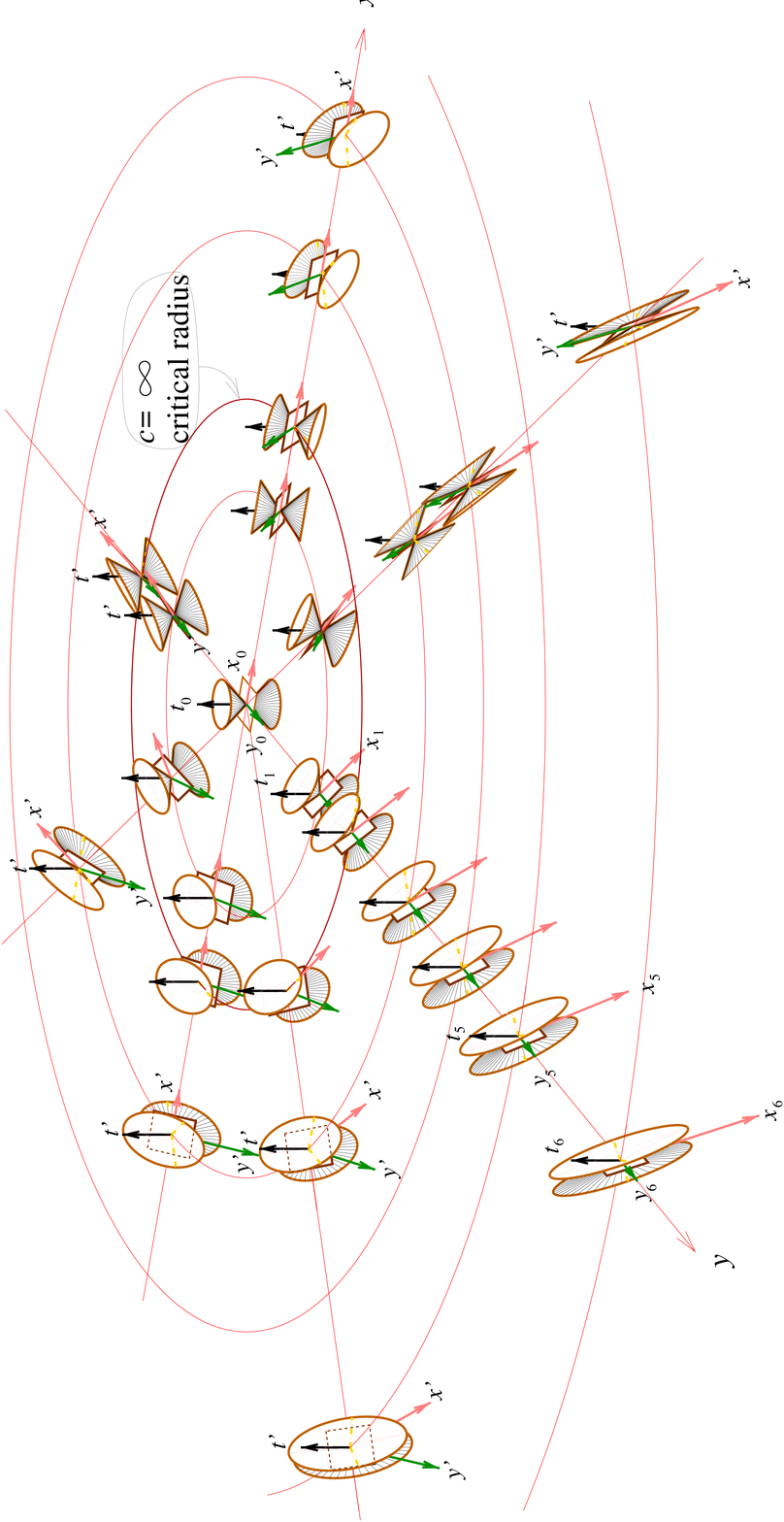}
\end{picture}
\end{center}
\caption{\label{Goder3-fig} Tilted Dervish World. $\omega=\pi/30$,
Map 2 applies.}
 \end{figure}


\begin{figure}[p]
\setlength{\unitlength}{0.065 truemm} \small
\begin{center}
\begin{picture}(2020,3260)(0,0)
\epsfysize =3260   \unitlength \epsfbox{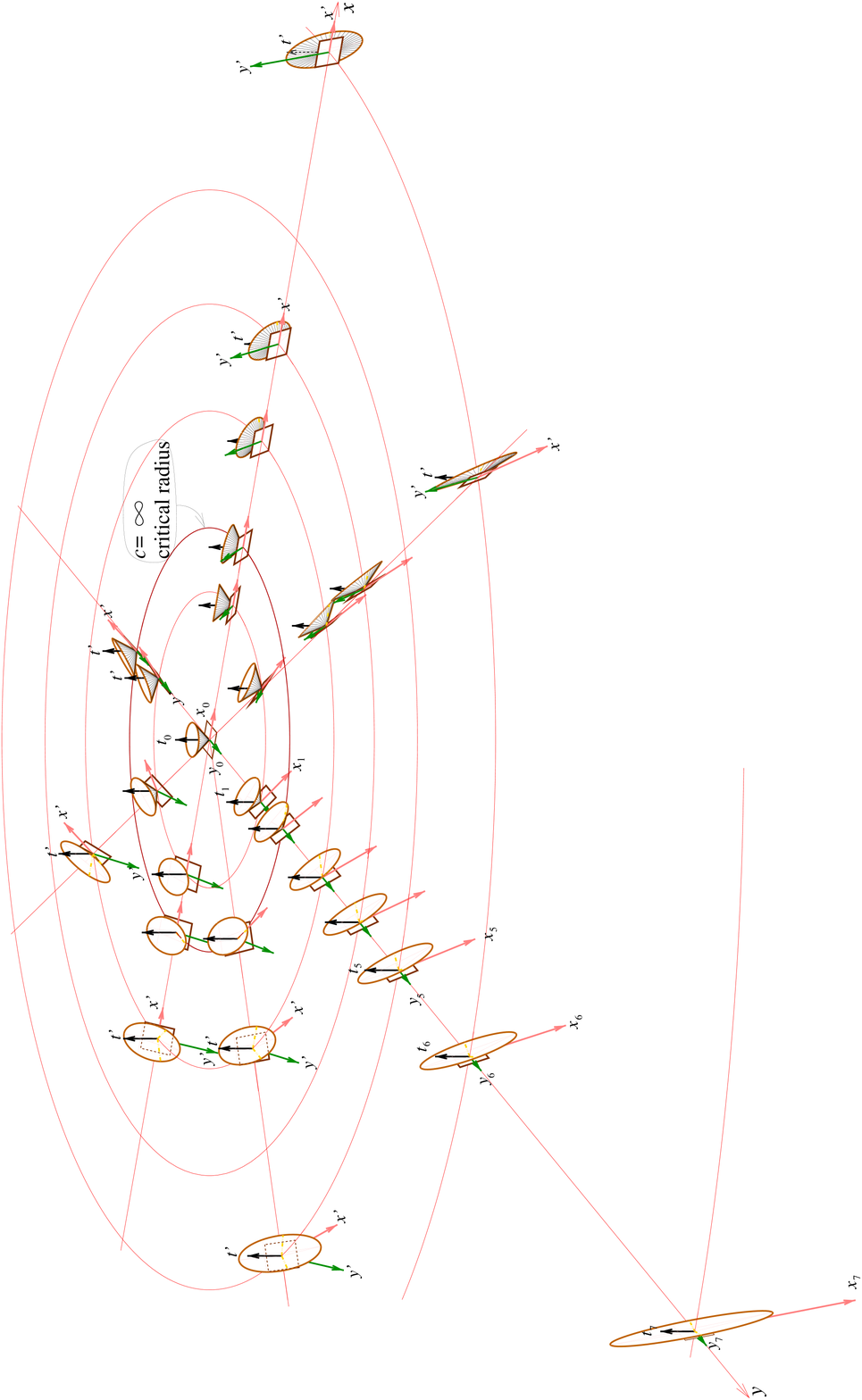}
\end{picture}
\end{center}
\caption{\label{1goder-fig} Tilted Dervish World. Compare with
Figure 61 on p.169 in Hawking-Ellis~\cite{Hawel} (cf.\ also
Fig.\ref{ujgodel-fig} herein). $\omega=\pi/30$, Map 2 applies.}
 \end{figure}


\begin{figure}[p]
\setlength{\unitlength}{0.061 truemm}
\small
\begin{center}
\begin{picture}(2560,3260)(0,0)
\epsfysize = 3260  \unitlength \epsfbox{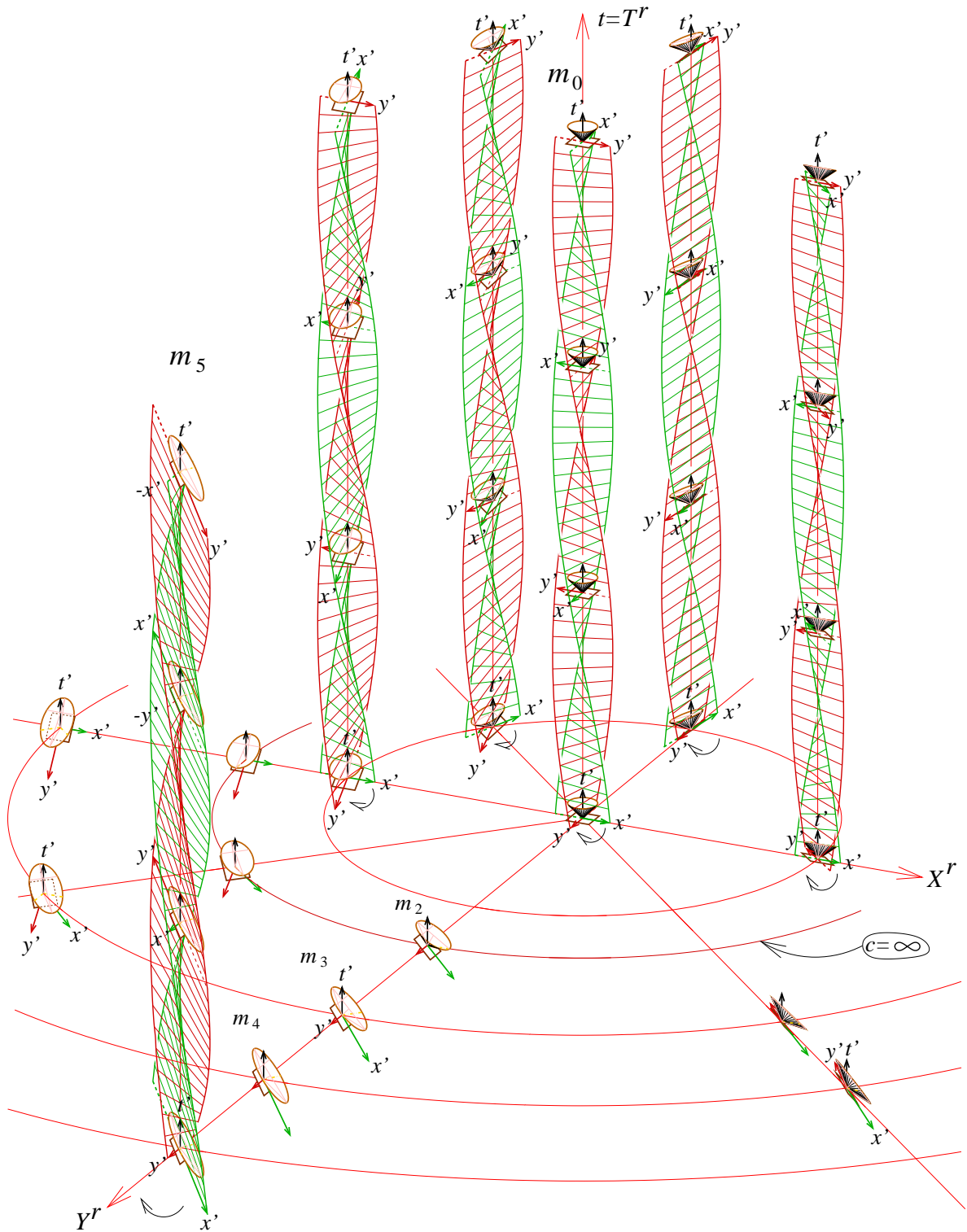}
\end{picture}
\end{center}
\caption{\label{5hastanc-fig} \label{6hastanc-fig} Tilted Dervish
World.
 ``$\omega$ of universe'' $=\ \pi/60$ (recalibrated version
of Map 2 applies). Spinning dervishes are artificially sped up
(``artificial $\omega$ of dervishes'' $=\ \pi/15$).}
 \end{figure}


\begin{figure}[p]
\setlength{\unitlength}{0.074 truemm} \small
\begin{center}
\begin{picture}(2268,2927)(0,0)
\epsfysize =  2927 \unitlength \epsfbox{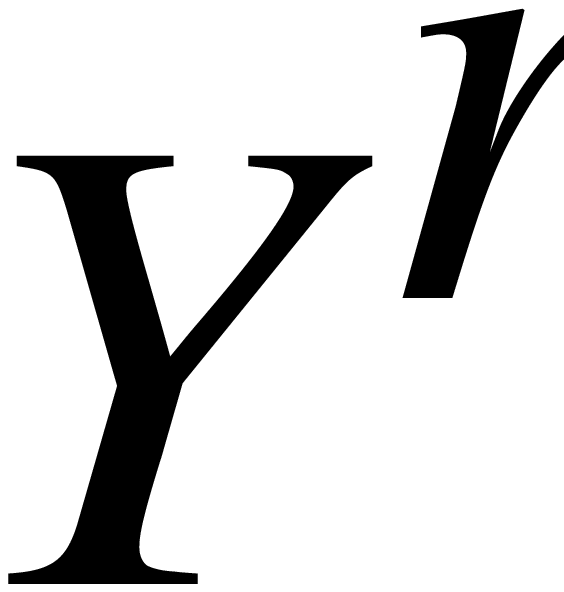}
\end{picture}
\end{center}
\caption{\label{gyors2-fig} Tilted Dervish World. $\omega=\pi/30$,
Map 2 applies.}
 \end{figure}

\label{goduniv2-vege}


\begin{figure}[!hp]
\setlength{\unitlength}{0.074 truemm} \small
\begin{center}
\begin{picture}(2268,2927)(0,0)
\epsfysize =  2927 \unitlength \epsfbox{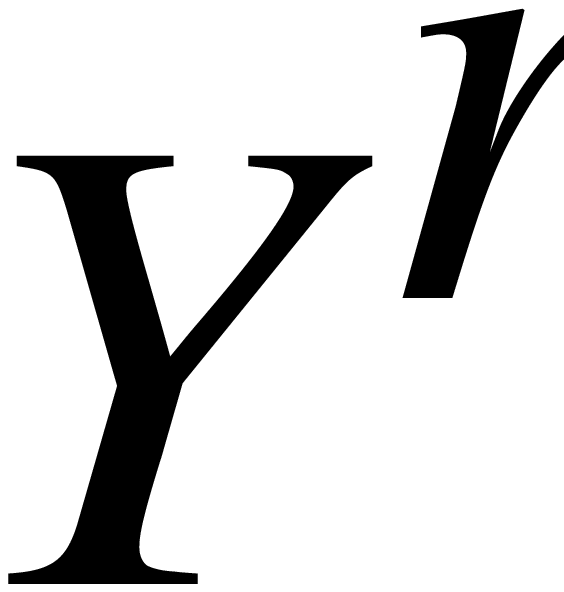}
\end{picture}
\end{center}
\caption{\label{gyors7-fig} Tilted Dervish World. $\omega=\pi/30$,
Map 2 applies.}
 \end{figure}


\begin{figure}[!hp]
\setlength{\unitlength}{0.072 truemm} \small
\begin{center}
\begin{picture}(2280,3020)(0,0)
\epsfysize =  3020 \unitlength \epsfbox{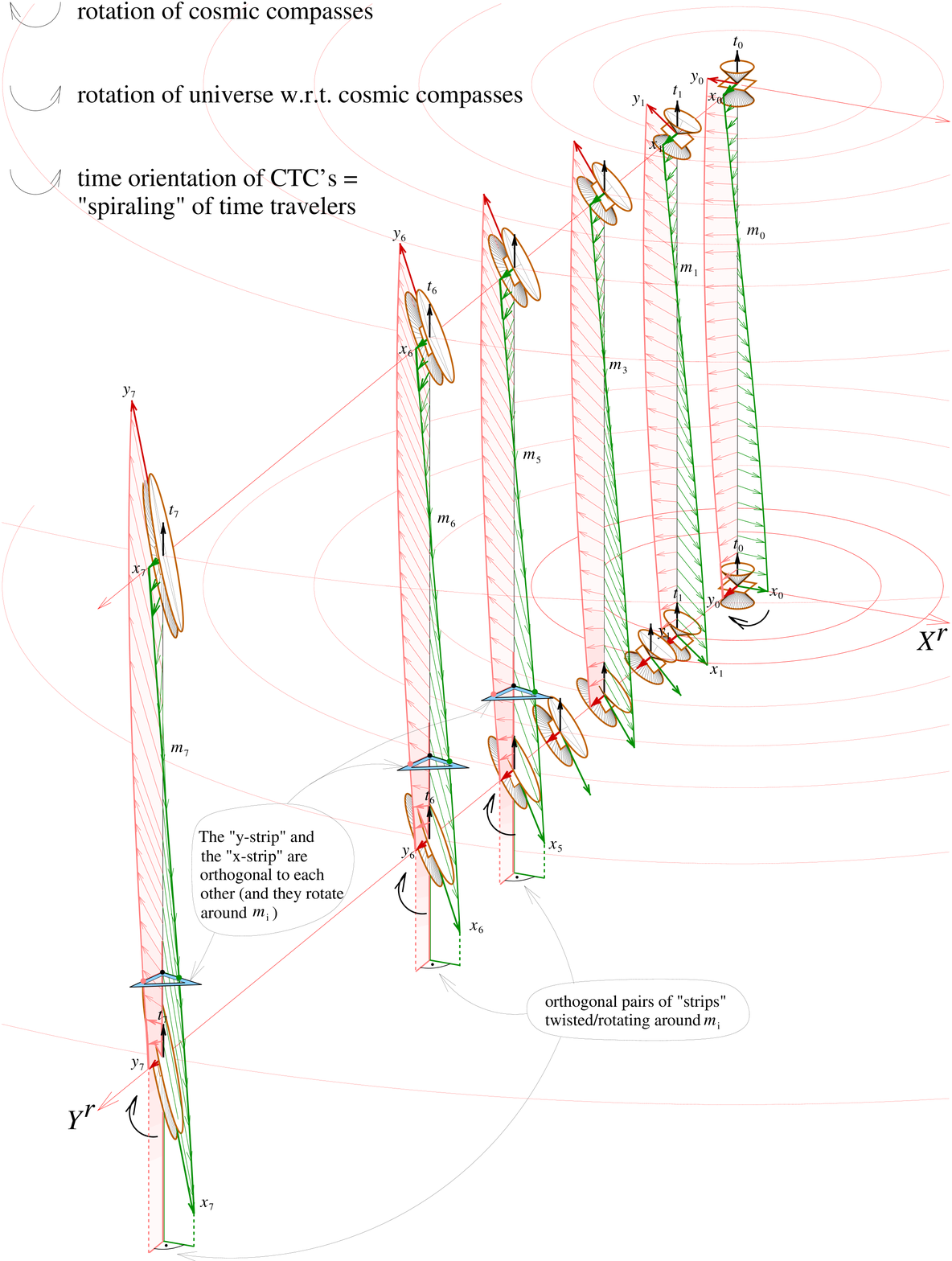}
\end{picture}
\end{center}
\caption{\label{has4a-fig} Tilted dervishes with original angular
velocity. $\omega=\pi/30$, Map 2 applies.}
 \end{figure}

\newpage

\section{Tilted Spiral World, i.e.\ Choice~1 Spiral World.}
\label{refined-section}


\begin{figure}[!hp]
\setlength{\unitlength}{0.16 truemm} \small
\begin{center}
\begin{picture}(880,1280)(0,0)
\epsfysize = 1280  \unitlength \epsfbox{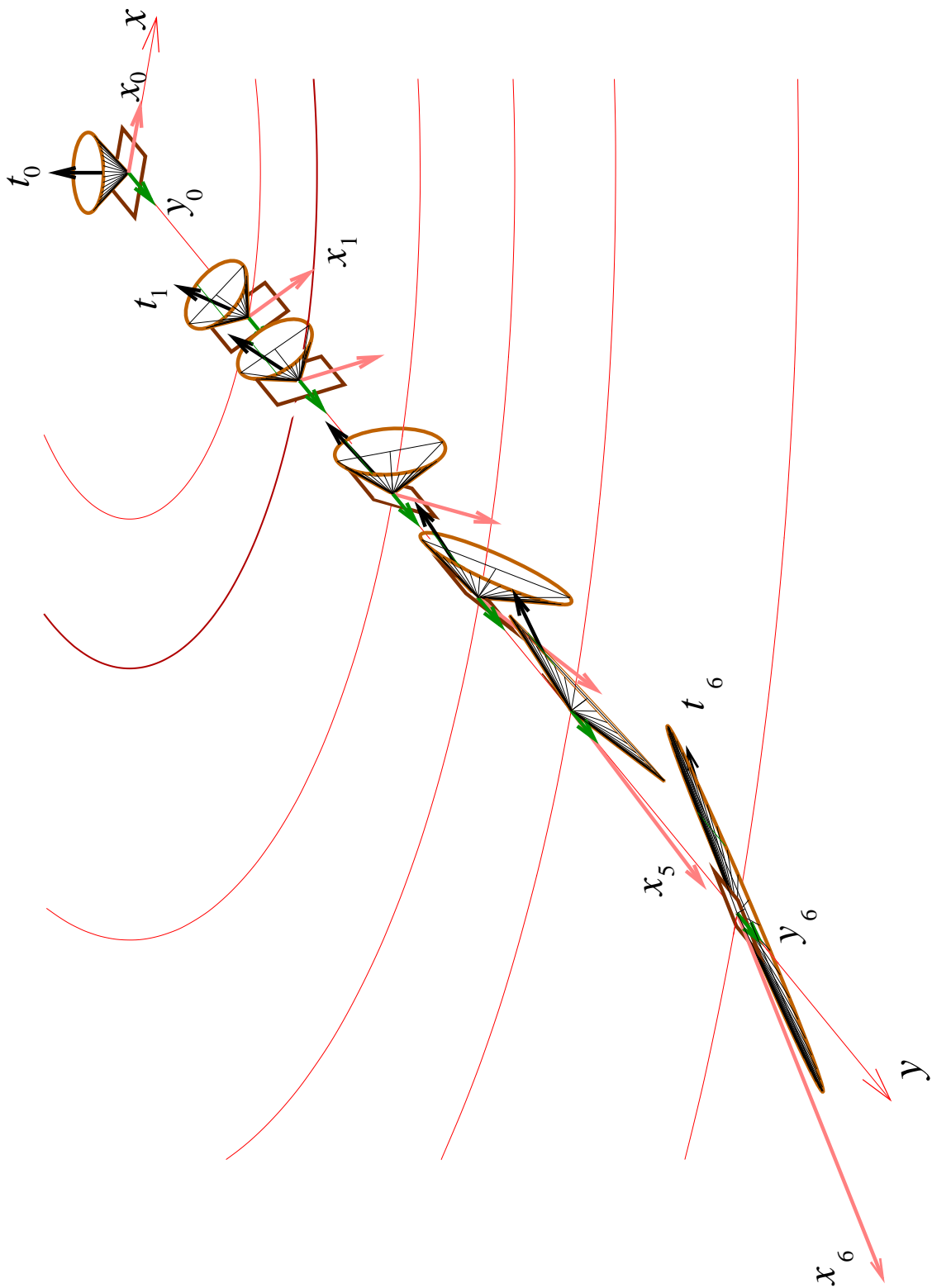}
\end{picture}
\end{center}
\caption{\label{vistorta5-fig} Tilted spiral world, i.e.\ Choice~1
Spiral World. Light-cones, unit-vectors along the $y$-axis.
$\omega=\pi/30$, Map 2 applies.}
\end{figure}


\begin{figure}[!hp]
\setlength{\unitlength}{0.073 truemm} \small
\begin{center}
\begin{picture}(1400,2860)(0,0)
\epsfysize =2860   \unitlength \epsfbox{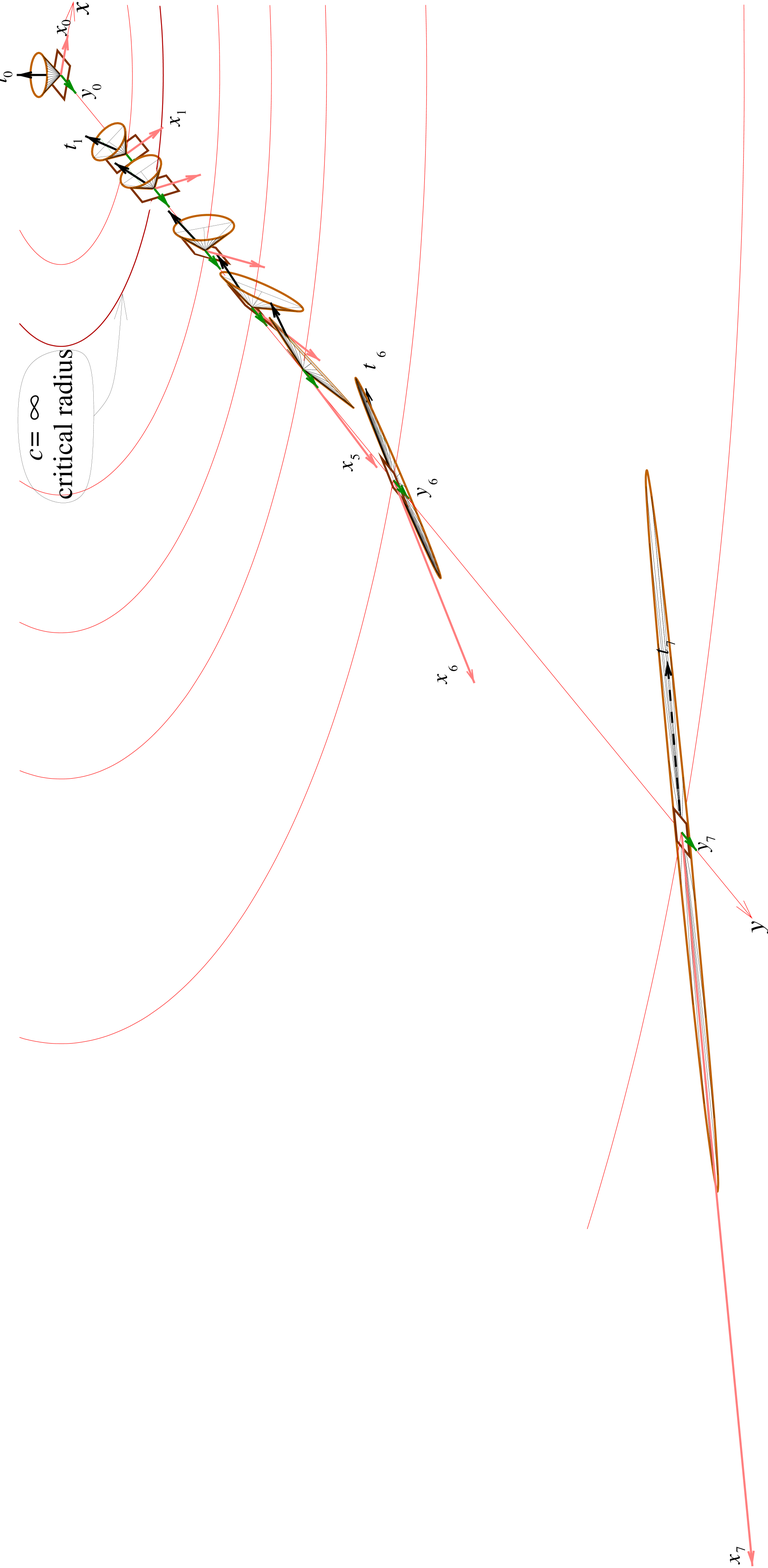}
\end{picture}
\end{center}
\caption{\label{vistorta4a-fig} Tilted Spiral World.
$\omega=\pi/30$, Map 2 applies.}
 \end{figure}



\begin{figure}[!hp]
\setlength{\unitlength}{0.052 truemm} \small
\begin{center}
\begin{picture}(1740,4060)(0,0)
\epsfysize =4060   \unitlength \epsfbox{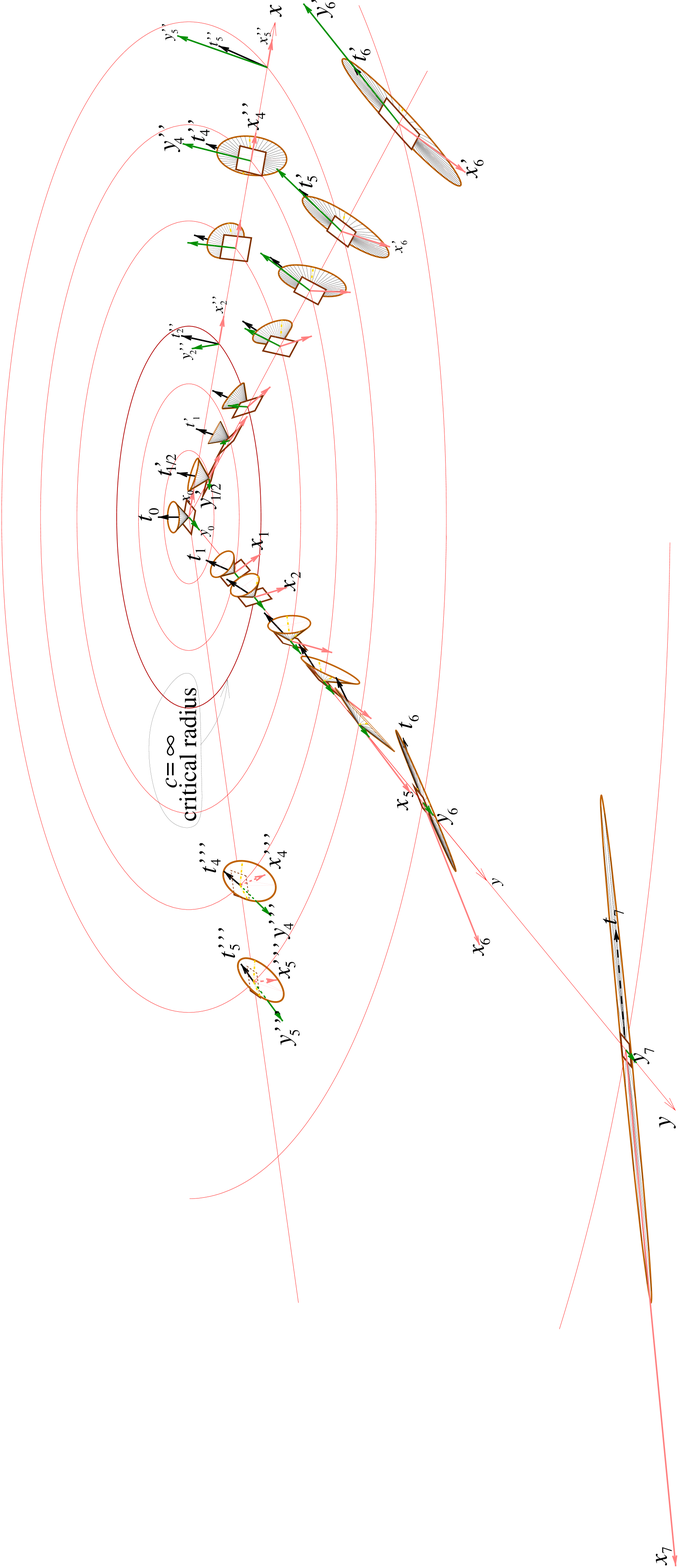}
\end{picture}
\end{center}
\caption{\label{nagyt1-fig} \label{nagyt0-fig} Tilted Spiral World.
$\omega=\pi/30$, Map 2 applies.}
 \end{figure}


\begin{figure}[!hp]
\setlength{\unitlength}{0.079 truemm} \small
\begin{center}
\begin{picture}(2120,2880)(0,0)
\epsfysize = 2880  \unitlength \epsfbox{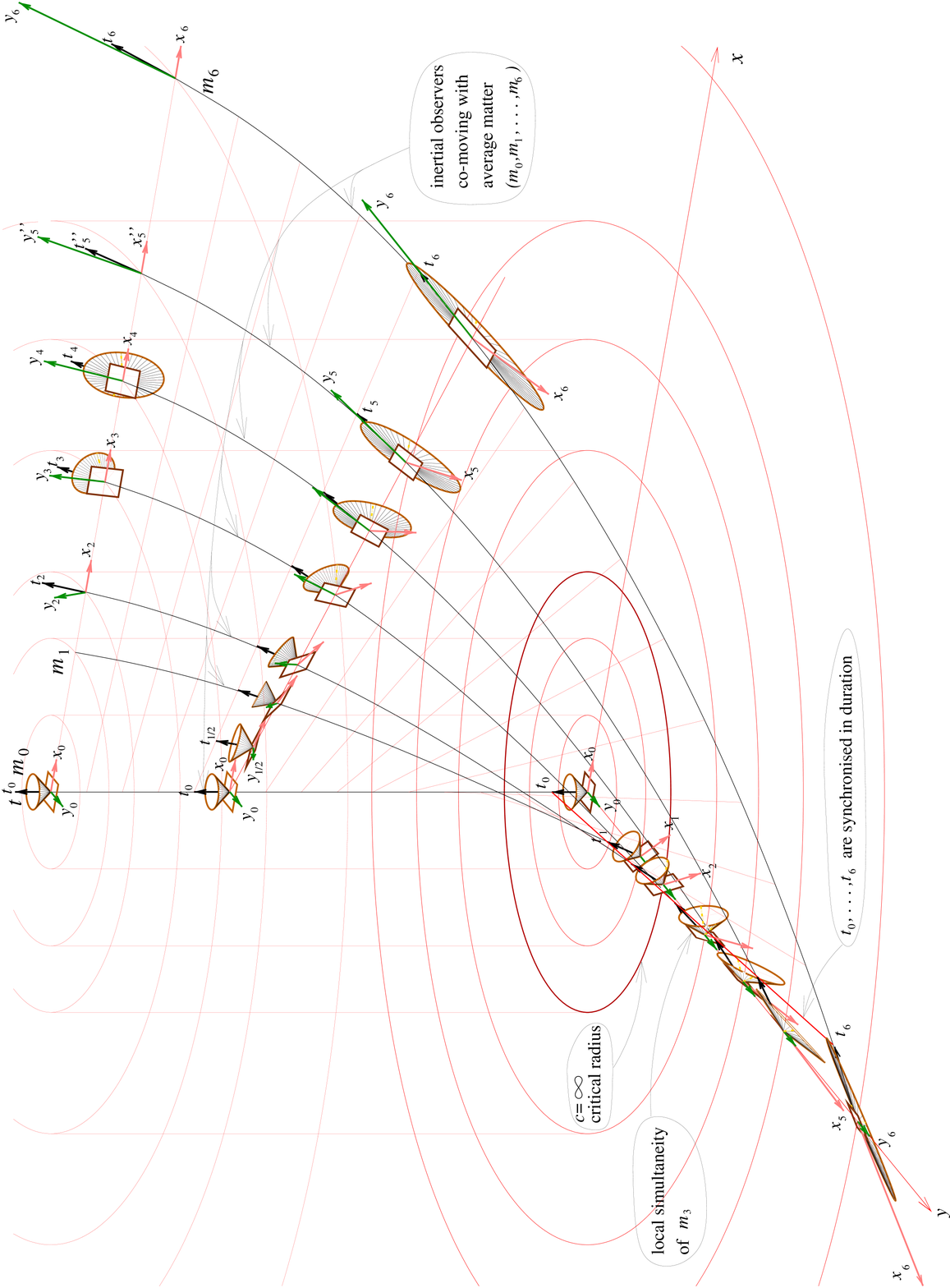}
\end{picture}
\end{center}
\caption{\label{vis-fig} Tilted Spiral World, full view.
Light-cones, life-lines, unit-vectors etc. Cf.\
Hawking-Ellis~\cite[Figure 61]{Hawel}. Cf.\ also
Figure~\ref{ujgodel-fig} herein. $\omega=\pi/30$, Map 2 applies.}
 \end{figure}


\begin{figure}[!hp]
\setlength{\unitlength}{0.05 truemm} \small
\begin{center}
\begin{picture}(2560,4480)(0,0)
\epsfysize =  4480 \unitlength \epsfbox{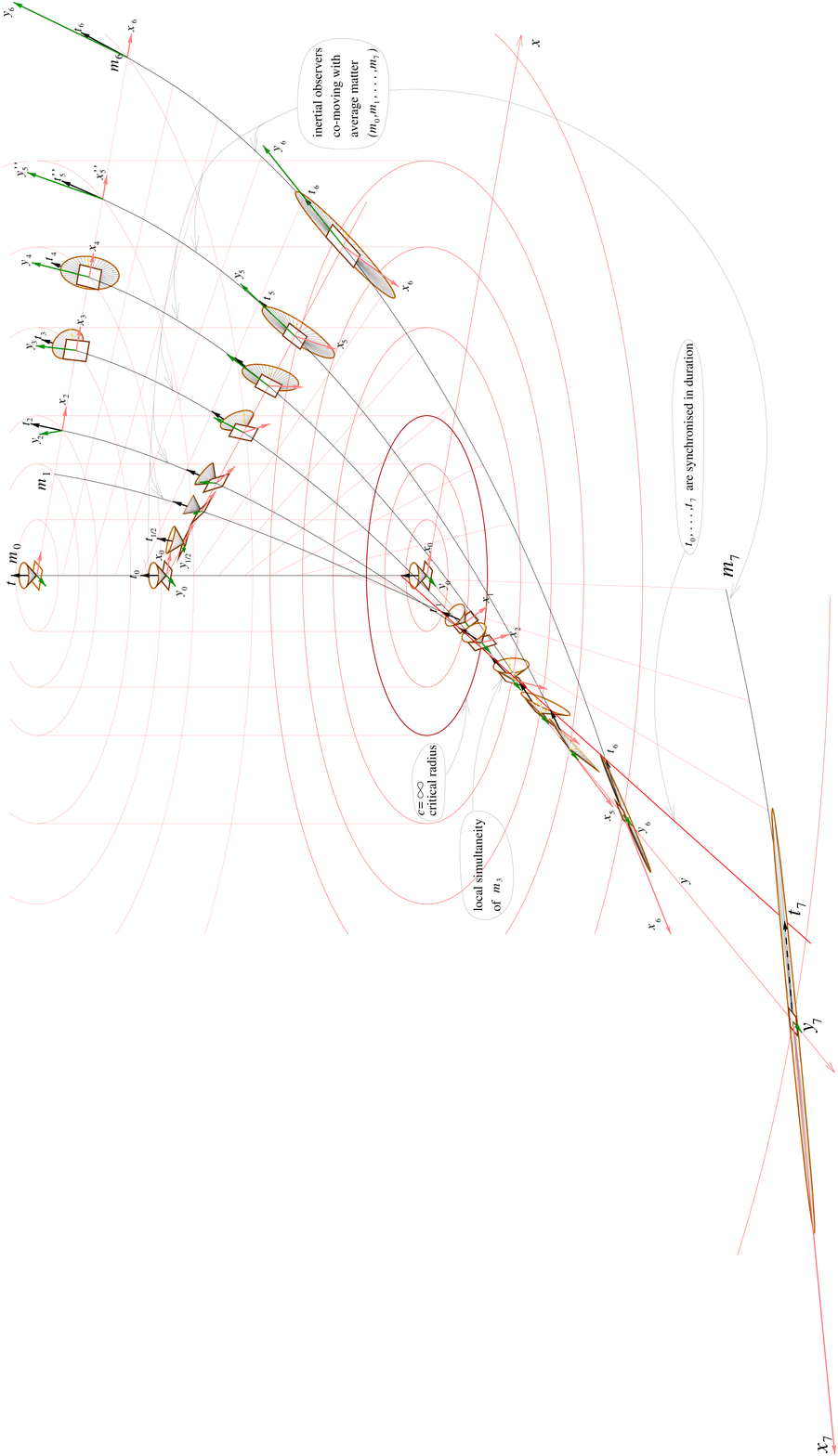}
\end{picture}
\end{center}
\caption{\label{visa-fig} Tilted Spiral World, full view.
$\omega=\pi/30$, Map 2 applies.}
 \end{figure}

\begin{figure}[!hp]
\setlength{\unitlength}{0.075 truemm} \small
\begin{center}
\begin{picture}(2120,2880)(0,0)
\epsfysize = 2880  \unitlength \epsfbox{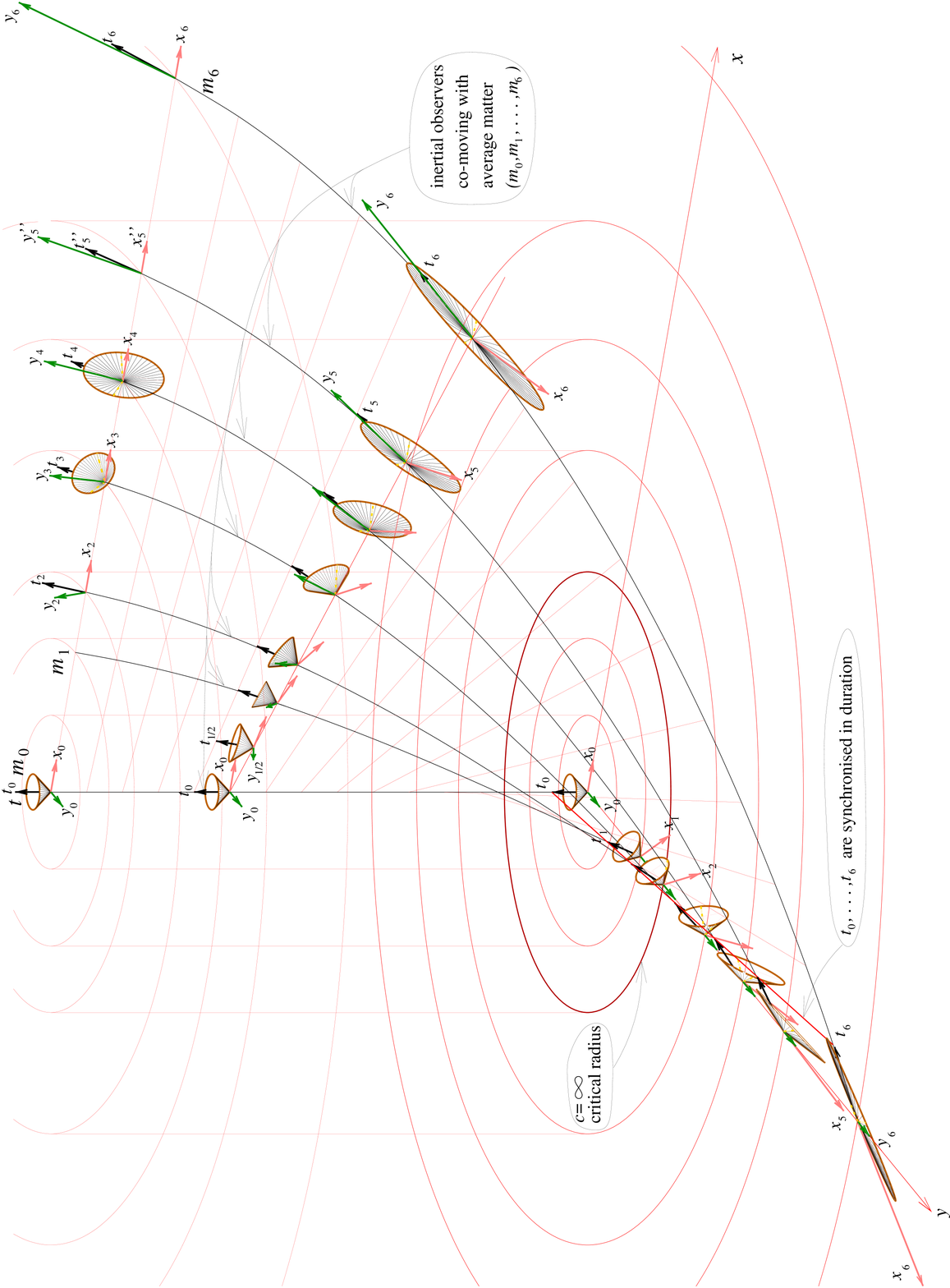}
\end{picture}
\end{center}
\caption{\label{2vis-fig} Full view of new spiral world. Cf.\
Hawking-Ellis~\cite{Hawel}, Figure~\ref{ujgodel-fig} herein and the
figure in Malament~\cite{Mal84}.
 $\omega=\pi/30$, Map 2 applies.}
 \end{figure}

\newpage

\section{Giving physical meaning to cosmic compasses. What rotates in
which direction (relative to whom).} \label{gyroscope-section}

\begin{figure}[!h]
\setlength{\unitlength}{0.61 truemm} %
\begin{center}
\begin{picture}(178,96)(0,0) 
\epsfysize = 96\unitlength   
\epsfbox{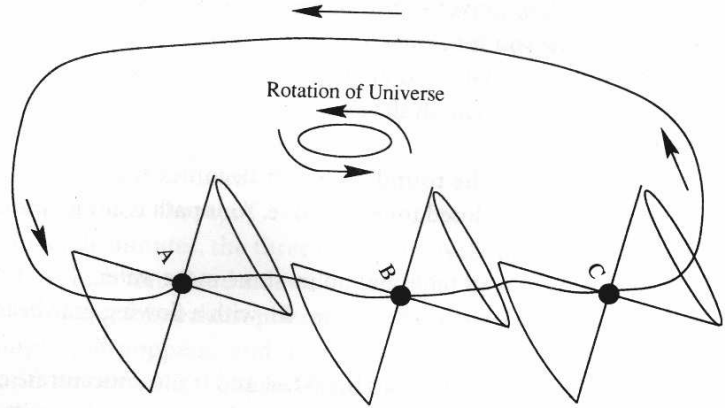}
\end{picture}
\end{center}
\caption{\label{Pickover-fig} What rotates in which direction? The
above is a picture from Pickover~\cite[p.185]{Picktime} from the
chapter on G\"odelian Universe implicitly offering a natural answer
to this question. This is also Figure 7.5 in
Gribbin~\cite[p.215]{Gri}.}
 \end{figure}

In our ``Tilted Spiral World''
(Figures~\ref{nagyt0-fig},\ref{vis-fig}) the light cones are very
strongly tilted forwards with increasing radius $r$. Therefore, if
$m_0$ throws a ball, say in the $y$ direction, the ball will start
moving in the $y$ direction but with increasing radius it will {\em
have} to turn in the $\varphi$ direction because the life-line of
the ball has to stay inside the light-cones (i.e.\ it has to be a
time-like curve). The same applies even to a photon in place of the
ball. This
effect is called the {\em gravitational drag effect}%
\footnote{What we call drag effect is often called {\em dragging of
inertial frames}. For references on gravitational drag effect see
p.\pageref{drag-p}.}
 and is illustrated e.g.\ in our Figure~\ref{ujgodel-fig} or
equivalently in Figure~31 of Hawking-Ellis~\cite{Hawel} as the
curving of the photon-geodesics. The drag effect affects those and
only those inertial bodies which are not at rest relative to one of
the $m_i$'s. This  drag effect is present in the Naiv GU, too, but
in a less dramatic way. To study the drag effect in our Tilted GU
(in Figures~\ref{vis-fig}, \ref{6hastanc-fig}), we notice that our
Tilted Dervish World (Figure~\ref{6hastanc-fig}) is structurally
very close to G\"odel's original universe described and studied in
G\"odel~\cite{Go96}, Hawking-Ellis~\cite[pp.168-170]{Hawel} and
later papers. Hence the results about the drag effect in G\"odel's
universe obtained in these works are applicable to our version of GU
in Figure~\ref{6hastanc-fig}. The drag effect can be analyzed and
described by studying the behavior of geodesics. Indeed,
Figure~\ref{ujgodel-fig} represents ``dragging'' of some
characteristic geodesics. Let us be in dervish world. Then
Figure~\ref{ujgodel-fig} indicates the following. A ball thrown by
$m_0$ will start out radially, then will make a big circle and will
come back to $m_0$ from a new direction. From now on, we will call
the circular motion or rotation traced out by this circle the {\em
drag rotation}. In Figure~\ref{ujgodel-fig} the direction of the
drag rotation coincides with the $\varphi$-direction which in turn
coincides with the direction of CTC's. All this remains true in our
Tilted Dervish World (Figure~\ref{6hastanc-fig}). In the Tilted
Spiral World, matter (the $m_i$'s) is seen to rotate in the same
direction $\varphi$. Therefore in the Tilted Spiral World what we
said above about the drag rotation, CTC's etc.\ remains true. Hence,
in the Tilted Spiral World the drag rotation is even stronger than
in the dervish world and points in the same direction $\varphi$ in
which the matter content of the universe rotates. Hence in the
Tilted Spiral World, we have an {\em increased drag effect}. As a
curiosity we note that in the Tilted Spiral World everything rotates
in the same direction $\varphi$.

Next we turn to replacing our cosmic compasses%
\footnote{which were ``abstract directions'' so far} with physically
tangible compasses of an ``observational'' kind (i.e.\ subject to
testing by thought experiment). In general relativity, the devices
used for this purpose are called {\em gyroscopes} or {\em compasses
of inertia}. The nonspecialist reader does not need to recall the
definition, what we write below is amply enough for the present
paper. The most important property (for us) of gyroscopes is that
their working is based on inertial motion, hence the behavior of
geodesics will also influence the behavior of gyroscopes. For the
non-physicist reader we note the following.

In Newtonian physics it is provable that certain devices called
gyroscopes preserve their directions despite of our moving them
around, in other words, they behave like ``cosmic
compasses''.\footnote{See e.g.\ Epstein~\cite[p.128]{Eps} for nice
illustration.} We do
not recall the definition of gyroscopes in detail.%
 However we note that
they can be made smaller and smaller in some sense such that their
Newtonian property of preserving direction (whatever this means)
remains true in general relativity (here the basic idea is that
general relativity agrees with Newtonian mechanics for small enough
speeds [with sufficient precision]).  The essential idea behind
gyroscopes is that a rigid body rotating fast enough tends to
preserve its axis of rotation (in Newtonian physics). If we make the
body small enough, then the tangential velocities of its parts will
tend to zero. Hence the tangential velocities involved can be made
small enough for the Newtonian approximation to be satisfactory.

It is natural to assume that the increased drag effect in Tilted GU
described above will ``drag'' the gyroscopes, too, in the $\varphi$
 direction. Indeed, an analysis of the geodesics of G\"odel's universe in
Lathrop-Teglas~\cite{Lath} suggests that this is so.

Our next goal is to find a new coordinatization $C^+$ for our Tilted
GU in which the gyroscope directions do not rotate.%
\footnote{Below by gyroscopes we always mean gyroscopes of $m_0$.}
One needs not regard this new coordinatization $C^+$ superior in
some sense to e.g.\ our Tilted Spiral World or more ``real'' than
Tilted, instead, $C^+$ is a coordinatization with some interesting
and useful properties. $C^+$ will be a (new) spiral world. We will
call this new spiral world Refined (or Choice 2) Spiral World. After
constructing $C^+$, it will be worthwhile to reconstruct the dervish
world in such a form that the {\em new local frames} (i.e.\
``veils'' or ``hands'' of the whirling dervishes) will be frames
co-rotating with the gyroscopes. Then the local frames will be what
are called {\em local inertial frames} in general relativity. A
representation of the dervish world with these new local inertial
frames represented as the ``veils'' of the dervishes will be called
Refined (or Choice~2) Dervish World. The two tilted spiral worlds
(Choices~1,2) and the two tilted dervish worlds (Choices~1,2)
represent the same space-time in different coordinates.

In the Refined Dervish World all the mass-carrier observers $m_i$
are at rest, they are evenly distributed and they are completely
alike, yet their compasses of inertia are rotating. This violates
Mach's principle that the state of zero rotation of an inertial
frame should coincide with the state of zero rotation with respect
to the distribution of matter in the universe. \label{Mach-page} For
Mach's principle see e.g.\ Barbour~\cite{Bar} and \cite{Mach}. For
more references on the drag effect and its connection with Mach's
principle see page~\pageref{drag-p}.

Above (p.\pageref{Pickover-fig}) we recalled a picture from
Pickover~\cite{Picktime} because it ``addresses'' the question of
what rotates in which direction. (E.g.\ does the universe rotate in
the same direction as the time-travelers (CTC's) do?) To make the
question meaningful, one has to tell relative to what coordinate
system is the question understood.%
\footnote{E.g.\ {\em relative} to the coordinates of our Tilted
Spiral World everything rotates in the same direction $\varphi$.} Of
course, one would like to name an ``observable" coordinate system
for asking such a question. A possibility is to choose that
coordinate system in which the gyroscopes do not rotate.%
\footnote{Technically, we have Fermi coordinates in mind.} This is
$C^+$ of our Choice~2 Spiral World. We will see that in $C^+$ the
directions of the various rotations are essentially different from
the ones in Pickover's picture. If one looks at $C^+$ without any
preparation, then the directions of rotations appear as ad hoc,
almost counter-intuitive. However, at least in our opinion, the
train of thought outlined in this paper may provide an explanation
for the arrangement of these directions. For more on this question
of counter-rotation in the case of rotating (Kerr-Newman) black
holes see \cite{ANW08}.

Let us return to our goal of finding a coordinatization $C^+$ of our
Spiral World in which gyroscope directions do not rotate.%
\footnote{This means that in $C^+$, gyroscopes of $m_0$ preserve
their directions (relative to the coordinate system).} We have
already observed that gyroscopes do rotate in our Tilted Spiral
World (Figure~\ref{vis-fig}). There are two equivalent ways for
finding $C^+$:

(i) We analyze the rotation of gyroscopes as seen from the Tilted
Dervish World, we observe that they rotate in the
$\varphi$-direction. This means that in the spiral world gyroscopes
rotate faster than the dervish world itself does (i.e.\ faster than
$\omega$). We choose the {\em refined spiral coordinates} to
co-rotate with these gyroscopes. Hence the ``gyroscope''-directions
will be fixed when viewed from the Refined Spiral World as we
wanted.

(ii) The following turns out to be equivalent with what we outlined
in (i) above. Let us go back to Section~\ref{tilting-section}
p.\pageref{choice1-p}, where we refined our Naive GU to get Tilted
GU. There, on p.\pageref{choice1-p}, we found two possible choices
(Choices~1,2) for the desired fine-tuning. Of the two, so far we
took the simpler one, Choice~1. Choice~2 consists of tilting the
light-cones in the dervish world {\em backwards} i.e.\ in a
direction {\em opposite} to that of $\varphi$ (in Choice~1 we tilted
them forwards). What we claim here is that the result of choosing
Choice~2 in Section~\ref{tilting-section} is equivalent with the
result of the refinements outlined in item (i) above. This is the
reason why we call our newest refined spiral and dervish worlds
outlined in item (i) above {\em Choice~2} worlds as well as Refined
worlds.

The new Choice~2  spiral and dervish worlds are illustrated and
elaborated (constructed) in the figures below. A natural question
comes up: If we had to refine our Choice~1 worlds because the drag
effect made the gyroscope directions rotate, how do we know that the
same problem will not come up in the new Choice~2 worlds? The answer
is two-fold. (1) The extremely strong drag effect in Choice~1 Spiral
World was caused by tilting the light-cones forwards extremely with
increasing radius $r$. Cf.\ Figure~\ref{nagyt0-fig} for this effect.
Now, in our Choice~2 Spiral World the light-cones are not tilted
forwards so much, actually recall that Choice~2 was obtained from
Choice~1 by tilting light-cones backwards (relative to our naive
GU). So, this very strong drag effect affecting even the gyroscopes
need not arise (more precisely, need not be strong enough for
affecting the gyroscopes). Indeed, as we said earlier, our dervish
world is very close structurally to G\"odel's original space-time
(GU). Therefore results about the original GU are applicable to our
versions (calibrated slightly differently). Now, the results in
Lathrop-Teglas~\cite{Lath} can be used to conclude that in our
Choice~2 Spiral World gyroscope directions are fixed, i.e.\ they do
not rotate. This can be seen by their characterization of geodesics
in basically%
\footnote{Our Choice~2 Spiral World is structurally very close to
the coordinatization $\langle t,r,\theta,z\rangle$ of GU given in
Lathrop-Teglas~\cite{Lath}.} Choice~2 Spiral World, as well as from
their claim that Choice~2 Spiral coordinates are so called Fermi
coordinates.

\begin{figure}[hbtp]
\setlength{\unitlength}{0.07 truemm} \small
\begin{center}
\begin{picture}(2100,2820)(0,0)

\epsfysize = 2820  \unitlength \epsfbox{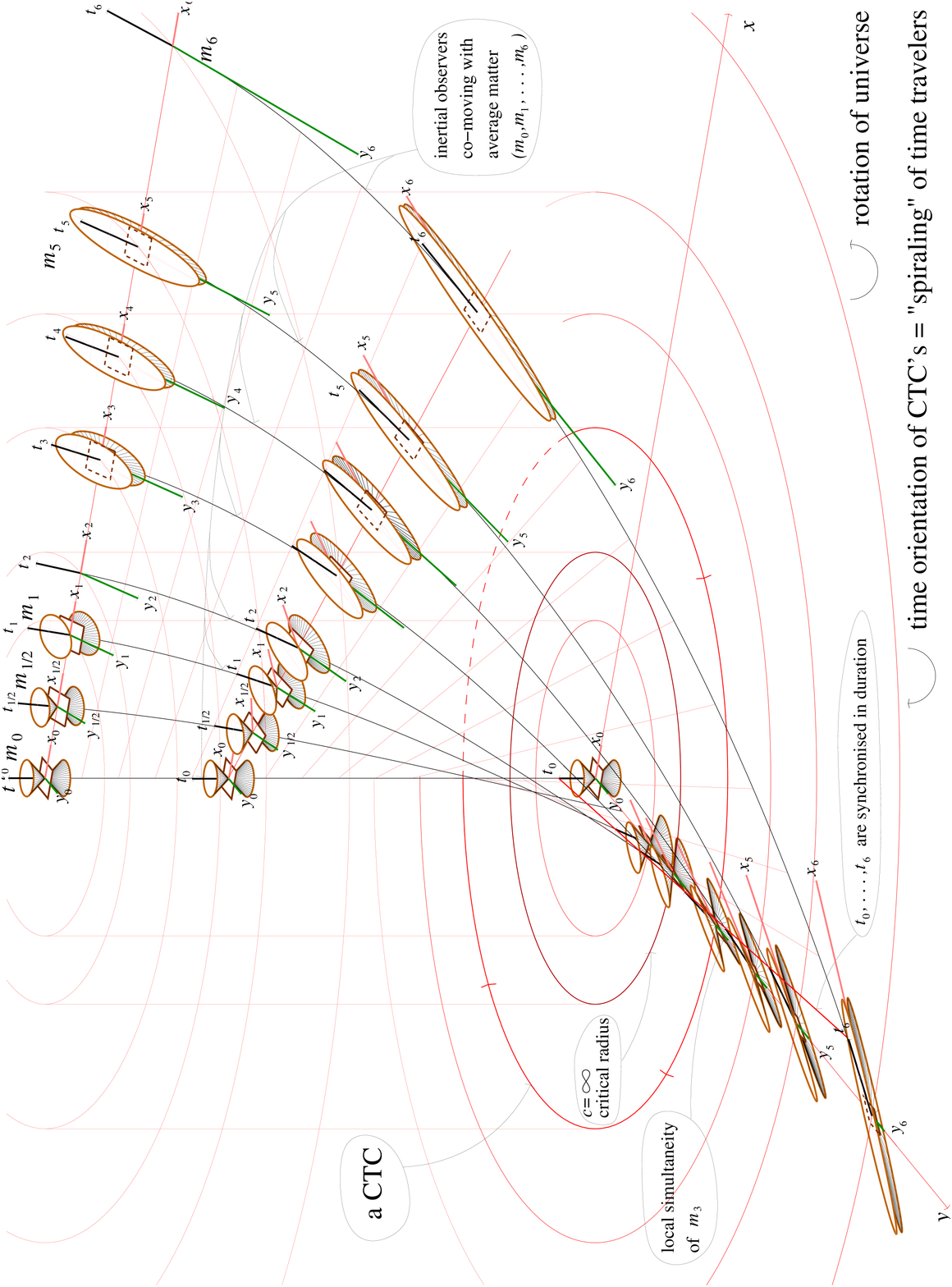}
\end{picture}
\end{center}
\caption{\label{ujvis-fig} Choice~2 GU spiral view (i.e., Refined
Spiral World). Here gyroscope directions are fixed (they do not
change).  (We are in Fermi coordinates in the sense of e.g.\
Lathrop-Teglas~\cite{Lath}.) $\omega=\pi/30$, Map 2 applies.}
 \end{figure}


\begin{figure}[hbtp]
\setlength{\unitlength}{0.07 truemm}
\small
\begin{center}
\begin{picture}(2100,2820)(0,0)

\epsfysize = 2820  \unitlength \epsfbox{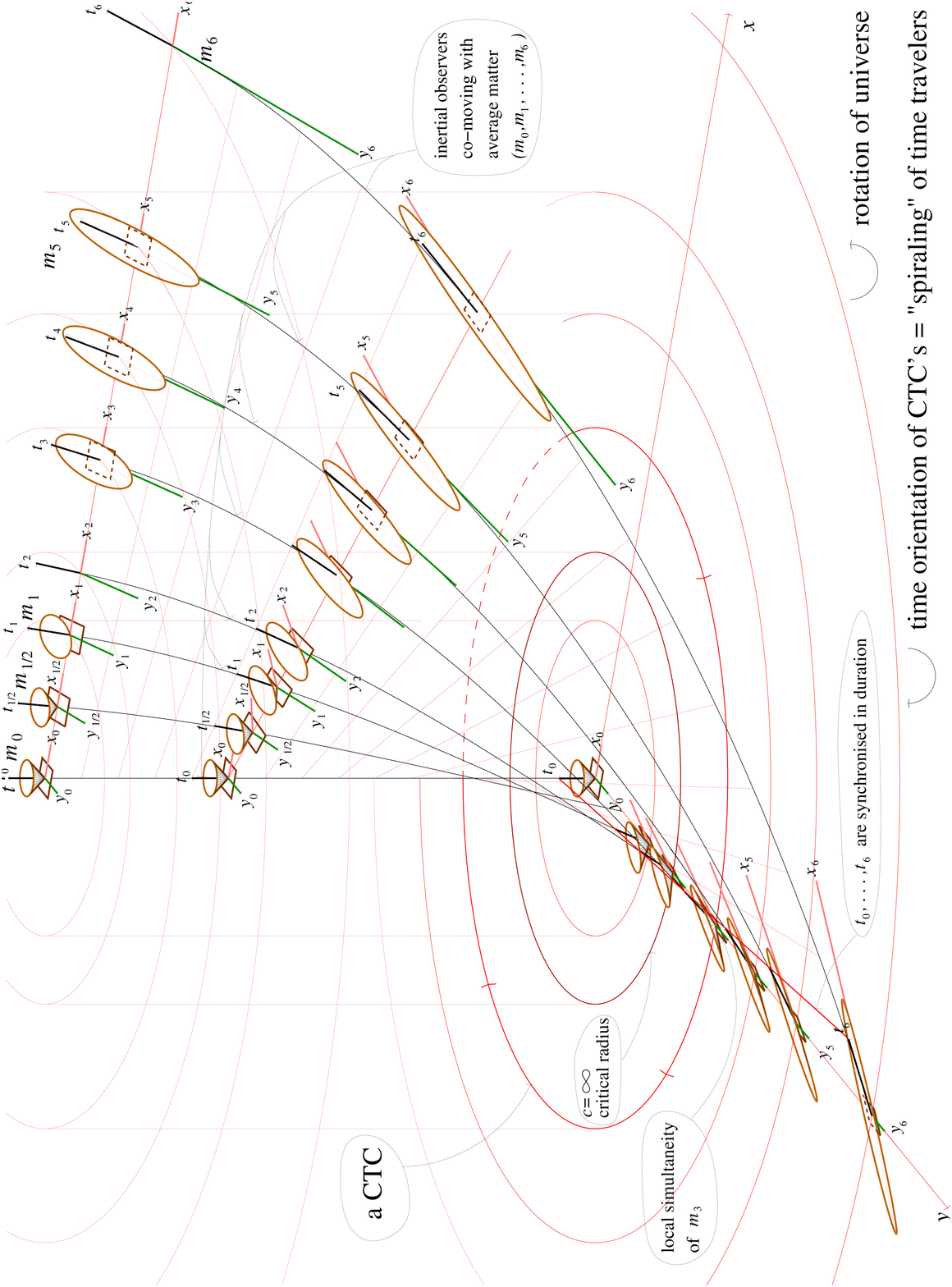}
\end{picture}
\end{center}
\caption{\label{ujvis1-fig} Choice~2 spiral view (Refined Spiral
World). Here gyroscope directions are fixed (they do not change).
(We are in Fermi coordinates in the sense of e.g.\
Lathrop-Teglas~\cite{Lath}.) $\omega=\pi/30$, Map 2 applies.}
 \end{figure}


\begin{figure}[hbtp]
\setlength{\unitlength}{0.09 truemm} \small
\begin{center}
\begin{picture}(1380,2360)(0,0)

\epsfysize = 2360  \unitlength \epsfbox{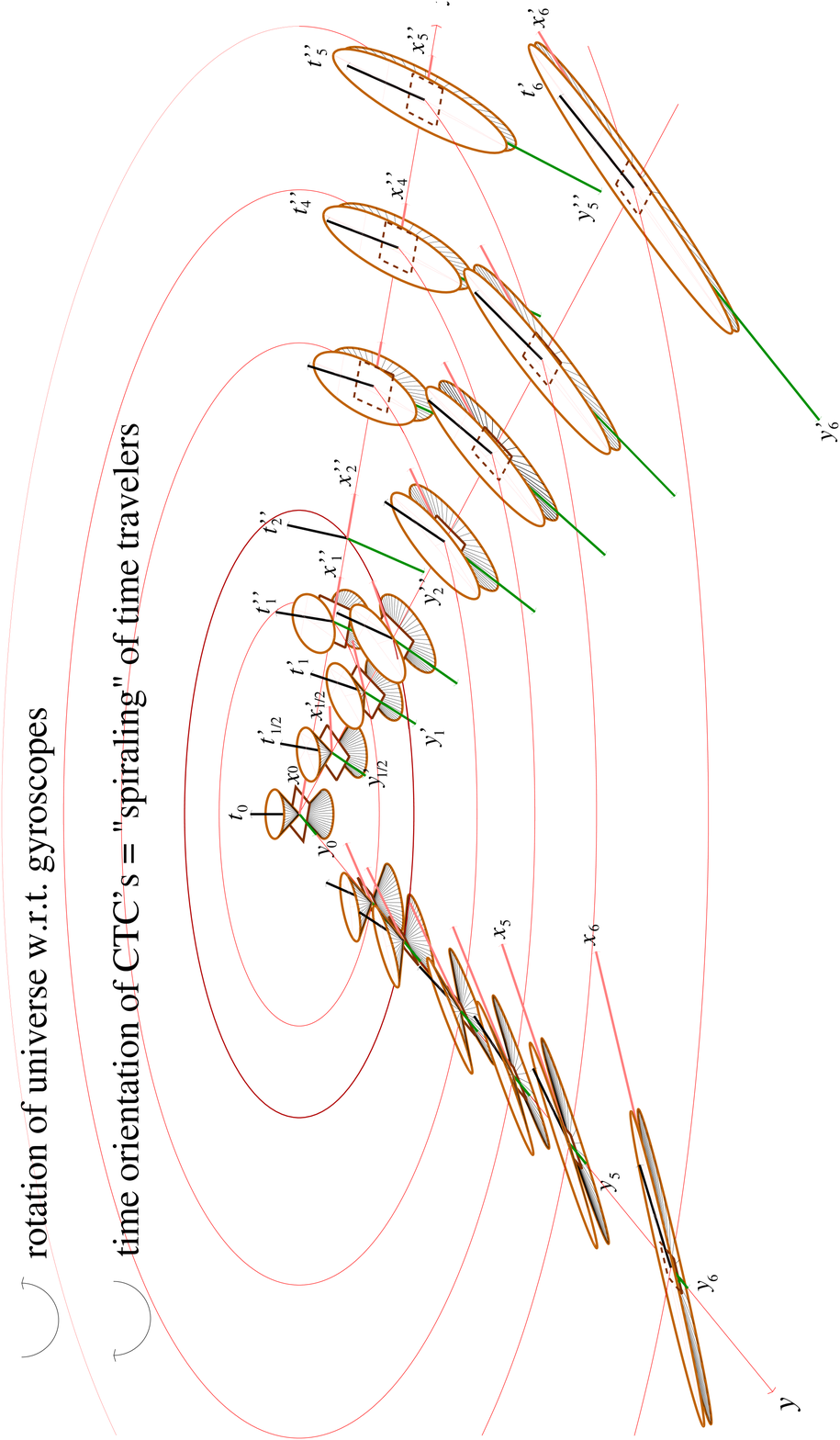}
\end{picture}
\end{center}
\caption{\label{ujt-fig} Choice~2 spiral view. Cf.\
Fig.\ref{ujvis-fig} for more information. $\omega=\pi/30$, Map 2
applies.}
 \end{figure}


\begin{figure}[hbtp]
\setlength{\unitlength}{0.09 truemm} \small
\begin{center}
\begin{picture}(1380,2360)(0,0)

\epsfysize = 2360  \unitlength \epsfbox{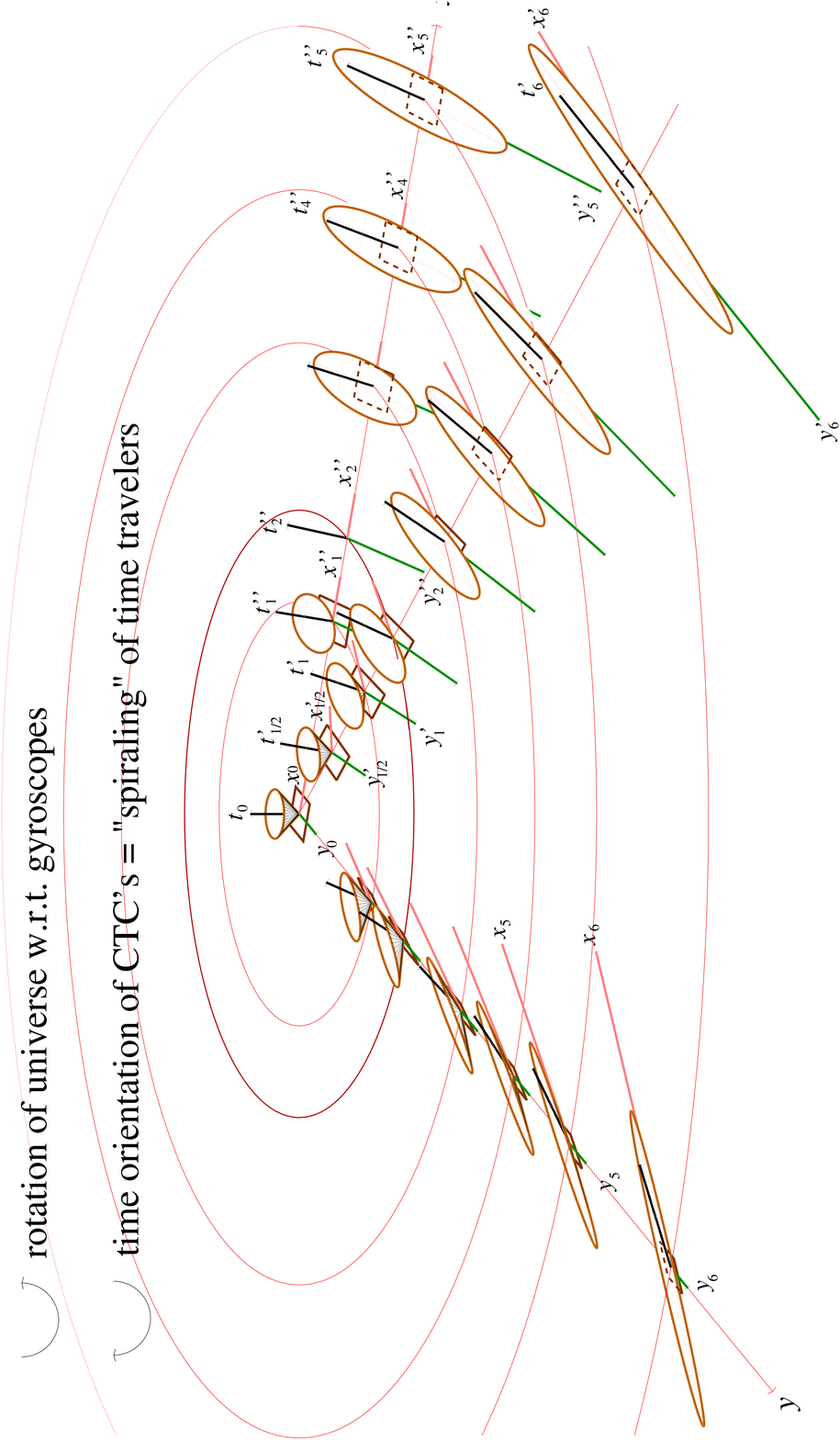}
\end{picture}
\end{center}
\caption{\label{ujt1-fig} Choice~2 spiral view. Cf.\
Fig.\ref{ujvis1-fig} for more information. $\omega=\pi/30$, Map 2
applies.}
 \end{figure}


\begin{figure}[hbtp]
\setlength{\unitlength}{0.081 truemm} \small
\begin{center}
\begin{picture}(1860,2580)(0,0)

\epsfysize = 2580  \unitlength \epsfbox{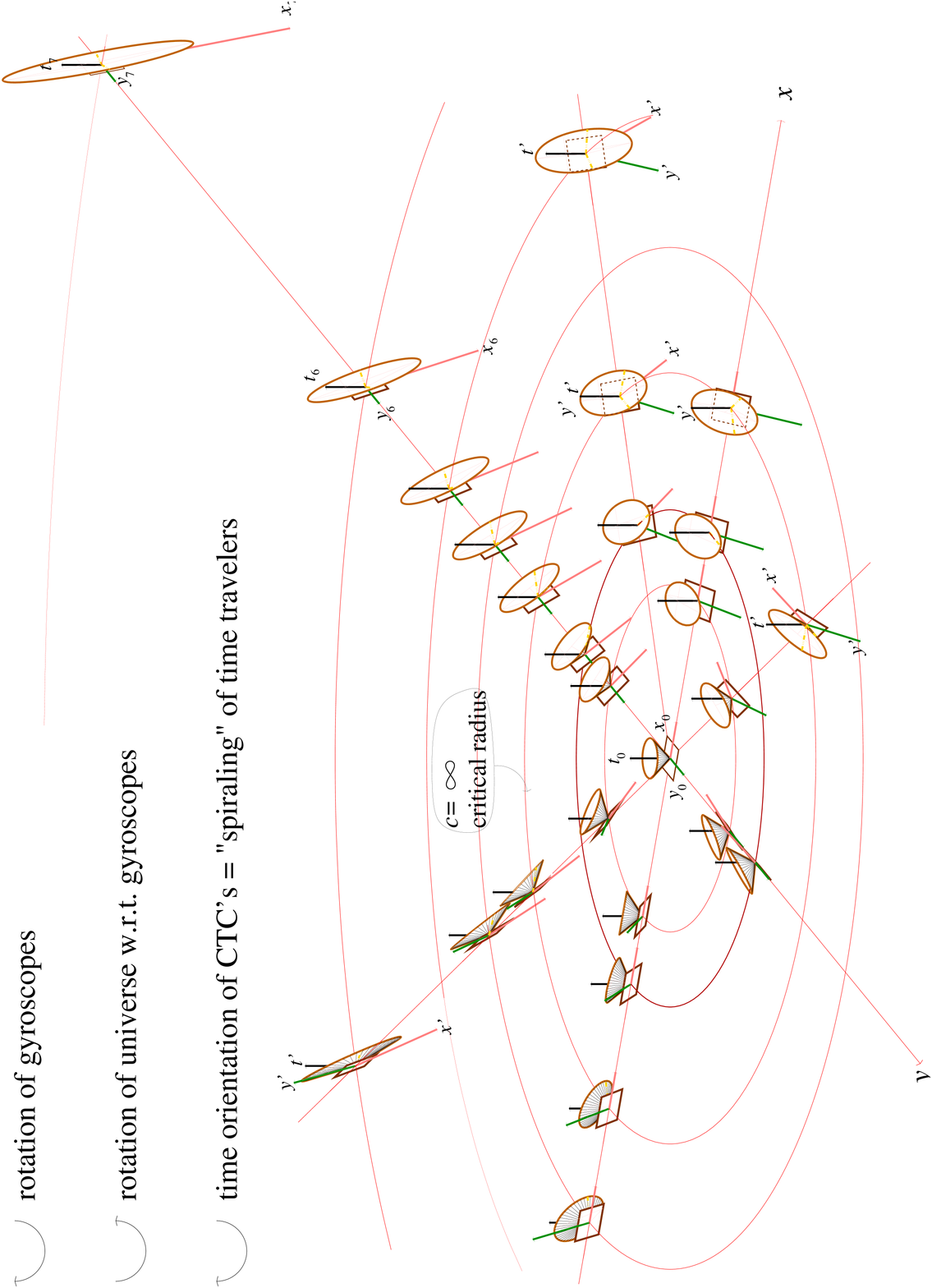}
\end{picture}
\end{center}
\caption{\label{2agoder-fig} Dervish view in dual GU (Choice~2).
Compare with Figure 61 on p.169 in Hawking-Ellis~\cite{Hawel} (cf.\
also Fig.\ref{ujgodel-fig} herein). $\omega=\pi/30$, Map 2 applies.}
 \end{figure}


\begin{figure}[hbtp]
\setlength{\unitlength}{0.073 truemm} \small
\begin{center}
\begin{picture}(2280,2980)(0,0)

\epsfysize =  2980 \unitlength \epsfbox{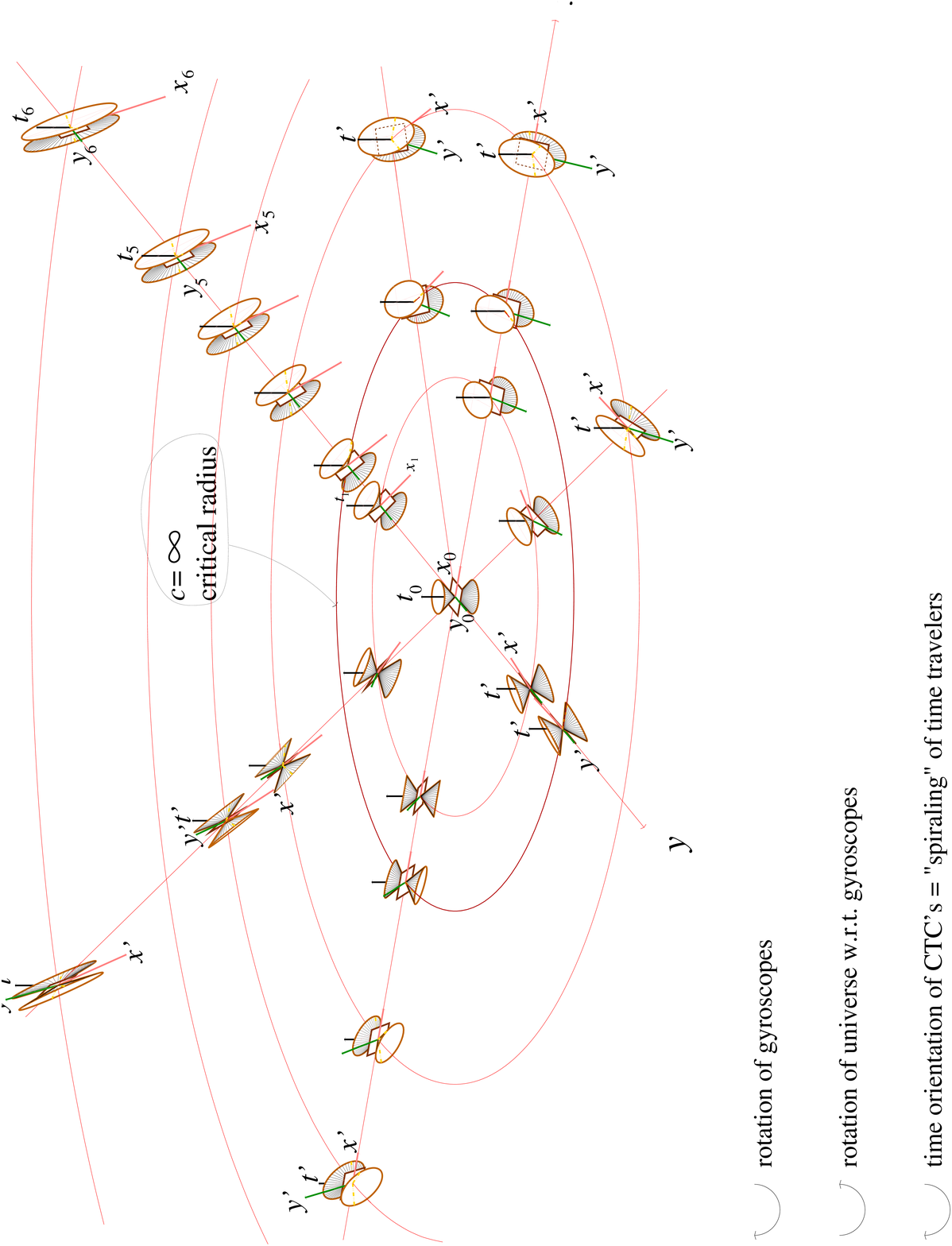}
\end{picture}
\end{center}
\caption{\label{goder3a-fig} Dervish view in dual GU (Choice~2).
$\omega=\pi/45$, Map 1 applies.}
 \end{figure}


\begin{figure}[hbtp]
\setlength{\unitlength}{0.072 truemm} \small
\begin{center}
\begin{picture}(2280,3040)(0,0)

\epsfysize =  3040 \unitlength \epsfbox{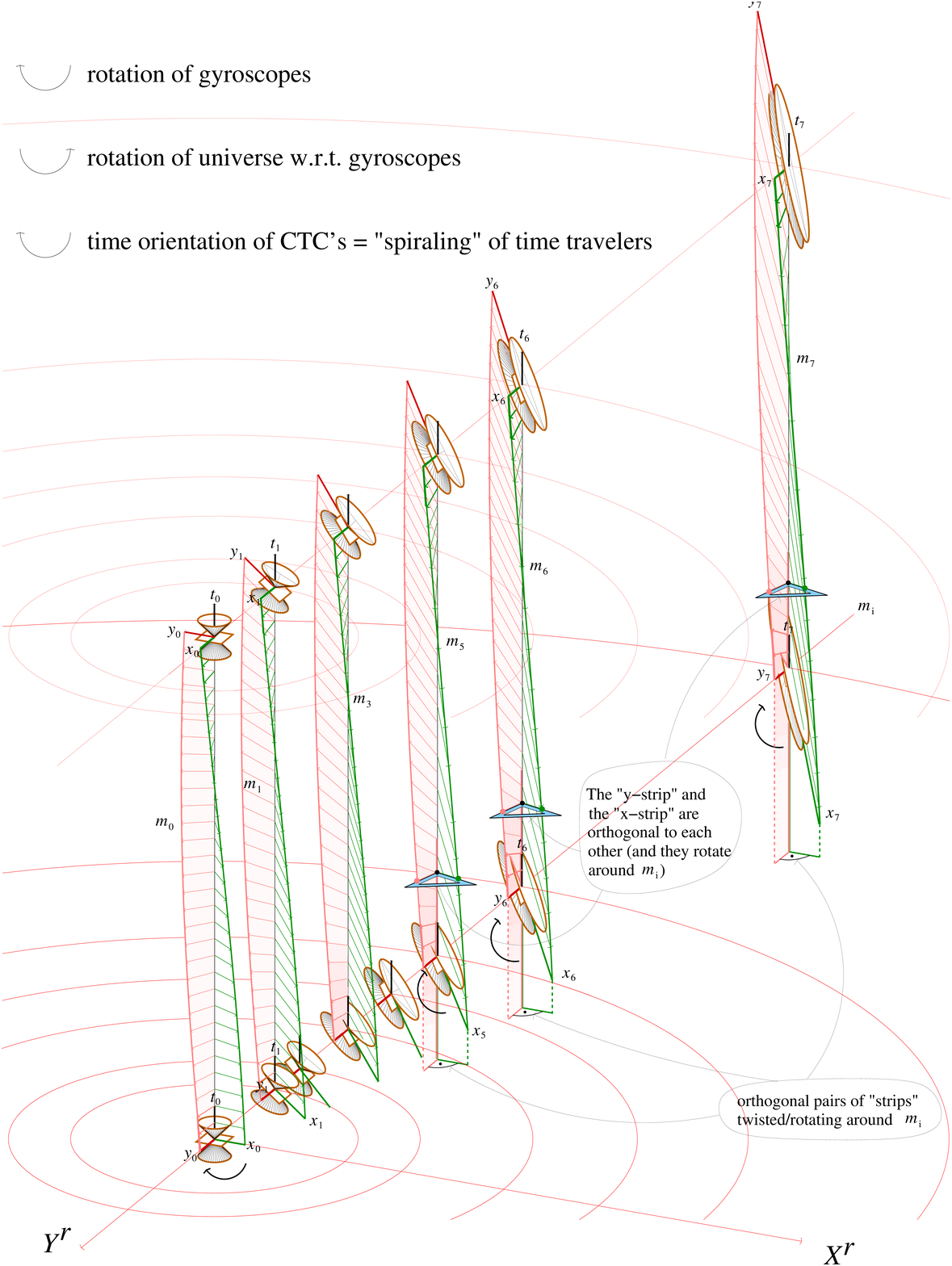}
\end{picture}
\end{center}
\caption{\label{4ahas-fig} Tilted-dervishes, Choice~2 with original
angular velocity. $\omega=\pi/30$, Map 2 applies. }
 \end{figure}

\begin{figure}[hbtp]
\setlength{\unitlength}{0.065 truemm} \small
\begin{center}
\begin{picture}(2580,3320)(0,0)

\epsfysize = 3320  \unitlength \epsfbox{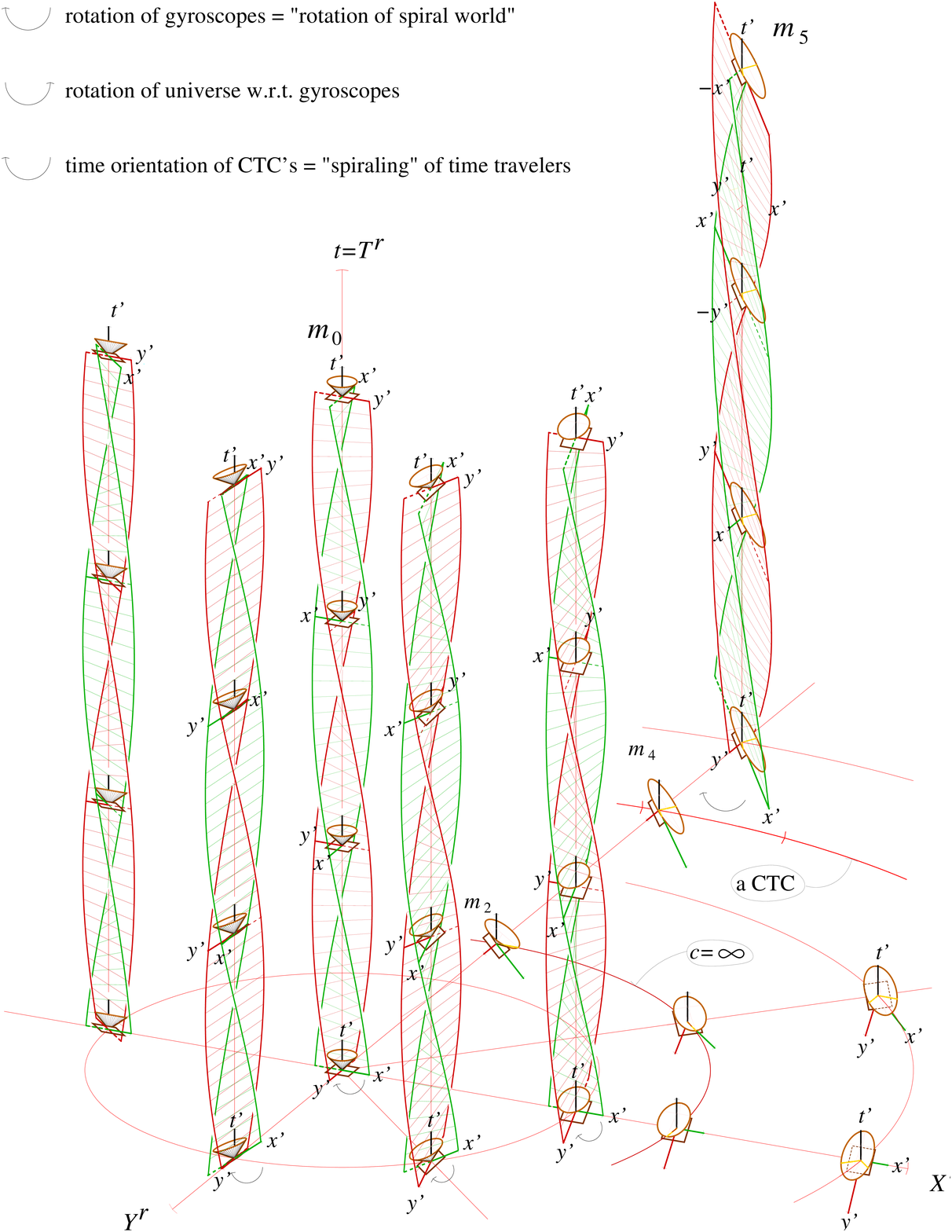}
\end{picture}
\end{center}
\caption{\label{5ahastanc-fig} Dervish view in dual GU (Choice~2).
``$\omega$ of universe'' $=\ \pi/60$ (recalibrated version of Map 2
applies). Spinning dervishes are artificially sped up (``artificial
$\omega$ of dervishes'' $=\ \pi/15$).}
 \end{figure}



\begin{figure}[hbtp]
\setlength{\unitlength}{0.065 truemm} \small
\begin{center}
\begin{picture}(2580,3320)(0,0)

\epsfysize = 3320  \unitlength \epsfbox{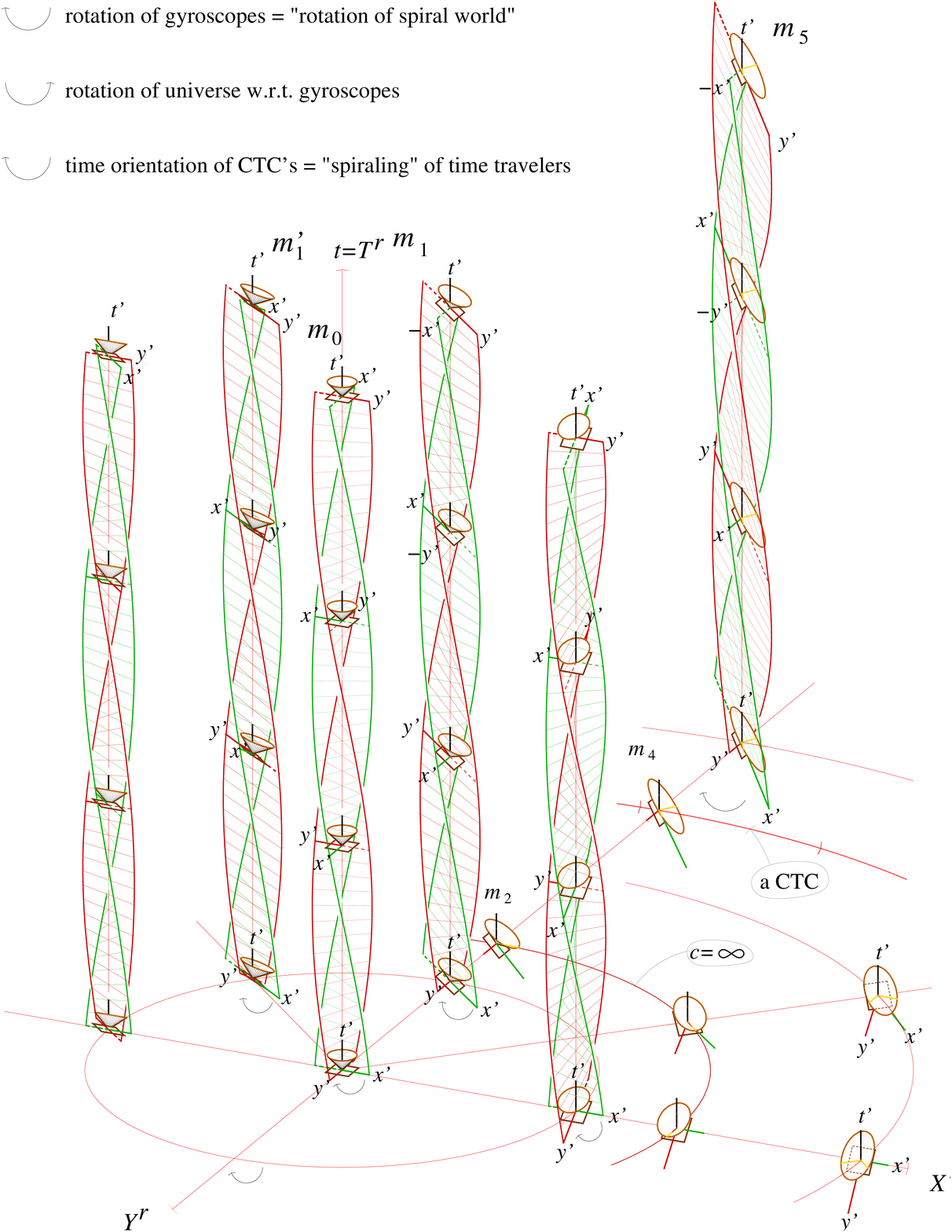}
\end{picture}
\end{center}
\caption{\label{5bhastanc-fig} Dervish view in dual GU (Choice~2).
``$\omega$ of universe'' $=\ \pi/60$ (recalibrated version of Map 2
applies). Spinning dervishes are artificially sped up (``artificial
$\omega$ of dervishes'' $=\ \pi/15$).}
\end{figure}


\begin{figure}[hbtp]
\setlength{\unitlength}{0.065 truemm} \small
\begin{center}
\begin{picture}(2380,3300)(0,0)

\epsfysize = 3300  \unitlength \epsfbox{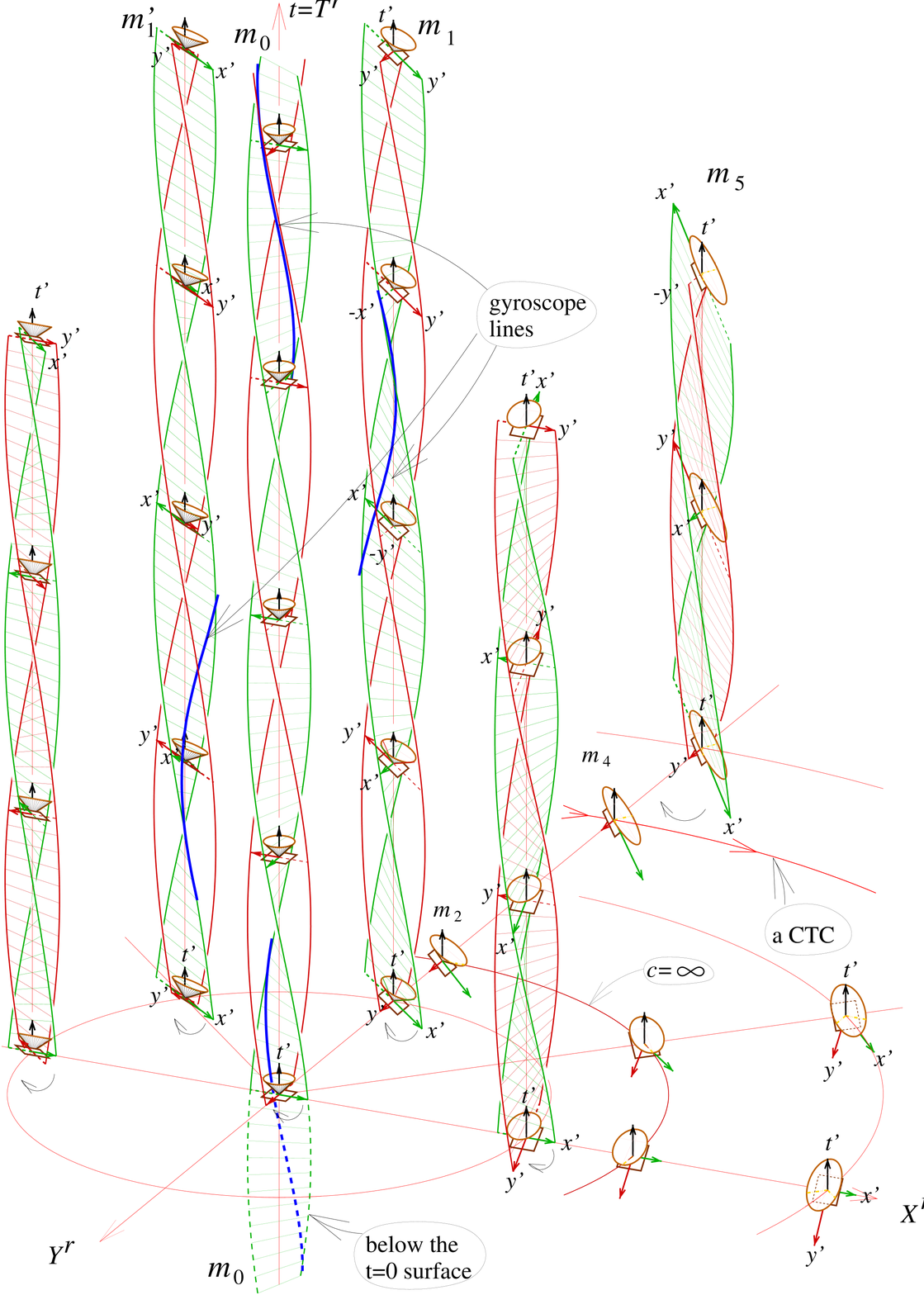}
\end{picture}
\end{center}
\caption{\label{ujlabda1-fig} Choice~2 dervish view. ``Fast''
gyroscope lines. ``$\omega$ of universe'' $=\ \pi/60$ (recalibrated
version of Map 2 applies). Spinning dervishes are artificially sped
up (``artificial $\omega$ of dervishes'' $=\ \pi/15$).}
 \end{figure}



\begin{figure}[!hp]
\setlength{\unitlength}{0.938 truemm} \small 
\begin{center}
\begin{picture}(180,200)(0,0)
\epsfysize = 200  \unitlength \epsfbox{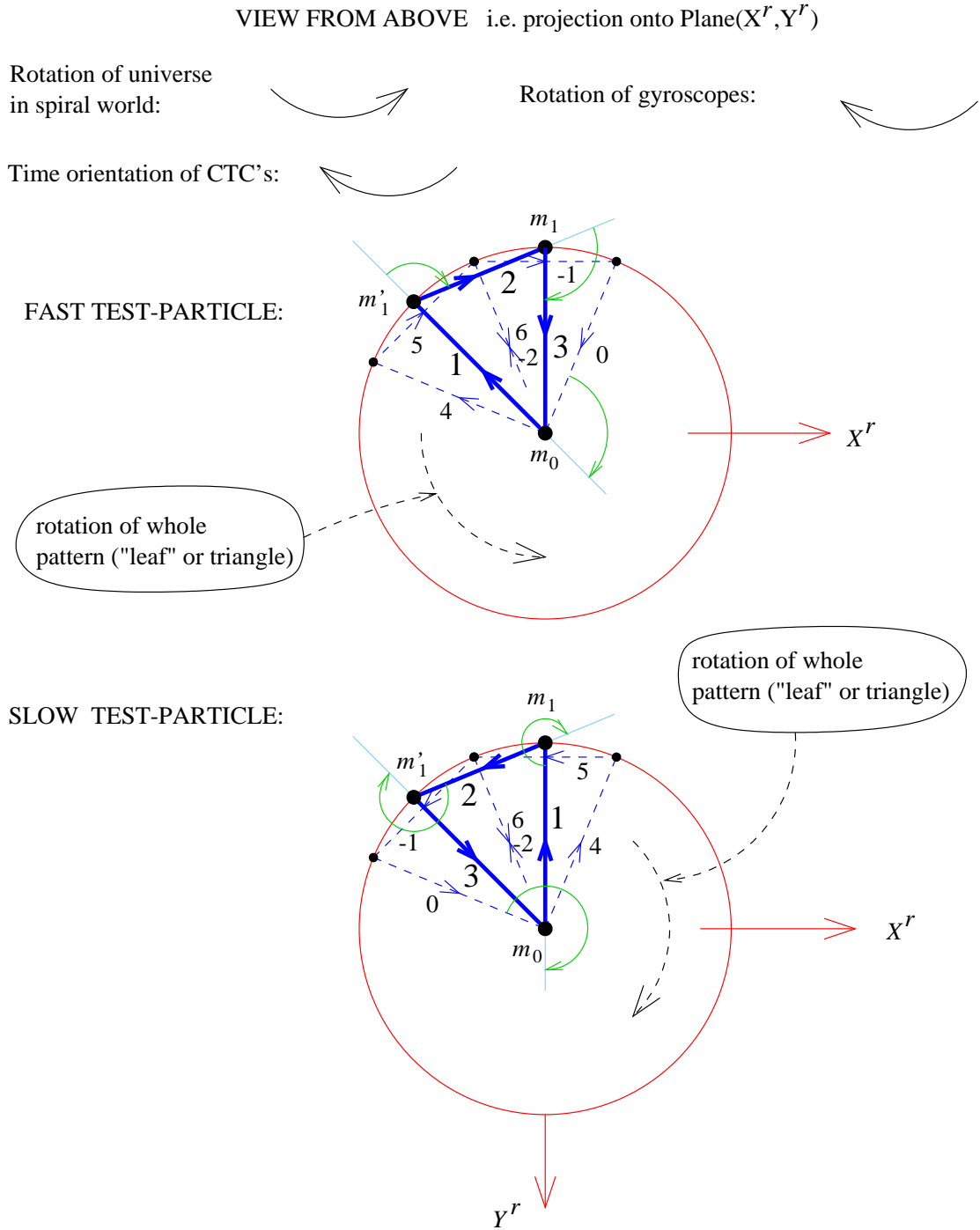}
\end{picture}
\end{center}
\caption{\label{ujlabda3-fig}
 This belongs to the previous two figures involving
gyroscope lines: Schematic paths of \underbar{gyroscopes-directed}
test-particles. Such particles can be visualized as small spaceships
whose pilots follow  (the direction shown by) their gyroscopes
strictly. Choice~2.}
 \end{figure}


\begin{figure}[!hp]
\setlength{\unitlength}{0.1 truemm} \small
\begin{center}
\begin{picture}(1600,1860)(0,0)
\epsfysize = 1860  \unitlength \epsfbox{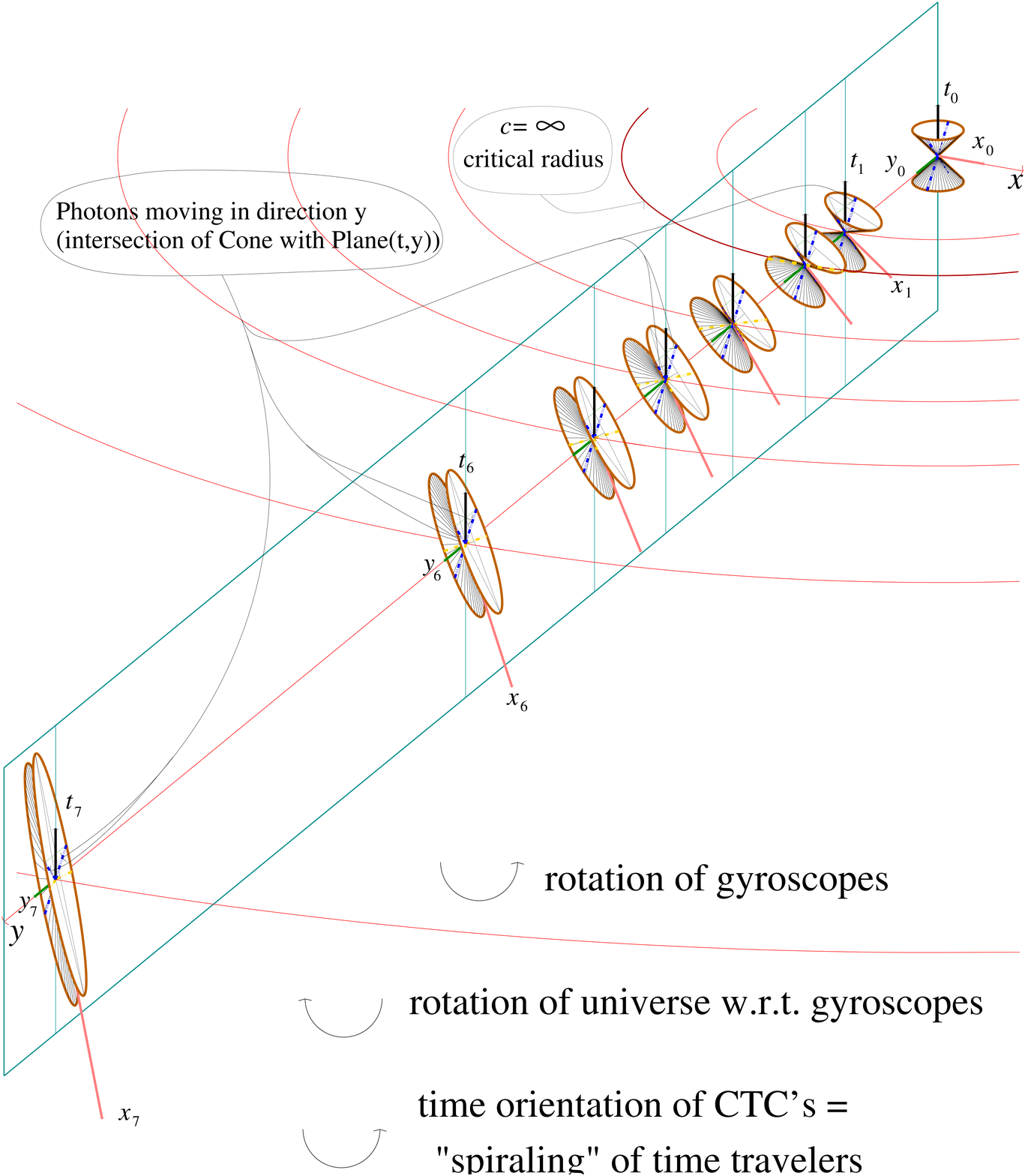}
\end{picture}
\end{center}
\caption{\label{2agodtorta-fig} GU, Choice~2. $\omega=\pi/30$, Map 2
applies.}
 \end{figure}


\vfill\eject
\newpage
\section{Metric tensors and some literature.}
\label{literature-section}

\subsection{The metric tensor of the Naive GU.}
The linear element in the Naive Spiral World is

$$
{\sf ds}^2=-\frac{1-r^2\omega^2}{(1+r^2\omega^2)^2}\,{\sf dt}^2+{\sf
dr}^2+{\sf dz}^2 +
\frac{r^2(1-r^2\omega^2)}{(1+r^2\omega^2)^2}\,{\sf d\varphi}^2
-\frac{4r^2\omega}{(1+r^2\omega^2)^2}\,{\sf d\varphi}{\sf dt}\,.
$$

\noindent Thus the components of the metric tensor {\sf g} of the
Naive GU in the Naive Spiral World are

$$
{\sf g}_{tt} = -\frac{1-r^2\omega^2}{(1+r^2\omega^2)^2},\quad {\sf
g}_{rr} = 1,\quad {\sf g}_{zz} = 1,\quad
{\sf g}_{\varphi\varphi} =
\frac{r^2(1-r^2\omega^2)}{(1+r^2\omega^2)^2},\quad {\sf g}_{\varphi
t} = {\sf g}_{t\varphi} = -\frac{2r^2\omega}{(1+r^2\omega^2)^2}\,,
$$

\noindent and the rest of the ${\sf g}_{ij}$'s are 0.
The nonzero Christoffel symbols ${\Gamma}^i_{jk}$ are

$$
{\Gamma}^{r}_{tt} =
\frac{r\omega^2(r^2\omega^2-3)}{(1+r^2\omega^2)^3}\,,\qquad
{\Gamma}^{t}_{tr} =
\frac{(1-r^2\omega^2)r\omega^2}{(1+r^2\omega^2)^2}\,,\qquad
{\Gamma}^{\varphi}_{tr} = \frac{-2\omega}{(1+r^2\omega^2)^2r}\,,
$$
$${\Gamma}^{r}_{t\varphi} =
\frac{2r\omega(1-r^2\omega^2)}{(1+r^2\omega^2)^3}\,,\qquad
{\Gamma}^{t}_{r\varphi} =
\frac{2r^3\omega^3}{(1+r^2\omega^2)^2}\,,\qquad
{\Gamma}^{\varphi}_{r\varphi} =
\frac{1-r^2\omega^2}{(1+r^2\omega^2)^2r}\,,\qquad
$$
$${\Gamma}^r_{\varphi\varphi} =
\frac{r(3r^2\omega^2-1)}{(1+r^2\omega^2)^3}\,,\qquad \mbox{ and the
${\Gamma}^i_{kj}={\Gamma}^i_{jk}$ for the nonzero ${\Gamma}^i_{jk}$
listed above}.
$$

\noindent The scalar curvature is
$$R = 2\omega^2\frac{(2r^2\omega^2-7)}{(r^2\omega^2+1)^2}\,. $$

Now, $\Gamma_{rr}=\bar 0 =\langle 0,0,0,0\rangle$ shows that the
radial straight lines in the $xy$-planes (i.e., the lines with
direction ``{\sf dr}") are geodesics. The life-lines of the galaxies
are of direction $\omega\mbox{\sf d}\varphi+\mbox{\sf dt}$, hence
$$
\omega^2\Gamma_{\varphi\varphi}+2\omega\Gamma_{\varphi
t}+\Gamma_{tt}=\bar 0
$$
shows that the life-lines of the distinguished observers $m_i$ are
geodesics in the Naive GU.
\smallskip

 G\"odel wanted the distinguished observers $m_0,\dots,m_i$ to be
fully ``equivalent'' with each other. This means that $m_i$ and
$m_0$ should be indistinguishable for any choice of $m_i$. This
means that there should exist an automorphism $h_{i,0} :\langle
\Reals,{\sf g}\rangle\longrightarrow\langle \Reals,{\sf g}\rangle$
such that $h_{i,0}$ takes the life-line of $m_i$ to that of $m_0$.
Since the scalar curvature is preserved by automorphisms, this
implies that the scalar curvature should not depend on $r$ (as it
really does not depend on $r$ in G\"odel's universe as we will see
soon). This implies  that in the Naive GU, the distinguished
observers $m_i$ are not fully equivalent with each other, because
the scalar curvature depends on $r$.
\smallskip

We note that the linear element in the Naive Dervish World is

$${\sf ds}^2=-\,{\sf dt}^2+{\sf dr}^2+{\sf dz}^2 +
\frac{r^2(1-r^2\omega^2)}{(1+r^2\omega^2)^2}\,{\sf d\varphi}^2 +
\frac{2r^2\omega}{(1+r^2\omega^2)}\,{\sf d\varphi}{\sf dt}\,.
$$

\bigskip
\subsection{The metric tensor of G\"odel's universe GU.}
G\"odel in \cite[p.275]{Go96}, \cite[p.195]{Gcwii}  and elsewhere
defines his universe by presenting the ``linear element'' (i.e.\ the
``metric tensor field'') as

\begin{description}
\item[$(\star)$]
${\sf ds}^2=\frac{2}{\omega^2}[-{\sf dt}^2+{\sf dr}^2+{\sf dz}^2 +
(\sinh^2r-\sinh^4r){\sf d\varphi}^2 + 2\sqrt{2}\sinh^2r{\sf
d\varphi}{\sf  dt}]\,.$
\end{description}
 This is understood in the
{\em cylindric-polar coordinates} $\langle t^d, r^d, \varphi^d,
z^d\rangle$ of the dervish world we discussed in
Sections~\ref{dervish-section},\ref{technical-section}. Cf.\
Figure~\ref{koor-fig}. Instead of $\frac{2}{\omega^2}$, G\"odel
writes $4a^2$ but in our notational system these two constants are
basically the same. (One can interpret G\"odel's $a$ as
$a=\frac{1}{\sqrt{2}}\omega$.%
\footnote{Cf.\ item (9) on p.191 in G\"odel~\cite{Gcwii}.} Anyway,
$a$ and $\omega$ are only ``parameters''.) Other differences are
that G\"odel used the $+---$ sign-convention and we also made a
$\varphi\to-\varphi$ coordinate transformation so as to use the same
form of G\"odel's metric that Lathrop-Teglas~\cite{Lath} uses. In
tensorial form, $(\star)$ can be written by specifying that
G\"odel's metric tensor field $\frac{2}{\omega^2}{\sf g}$ is defined
by

\begin{description}
\item[]
${\sf g}_{tt} = 1\,,\qquad {\sf g}_{rr} = -1$\,,\qquad
${\sf g}_{\varphi\varphi} = (\sinh^4r - \sinh^2r)$\,,\qquad ${\sf
g}_{\varphi t}=\sqrt{2}\sinh^2r$\,,\qquad ${\sf g}_{zz}=-1$\,,
\item[]
${\sf g}_{t\varphi}={\sf g}_{\varphi t}$, and the rest of the ${\sf
g}_{ij}$'s are 0.
\end{description}

Clearly, ${\sf g}(p)$ is a function of $p=\langle
t,r,\varphi\rangle$, but only ${\sf g}_{\varphi\varphi}$ and ${\sf
g}_{\varphi t}$ depend on $p$\,. Further, of the parts of $p$, they
depend only on $r_p$ and on $\varphi_p$\,. This is caused by the
symmetries of our space-time, i.e.\ rotation along $\varphi$ and
translation along $t$ are automorphisms of GU (both for all versions
of GU herein as well as in G\"odel's
quoted%
\footnote{There are papers of G\"odel in which these symmetries fail
(for rotating universes), cf.\ e.g.\ \cite[p.208]{Gcwii}.}
 papers).
Notice that in the Naive Dervish World, both {\sf
g}$_{\varphi\varphi}$ and {\sf g}$_{t\varphi}$ tend to constants as
$r$ tends to infinity while in G\"odel's Dervish World they both
tend to infinity as $r$ tends to infinity. This is why we refined
our Naive GU to obtain the Tilted GU.

Lathrop-Teglas~\cite{Lath} presents G\"odel's universe in so-called
Fermi coordinates. This means that the t axis as well as the radial
lines are geodesics and the gyroscopes (i.e., compasses of inertia)
of $m_0$ are not rotating. This is a spiral world where the cosmic
compasses are replaced with compasses of inertia. It is very similar
to Refined (Choice 2) Spiral World depicted in
Figure~\ref{ujvis1-fig}. Indeed, \cite{Lath} obtains this metric
from $(\star)$ above by the following coordinate transformation.
Below $t',r',z',\varphi'$ are the new coordinates, $t,r,z,\varphi$
are the coordinates used in ($\star$) and
$c=\frac{\sqrt{2}}{\omega}$.

$$t'=ct,\qquad r'=cr,\qquad z'=cz,\qquad \varphi'=\omega
t'-\varphi\,.$$

This is the transformation from forward tilted (Choice 1) Dervish
World to backward tilted (Choice 2) Spiral World (apart from
multiplying with a constant $c$). From now on, for simplicity, we
write $t,r,\varphi,z$ for $t',r',\varphi',z'$.
\newcommand{\sh}{\mbox{sh}}
\newcommand{\ch}{\mbox{ch}}
Let us use the notation
$$\sh=\sinh(\frac{\omega}{\sqrt{2}}\,r)\qquad\mbox{and}\qquad
\ch=\cosh(\frac{\omega}{\sqrt{2}}\,r)\,.
$$ Now, the ``linear element'' (i.e.\ the ``metric tensor field'') of
G\"odel's universe in Fermi coordinates is

$${\sf ds}^2=-(1+2\sh^2\ch^2){\sf dt}^2+{\sf dr}^2+{\sf dz}^2 +
\frac{2}{\omega^2}\,\sh^2(1-\sh^2){\sf d\varphi}^2 +
\frac{4}{\omega}\,\sh^4{\sf d\varphi}{\sf  dt}\,.$$

\goodbreak
The nonzero Christoffel symbols ${\Gamma}^i_{jk}$ are

$$
{\Gamma}^{r}_{tt} = \omega\sqrt{2}\sh\ch((2\ch^2-1)\,,\qquad
{\Gamma}^{t}_{tr} = \omega\sqrt{2}\sh\ch\,,\qquad
{\Gamma}^{\varphi}_{tr} = \omega^2\sqrt{2}\sh\ch\,,
$$
$${\Gamma}^{r}_{t\varphi} = -2\sqrt{2}\sh^3\ch\,,\qquad
{\Gamma}^{t}_{r\varphi} = \frac{\sqrt{2}sh^3}{\ch}\,,\qquad
{\Gamma}^{\varphi}_{r\varphi} =
\frac{-\omega(2\ch^4-4\ch^2+1)}{\sqrt{2}\sh\ch}\,,\qquad
$$
$${\Gamma}^r_{\varphi\varphi} =
\frac{\sqrt{2}\sh\ch(2\ch^2-3)}{\omega}\,,\qquad \mbox{ and the
${\Gamma}^i_{kj}={\Gamma}^i_{jk}$ for the nonzero ${\Gamma}^i_{jk}$
listed above}.
$$

\noindent The scalar curvature is
$$R = 2\omega^2\,. $$

\bigskip A sample of papers investigating G\"odel's universe is
Chakrabarti-Geroch-Liang~\cite{Chak},
Chandrasekhar-Wright~\cite{Chan}, Dorato~\cite{Dorato},
G\"odel~\cite{Go49}, \cite{Go96}, Heckmann-Sch\"ucking~\cite{Heck},
Kundt~\cite{Kundt}, Lathrop-Teglas~\cite{Lath},
Malament~\cite{Mal84}, Obukhov~\cite{Obuk}, Plaue-Scherfner-de
Sousa~\cite{PlaueSS}, Sklar~\cite{Sklar}, Stein~\cite{Stein}. A
sample of books about general relativity and time (especially
relevant to the present paper) is Earman~\cite{Earm},
Gibilisco~\cite{Gib}, Gott~\cite{Gott}, Horwich~\cite{Horw},
Novikov~\cite{Novi}, O'Neil~\cite{O95}, Pickover~\cite{Picktime},
Yourgrau~\cite{Yourg}.

For more on the drag effect and its connections with Mach's
principle cf.\ e.g.\ Wald~\cite[p.89 item 3.(c), p.187 Problem 3(b),
p.319 immediately below item (12.3.17)]{Wal}. For more detail on
``drag'' and Mach cf.\ Misner-Thorne-Wheeler~\cite[\S 21.12
(entitled ``Mach's...'') and especially pp.546-548, also item B on
p.879, pp.1117, 699, 893, 1120]{MTW}. Cf.\ also d'Inverno~\cite[\S
9.2 (pp.121-124)]{Din}, Gibilisco~\cite[pp.19-123 (subtitle: Alone
in the universe)]{Gib}. Cf.\ also \cite[pp.880-1]{MTW} for nice
drawings of rotating black holes.

\label{drag-p} For the gravitational drag effect we refer to
Rindler~\cite[pp.10-13, \S\S 1.15, 1.16]{Rin}, Wald
\cite[pp.9,71,89,183,319]{Wal}, Wald~\cite[pp.32-33]{Wal77},
together with Misner-Thorne-Wheeler[\S 40.7 (pp.1117-1120), \S 33.4
(p.892), \S 21.12 (in particular p.547), p.1120 (footnote)]{MTW}.
The gravitational drag effect is related to Mach's principle as is
explained e.g.\ in \cite[\S 21.12]{MTW} and in \cite[\S 1.15 (e.g.\
p.12)]{Rin}.

Figure~\ref{ujgodel-fig} is a slightly corrected version of
Figure~31 in \label{corr-p} Hawking-Ellis~\cite{Hawel}. This picture
can also be found in Yourgrau~\cite{Yourg}.
Malament~\cite[p.99]{Mal84} pointed out that the light-cones on that
figure are tilted so much that they do not contain the vertical
lines which are the life-lines of the distinguished observers in the
dervish-world (which the figure represents). Below we include the
Figure from Malament's paper (in which the light-cones are corrected
already).

\begin{figure}[hbtp]
\setlength{\unitlength}{0.68 truemm} %
\begin{center}
\begin{picture}(209,93)(0,0) 
\epsfysize = 93\unitlength   
\epsfbox{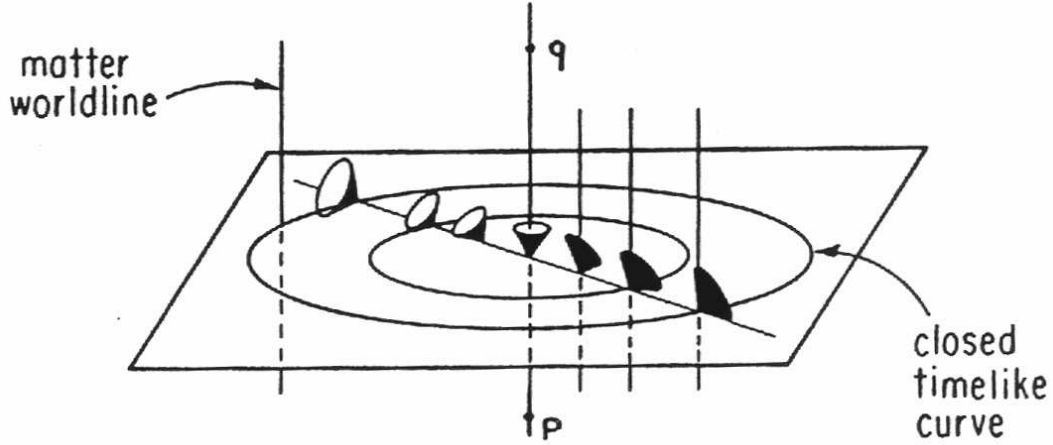}
\end{picture}
\end{center}
\caption{\label{Malament-fig} Figure from Malament's paper
\cite{Mal84}.}
 \end{figure}

The present work is part of a broader effort for what we could
bluntly call demystifying general relativity theory and its
relatives like wormhole-theory and cosmology. More concretely, we
try to provide a purely logic based conceptual analysis for general
relativity and its relatives. One of the aims is to provide a
technically correct but easily understandable introduction to
general relativity including its most exotic reaches for the
questioning mind of the nonspecialist. A sample of works in this
general direction is \cite{AMN07}, \cite{ANW08}, \cite{Maddis},
\cite{Szdis}, \cite{SzGVC08}.\bigskip

\noindent {\bf Acknowledgements.} Thanks are due to too many people
for their encouragement and help to be listed here. Special thanks
go to Mark Hogarth, Endre Szab\'o and Csaba T\H oke. We want to
express very special thanks to Csilla N\'emeti who gave invaluable
help in getting this project going, e.g.\ she put a lot of energy,
enthusiasm, and care into drawing, discussing, and redrawing the
first versions of the first few figures of this project. Research
supported by National Fund for Basic Research OTKA No.\ 73601. A.\
Andai was supported by Japan Society for the Promotion of Science,
contract number P 06917.
\bigskip

\hfill\eject

\section{Appendix: technical details for the constructions.}
\label{technical-section}

\noindent {\bf Connections between our spiral coordinate system}
$\langle t,x,y,z\rangle=\langle t^s,\dots,z^s\rangle$ {\bf and
co-rotating (dervish) coordinate system} $\langle t',x',y',z'\rangle
= \langle T^r,X^r, Y^r, Z^r\rangle$:
\bigskip
\bigskip
\label{coord1}

By definition, $t'=t$ and $z'=z$. Throughout we suppress the
irrelevant spatial coordinate $z$. Below, instead of the Cartesian
systems $\langle t,\dots,y\rangle,\langle t',\dots,y'\rangle$ we use
their cylindric-polar-coordinates variants $\langle
t,\varphi,r\rangle$
and $\langle t',\varphi',r'\rangle$.%
\footnote{Cf.\ e.g.\ d'Inverno~\cite[Fig.19.2 (p.253)]{Din}.} The
connections are the usual standard ones, e.g.\ $r=\sqrt{x^2+y^2}$,
$y=r\cdot\cos(\varphi)$, $x=r\cdot\sin(\varphi)$,
$\varphi=\arctan(x/y)$. In more detail, $r(p)=\sqrt{x(p)^2+y(p)^2}$
etc. $\langle t^s,\varphi^s,r^s\rangle:=\langle t,\varphi,r\rangle$
and $\langle T^r,\varphi^r,r^r\rangle=\langle
t^{der},\varphi^{der},r^{der}\rangle=\langle t',\varphi',r'\rangle$.
Here $s$ abbreviates ``spiral'' and ``der'' abbreviates ``dervish''.

\begin{figure}[!hp]
\setlength{\unitlength}{0.7 truemm} \small
\begin{center}
\begin{picture}(240,280)(0,0)

\put(92,144){\makebox(0,0)[lb]{$t^{\rm d}(p)=t^{\rm s}(p)$}}
\put(160,123){\makebox(0,0)[l]{$p$}}
\put(76,71){\makebox(0,0)[rb]{$1_{r}^{\rm d}=1_{r}^{\rm s}$}}
\put(88,80){\makebox(0,0)[rb]{$\bar 0$}}
\put(157,60){\makebox(0,0)[lt]{$\f^{\rm s}(p)$}}
\put(88,48){\makebox(0,0)[lt]{$\f^{\rm d}(p)$}}
\put(70,42){\makebox(0,0)[t]{$\f^{\rm d}(p)$}}
\put(110,34){\makebox(0,0)[l]{$\f^{\rm s}(p)$}}
\put(176,138){\makebox(0,0)[l]{\shortstack[l]{life-line of $m_i$\\(a
galaxy)}}} \put(86,160){\makebox(0,0)[r]{$1_{t}^{\rm s}=1_{t}^{\rm
d}$}} \put(95,280){\makebox(0,0)[lt]{$t^{\rm s}=t^{\rm d}$}}
\put(26,15){\makebox(0,0)[l]{\shortstack[l]{direction of expected
                                             rotation of \\cosmic compasses\\
in the $\la t^{\rm d},\f^{\rm d},r^{\rm d}\ra$
 coordinate system}}}

\put(160,15){\makebox(0,0)[l]{\shortstack[l]{direction of rotation
of universe\\ (i.e. of
                                            distant galaxies) w.r.t.\\
                                     cosmic compasses i.e.\ in $\la t^{\rm s},
\f^{\rm s},r^{\rm s}\ra$}}}
\put(2,278){\makebox(0,0)[lt]{\shortstack[l]{View \underbar{from}
the
\underbar{spiral coordinate}\\
\underbar{system} $\la t^{\rm s},\f^{\rm s},r^{\rm s}\ra$:}}}
\put(122,184){\makebox(0,0)[b]{\framebox{\large $\f^{\rm d}=0$}}}
\put(35,126){\makebox(0,0)[r]{\framebox{\normalsize $\f^{\rm
s}=0$}}} \put(182,95){\makebox(0,0)[l]{$t^{\rm s}(p)$}}
\put(198,40){\makebox(0,0)[r]{$\f^{\rm d}(p)=\f^{\rm
s}(p)-\omega\cdot t^{\rm s}(p)$}}

\put(132,57){\makebox(0,0)[b]{$1_{\f}^{\rm s}$}}
\put(138,138){\makebox(0,0)[b]{$\omega$}}
\put(35,35){\makebox(0,0)[r]{$r$}} \put(53,236){\makebox(0,0)[r]{the
$\f^{\rm s}=0$ plane}} \epsfysize = 280  \unitlength
\epsfbox{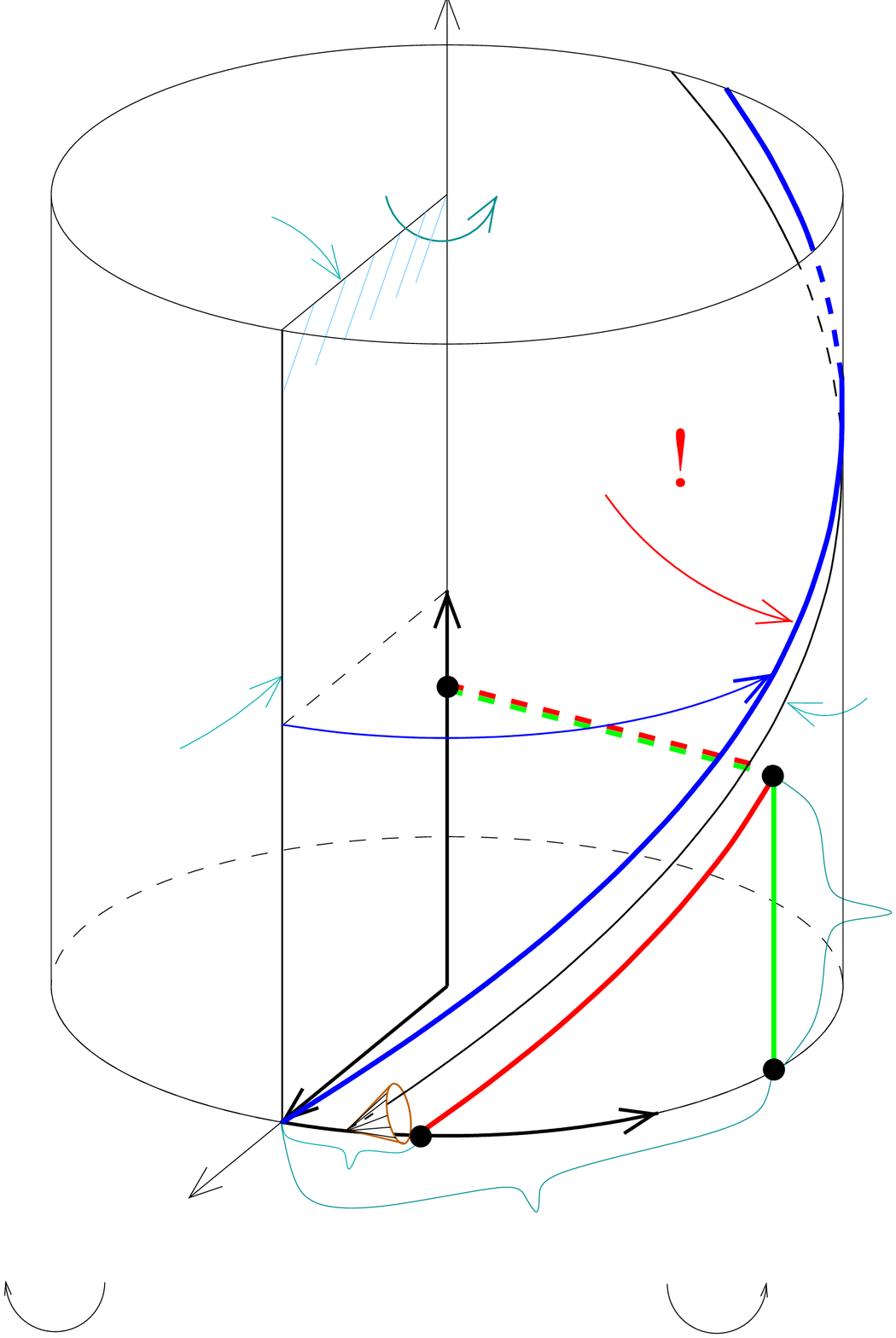}
\end{picture}
\end{center}
\caption{\label{koor-fig}
 As throughout this work, here too, the irrelevant spatial coordinates
$z^{\rm d}=z^{\rm s}=z_{\rm i}=z$ are suppressed.}
 \end{figure}

\begin{figure}[!hp]
\setlength{\unitlength}{0.069 truemm} \small  
\begin{center}
\begin{picture}(2420,2980)(0,0)

\put(1270,1300){\makebox(0,0)[rb]{$p$}}
\put(20,2960){\makebox(0,0)[lt]{\shortstack[l]{View \underbar{from}
the \underbar{dervish
coordinate}\\
\underbar{system} $\la t^{\rm d},\f^{\rm d},r^{\rm d}\ra$:}}}
\put(1380,2970){\makebox(0,0)[lt]{$t_0=t^{\rm d}=t^{\rm s}$}}
\put(1620,2560){\makebox(0,0)[l]{\shortstack[l]{direction of rotation\\
                                                 of dervishes i.e.\ of\\
                                                 cosmic compasses}}}

\put(1590,1760){\makebox(0,0)[l]{\framebox{\shortstack[l]{a dervish
co-rotating  \\ with ``$\f^{\rm s}=0$ surface''\\ i.e. with spiral\\
coordinate system}}}}

\put(340,1190){\makebox(0,0)[r]{\framebox{\large $\f^{\rm s} =0$}}}
\put(870,1460){\makebox(0,0)[r]{\framebox{\normalsize $\f^{\rm
d}=0$}}} \put(1145,1000){\makebox(0,0)[r]{$t^{\rm d}(p)=t^{\rm
s}(p)$}} \put(1450,850){\makebox(0,0)[l]{$1_{r}^{\rm d}=1_{r}^{\rm
s}$}} \put(1790,760){\makebox(0,0)[lb]{$1_{\f}^{\rm d}$}}
\put(2040,780){\makebox(0,0)[lt]{$\f^{\rm s}(p)$}}
\put(1310,660){\makebox(0,0)[lt]{$\f^{\rm d}(p)$}}
\put(820,540){\makebox(0,0)[r]{$r$}}
\put(1150,560){\makebox(0,0)[t]{$\f^{\rm d}(p)$}}
\put(1560,430){\makebox(0,0)[t]{$\f^{\rm s}(p)$}}
\put(260,400){\makebox(0,0)[l]{\shortstack[l]{direction of \\
                                            rotation of \\
                                        cosmic compasses
in the \\$\la t^{\rm d},\f^{\rm d},r^{\rm d}\ra$
 coordinate system}}}
\put(260,120){\makebox(0,0)[l]{\shortstack[l]{direction of rotation
                                              of universe\\
        (i.e.\ of distant galaxies) w.r.t.\ cosmic compasses \\ i.e.\
in $\la t^{\rm s},\f^{\rm s},r^{\rm s}\ra$}}}

\put(1270,2810){\makebox(0,0)[r]{\large $x_0$}}
\put(1410,2820){\makebox(0,0)[lb]{\large $y_0$}}
\put(1280,1890){\makebox(0,0)[r]{\large $y_0$}}
\put(1320,1850){\makebox(0,0)[rt]{\large $x_0$}}
\put(1360,1080){\makebox(0,0)[r]{$t_0$}}
\put(1450,965){\makebox(0,0)[l]{\large $x_0$}}
\put(1330,940){\makebox(0,0)[t]{\large $y_0$}}
\put(1380,1740){\makebox(0,0)[lb]{$1_t$}}
\put(1370,1545){\makebox(0,0)[lb]{$m_0$}}
\put(1370,2545){\makebox(0,0)[lb]{$m_0$}}
\put(920,1800){\makebox(0,0)[r]{life-line of $m_i$}}
\put(1800,500){\makebox(0,0)[l]{$\f^{\rm s}(p)=\f^{\rm
d}(p)+\omega\cdot t^{\rm d}(p)$}}

\put(760,1600){\makebox(0,0)[lb]{$-\omega$}}

\epsfysize =2980   \unitlength \epsfbox{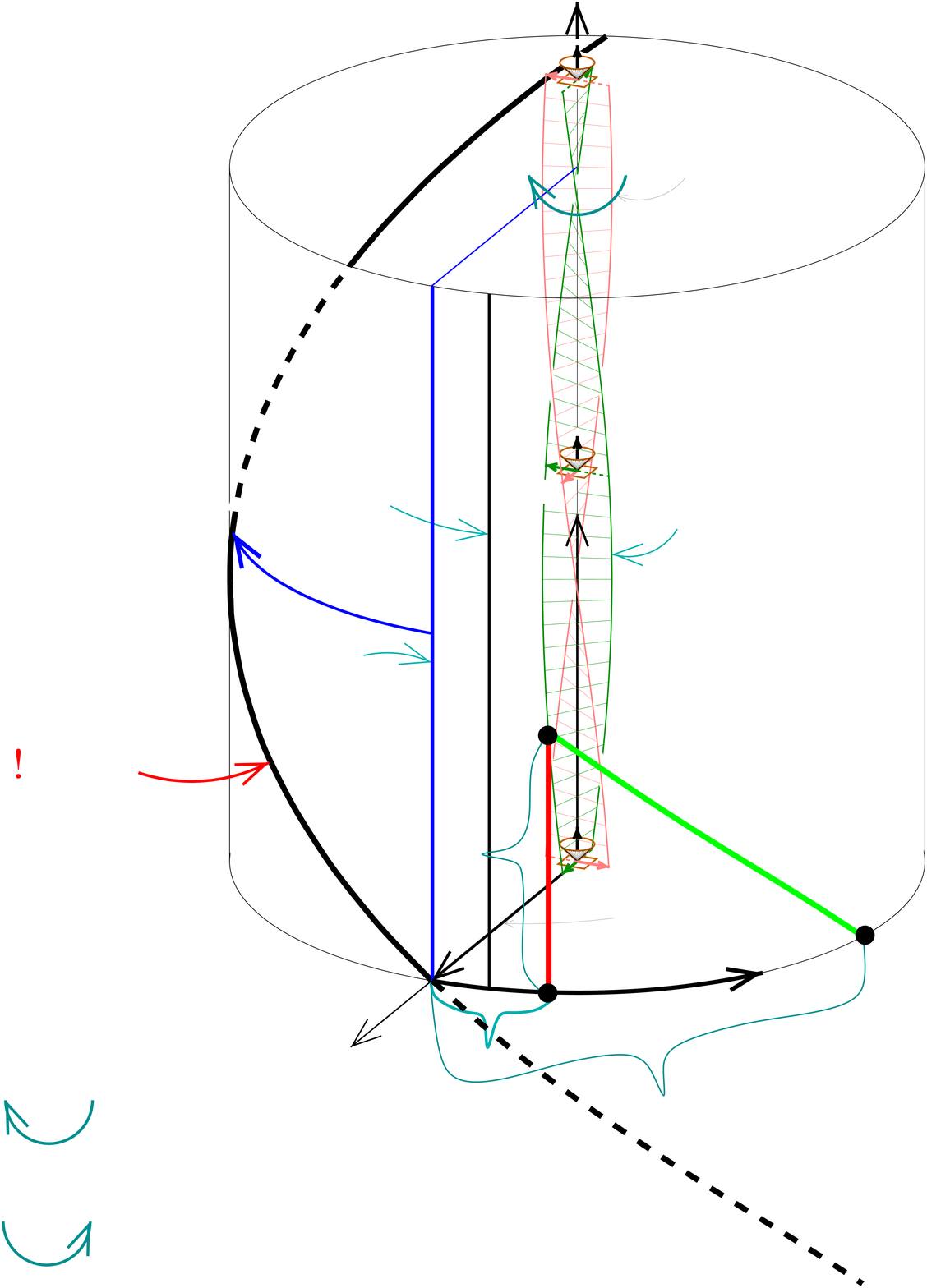}
\end{picture}
\end{center}
\caption{\label{koor5-fig}  Dervish view of spiral world, i.e.\
backward transformation $\la t^{\rm der},\ldots\ra\longrightarrow\la
t^{\rm spi},\ldots\ra$. Notice that the $t=0$ plane in this figure
coincides with that of previous figure (e.g.\ marked points are the
same on the two).}
 \end{figure}

\begin{figure}[!hp]
\setlength{\unitlength}{0.57 truemm} \small
\begin{center}
\begin{picture}(280,370)(0,0)
\epsfysize =   370\unitlength \epsfbox{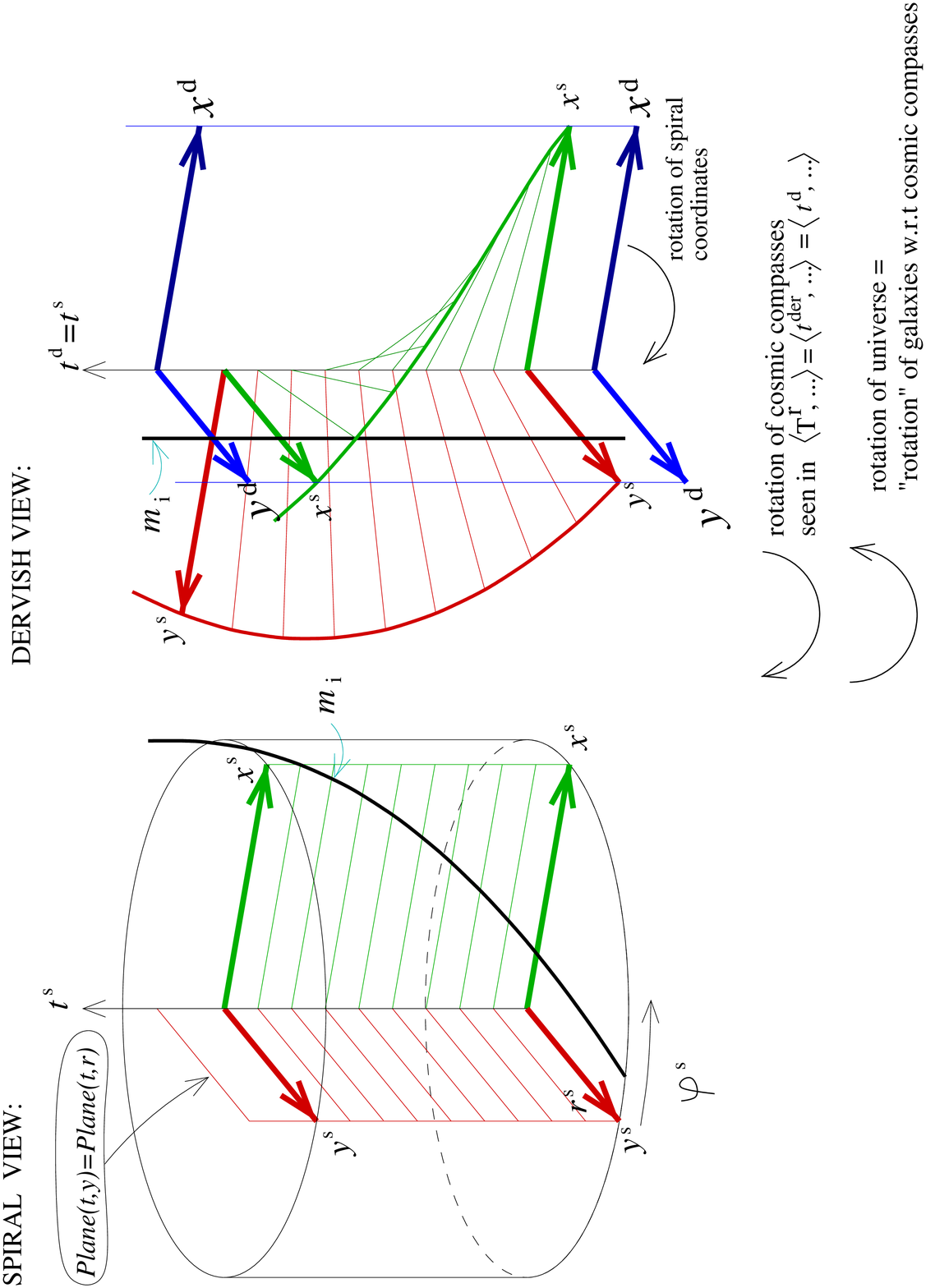}
\end{picture}
\end{center}
\caption{\label{koor4-fig} \label{coord2} }
 \end{figure}

The ``galaxies'' $m_1, m_2,\dots, m_i$ appear as rotating around
$m_0$ in direction $\varphi$ with angular velocity $\omega$ in
$\langle t^s,\dots\rangle$ while their cosmic compasses $x_i, y_i$
appear fixed (non rotating). As a contrast, $\langle
T^r,\dots\rangle$ shows $m_1,\dots, m_i$ as motionless, while it
shows their cosmic compasses
 as rotating in direction $-\varphi$ with angular velocity
$\omega$.  We use $p$ to denote an arbitrary point which has
coordinates $t(p), \varphi(p), r(p)$ etc. We represent these simple
connections in Figures~\ref{koor-fig}--\ref{koor4-fig}. As we said,
we suppress coordinate $z$. In Figure~\ref{koor-fig} below
(p.\pageref{koor-fig}) we regarded only such points $p$ which are on
the cylinder $r(p)=1$. Generalizing to arbitrary points is trivial
since $r$ does not change. As it is obvious from the picture, the
transformation ``spiral'' $\mapsto$ ``dervish'' is

\begin{description}
\item[]
$\varphi^d(p) = \varphi^s(p) - \omega\cdot t^s(p)$
\item[]
$r^d(p) = r^s(p)$
\item[]
$t^d(p) = t^s(p)$
\item[]
$z^d(p) = z^s(p)$.  Clearly,
\item[]
$\varphi^s(p) = \varphi^d(p) + \omega\cdot t^d(p)$.
\end{description}

The angular velocity of the rotation of the universe as seen by
$\langle t^s,\dots\rangle$ is $\omega$.

\newpage

\vfill\eject\newpage

\vfill\eject\newpage

\vfill\eject\newpage


\begin{figure}[!hbtp]
\setlength{\unitlength}{0.3 truemm} \small
\begin{center}
\begin{picture}(540,680)(0,0)

\epsfysize =  680 \unitlength \epsfbox{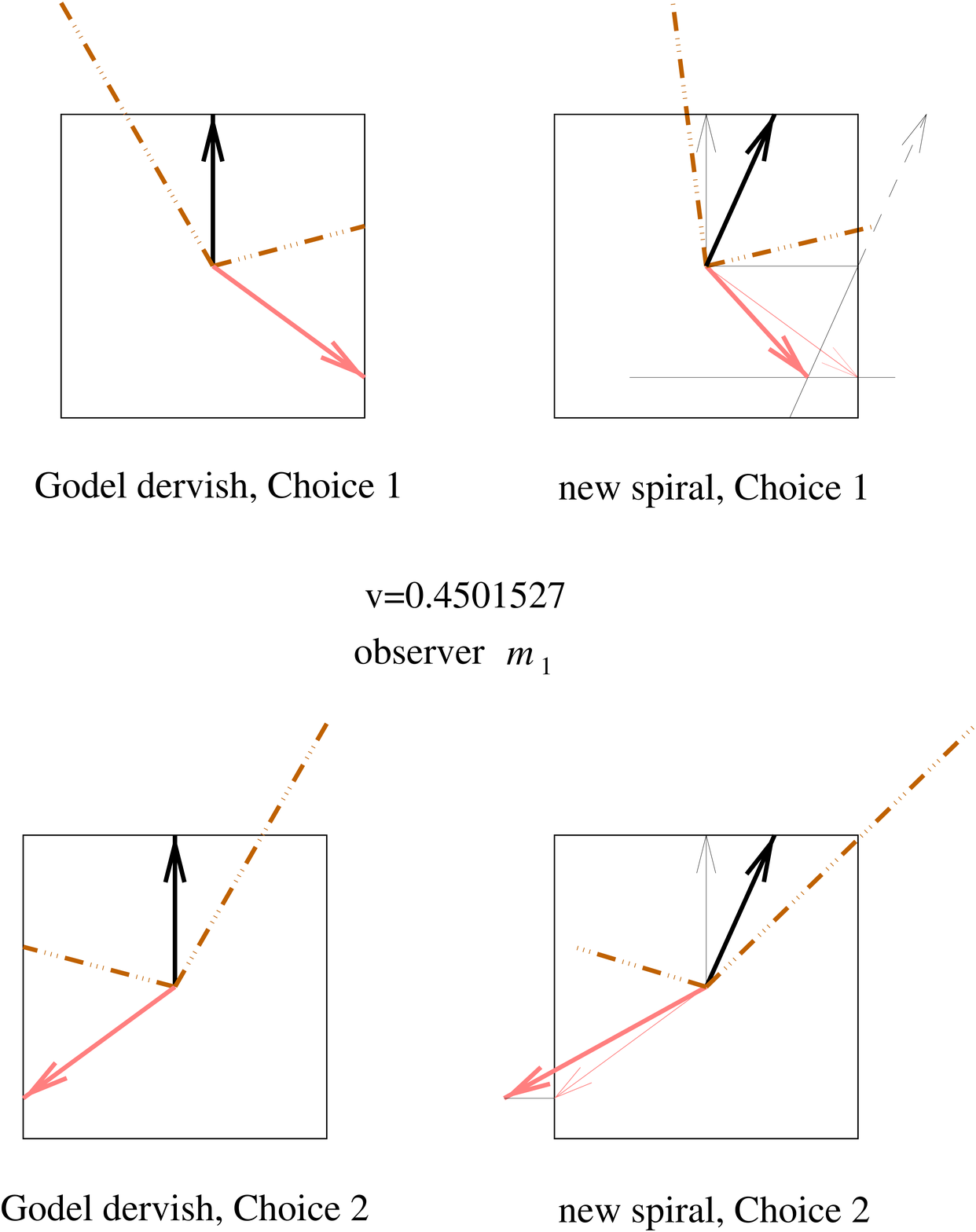}
\end{picture}
\end{center}
\caption{\label{1negy-fig} Details of observer $m_1$.}
 \end{figure}

\vfill\eject\newpage


\begin{figure}[!hbtp]
\setlength{\unitlength}{0.3 truemm} \small
\begin{center}
\begin{picture}(540,700)(0,0)

\epsfysize = 700  \unitlength \epsfbox{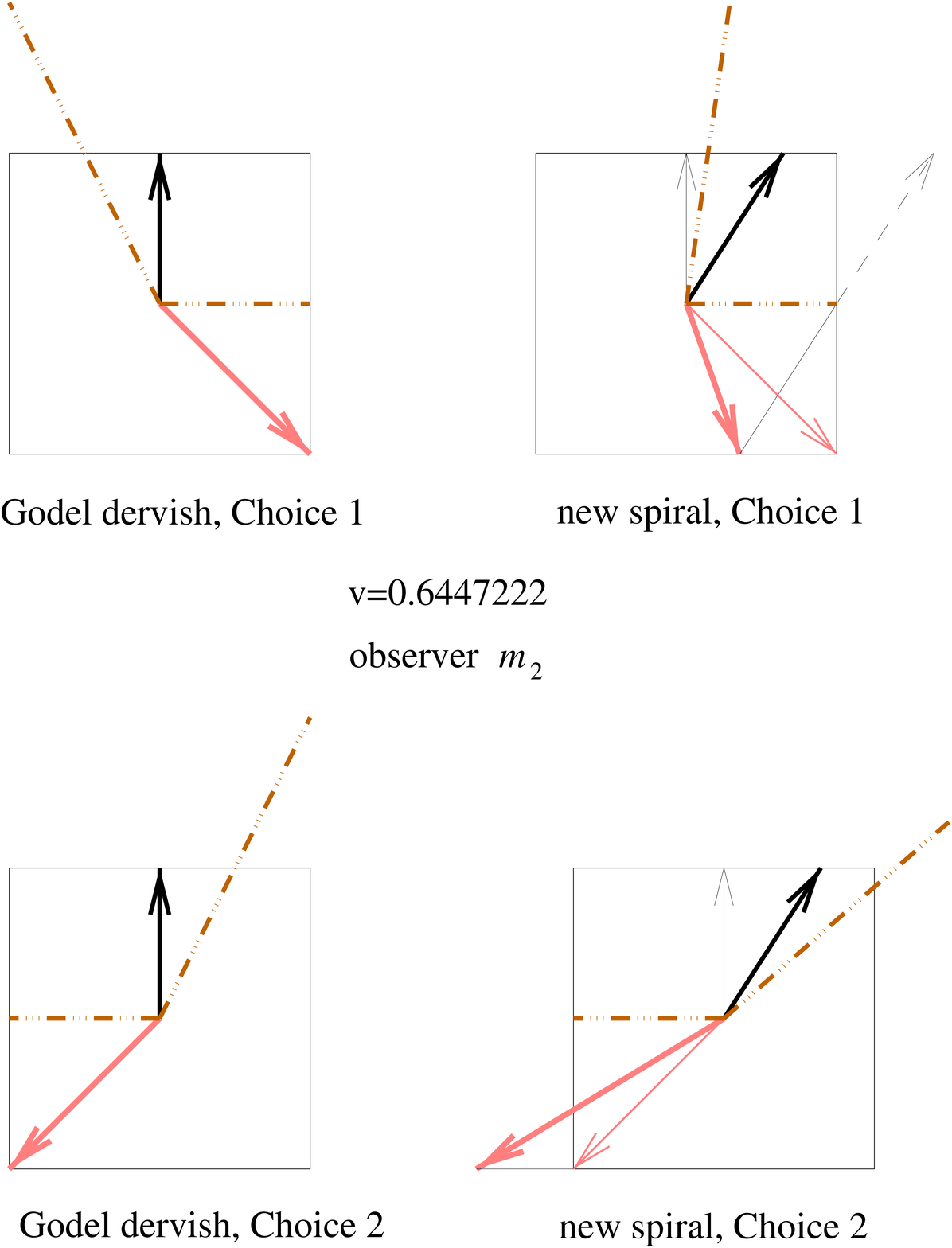}
\end{picture}
\end{center}
\caption{\label{10negy-fig} Details of observer $m_2$. }
 \end{figure}

\vfill\eject\newpage


\begin{figure}[!hbtp]
\setlength{\unitlength}{0.24 truemm} \small
\begin{center}
\begin{picture}(680,880)(0,0)

\epsfysize = 880  \unitlength \epsfbox{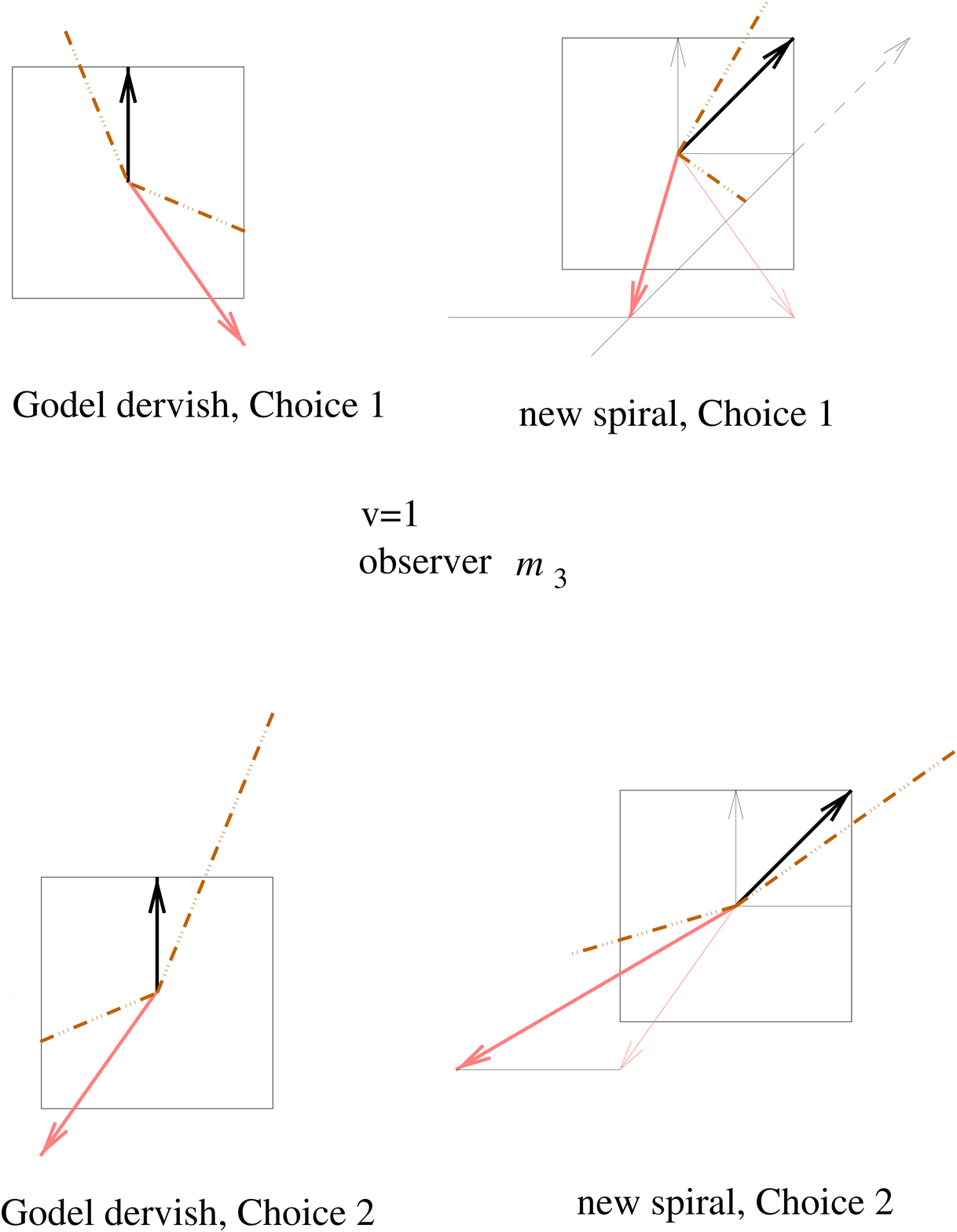}
\end{picture}
\end{center}
\caption{\label{unegyzet-fig} Details of observer $m_3$.}
 \end{figure}

\vfill\eject\newpage


\begin{figure}[!hbtp]
\setlength{\unitlength}{0.248 truemm} \small 
\begin{center}
\begin{picture}(680,720)(0,0)

\epsfysize =  720 \unitlength \epsfbox{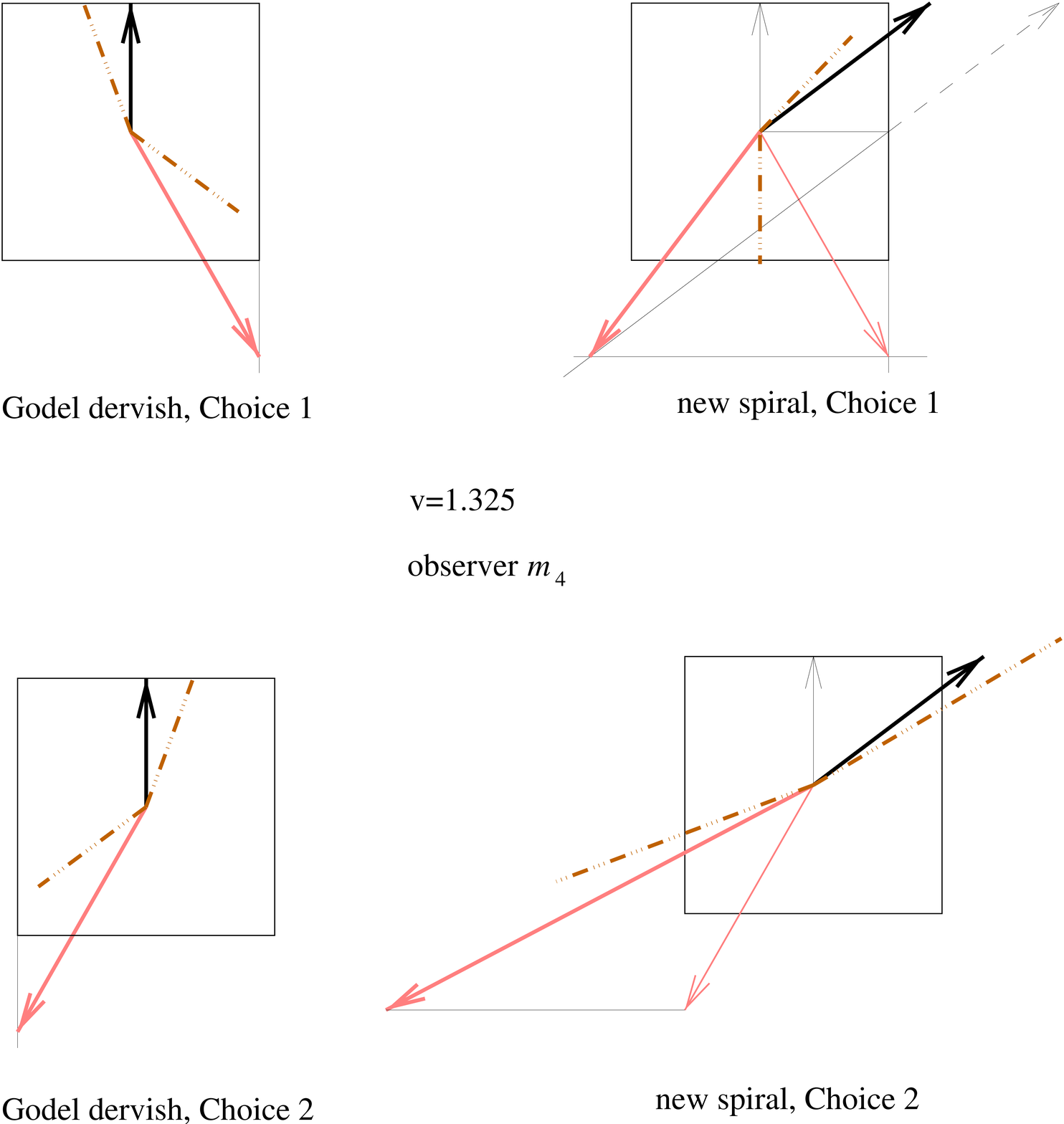}
\end{picture}
\end{center}
\caption{\label{13negy-fig} Details of observer $m_4$.}
 \end{figure}

\vfill\eject\newpage


\begin{figure}[!hbtp]
\setlength{\unitlength}{0.2 truemm} \small
\begin{center}
\begin{picture}(840,960)(0,0)

\epsfysize =960   \unitlength \epsfbox{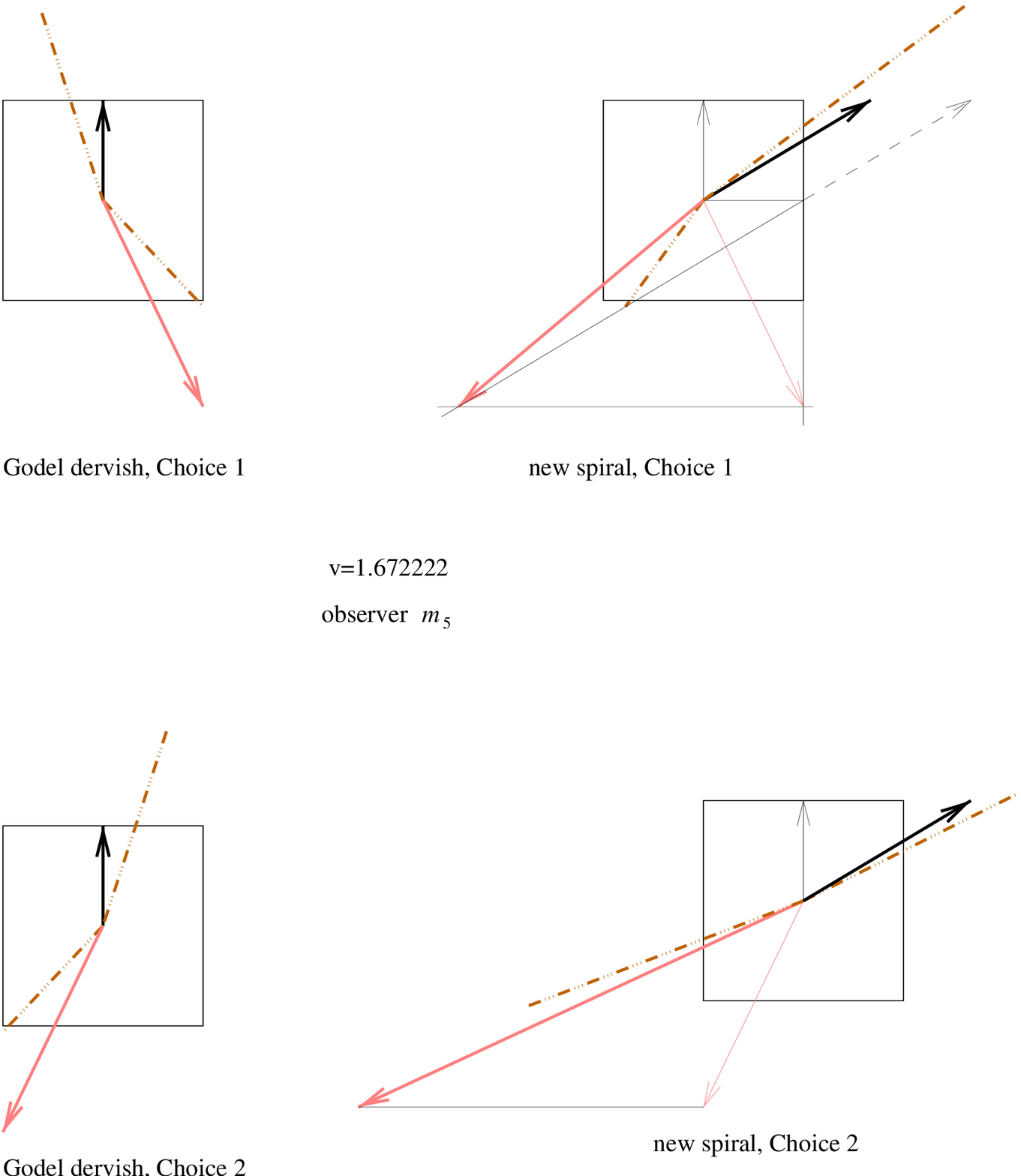}
\end{picture}
\end{center}
\caption{\label{3negy-fig} Details for observer $m_5$.}
 \end{figure}

\vfill\eject\newpage


\begin{figure}[!hbtp]
\setlength{\unitlength}{0.17 truemm} \small
\begin{center}
\begin{picture}(960,1200)(0,0)

\epsfysize =  1200 \unitlength \epsfbox{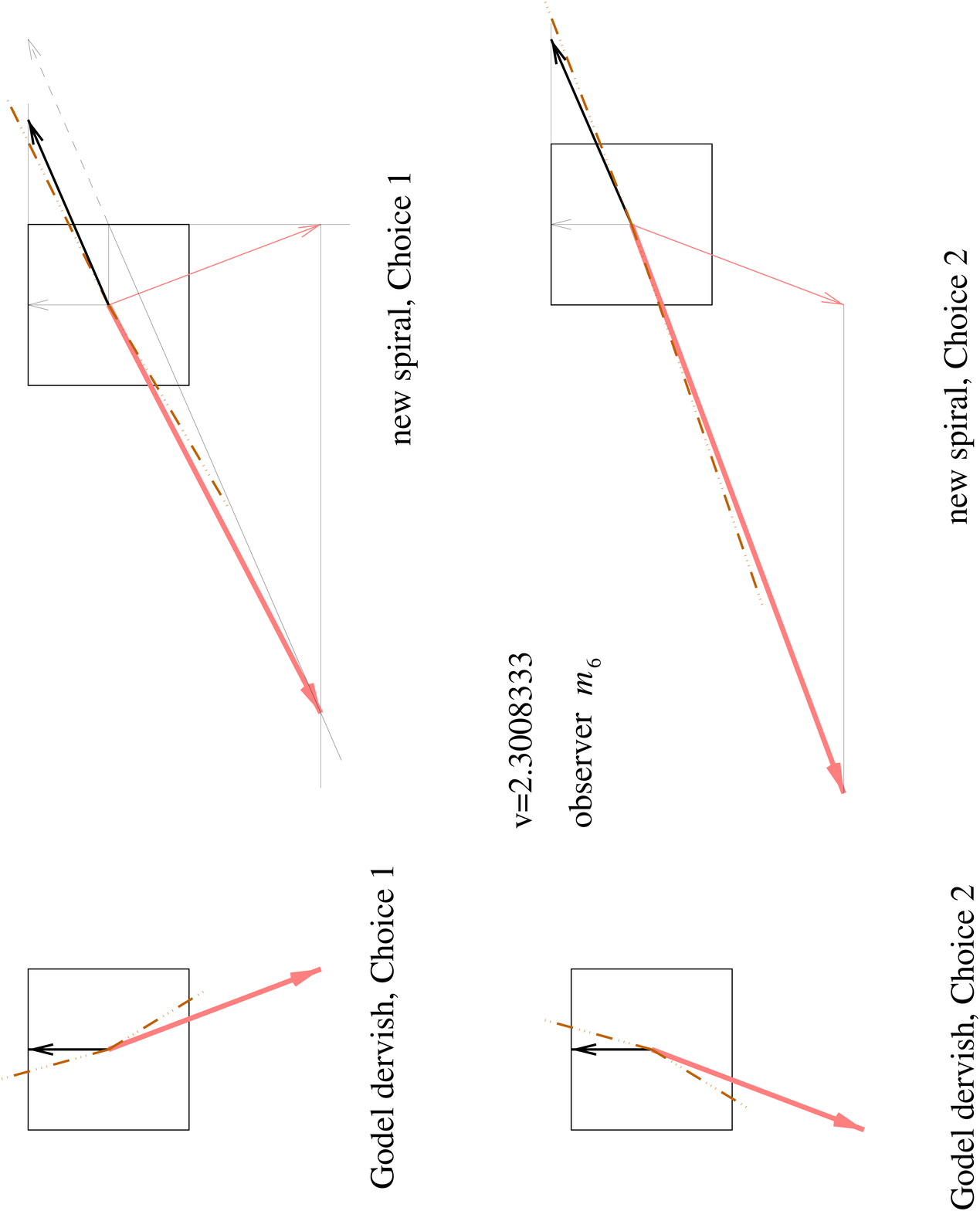}
\end{picture}
\end{center}
\caption{\label{4negy-fig} Details for observer $m_6$.}
 \end{figure}

\begin{figure}[!hp]
\setlength{\unitlength}{0.21 truemm} \small
\begin{center}
\begin{picture}(390,985)(0,0)
\epsfysize = 985  \unitlength \epsfbox{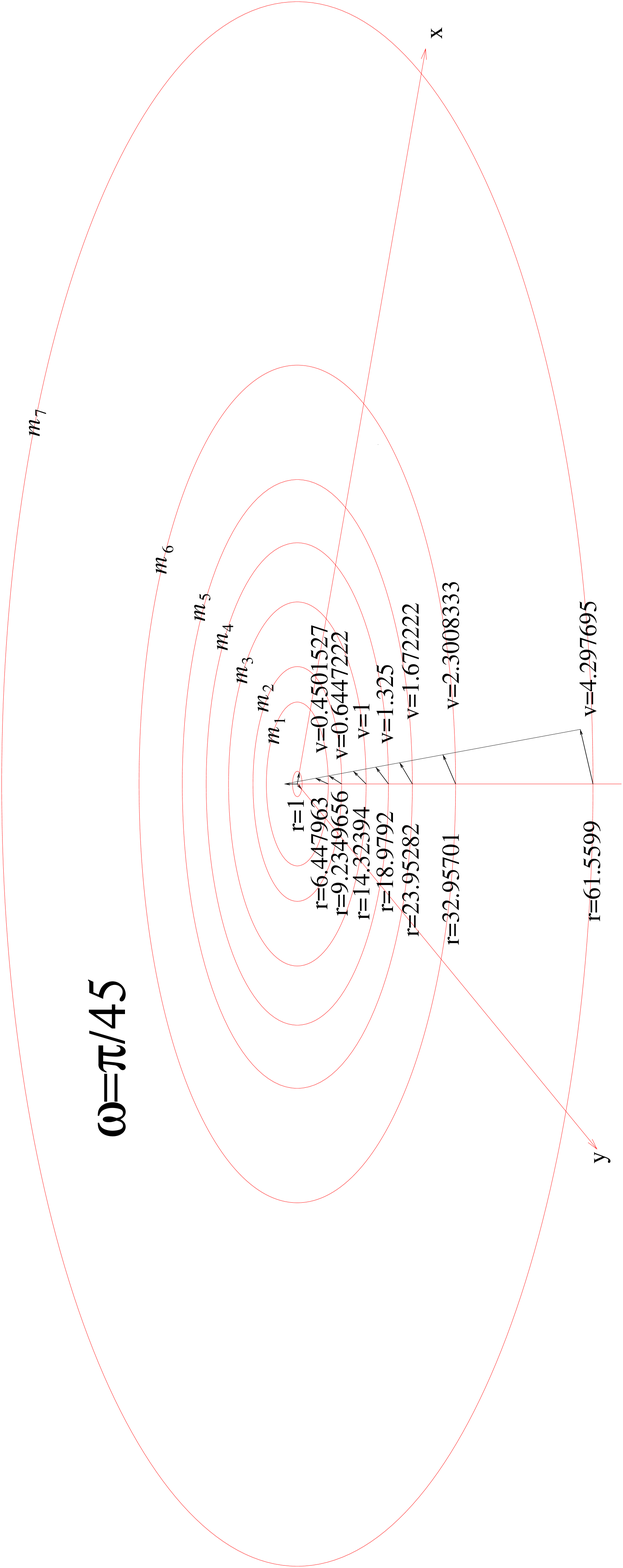}
\end{picture}
\end{center}
\caption{\label{map1-fig} Map 1}
 \end{figure}

\vfill\eject\newpage


\begin{figure}[!hp]
\setlength{\unitlength}{0.31 truemm} \small
\begin{center}
\begin{picture}(260,656)(0,0)
\epsfysize =656   \unitlength \epsfbox{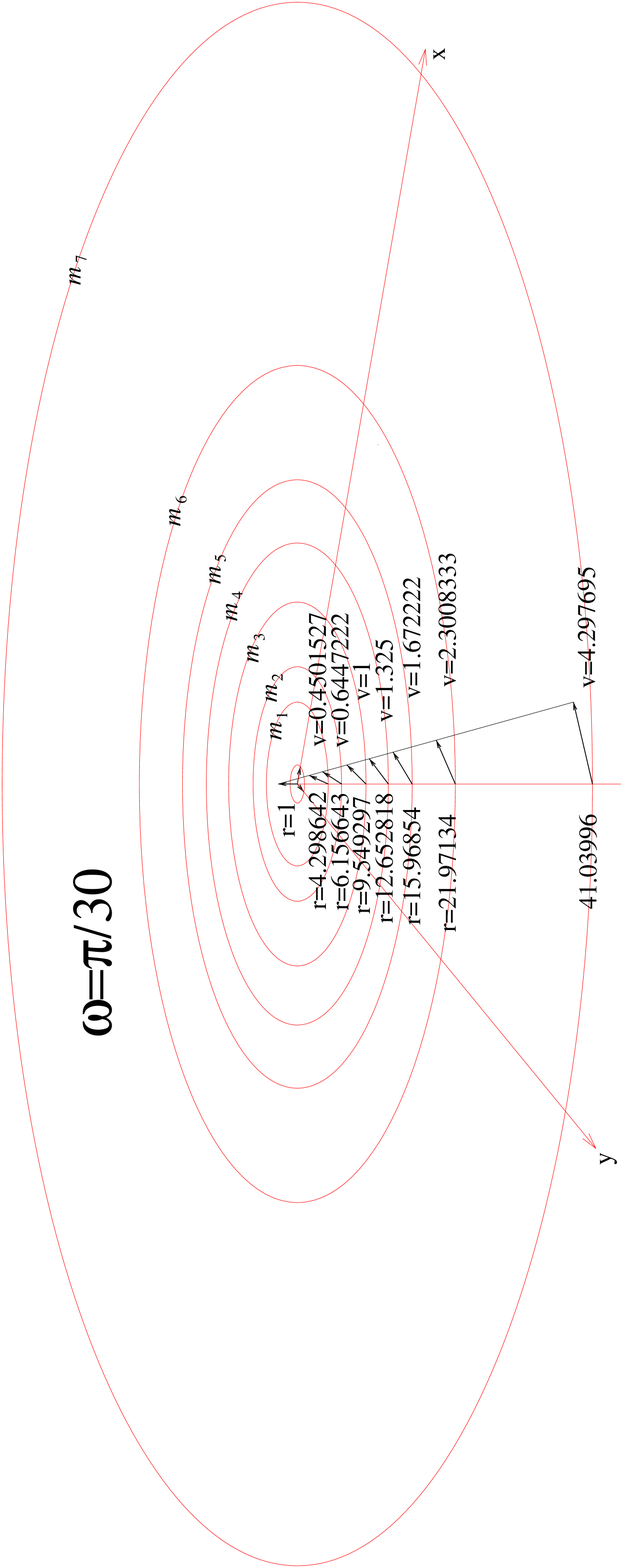}
\end{picture}
\end{center}
\caption{\label{map2-fig} Map 2}
 \end{figure}

\vfill\eject\newpage


\bibliographystyle{plain}
\label{references}
\bibliography{bibtex97-1}

\bigskip\bigskip
\noindent Addresses:

\noindent H.\ Andr\'eka, J.\ X.\ Madar\'asz and I.\ N\'emeti\\
R\'enyi Institute of Hungarian Academy of Sciences\\
Re\'altanoda street 13-15, H-1053 Hungary
\bigskip

\noindent A.\ Andai\\
Amari Research Unit, BSI, RIKEN\\
2-1 Hirosawa, Wako, Saitama, 351-0198 Japan.

\end{document}